\documentclass[prl,twocolumn,floatfix,epsfigs,nofootinbib]{revtex4}
\usepackage{amssymb,amsmath,amsfonts}
\usepackage{graphics,dcolumn,bm,fleqn,epic,eepic,float}
\usepackage{graphicx} 
\usepackage{epstopdf} 
\usepackage{bm} 
\usepackage{subfigure}
\usepackage{color} 

\newcommand{\RNum}[1]{\uppercase\expandafter{\romannumeral #1\relax}}

\begin{document}
\title{Behavioral gender differences are reinforced during the COVID-19 crisis}

\author{
Tobias Reisch$^{1,2,}\footnote{equal contribution}$,
Georg Heiler$^{1,2,*}$,
Jan Hurt$^{2}$,
Peter Klimek$^{1,2}$,
Allan Hanbury$^{2,3}$,
Stefan Thurner$^{1,2,4}\footnote{corresponding author}$
}

\affiliation{%
  $^1$~Section for Science of Complex Systems, Medical University of
  Vienna, Spitalgasse 23, 1090 Vienna, Austria}
   \affiliation{%
  $^2$~Complexity Science Hub Vienna, Josefst\"adterstrasse 39, 1080 Vienna, Austria}
 \affiliation{%
 $^3$~Institute of Information Systems Engineering, TU Wien, A-1040 Vienna, Austria}
 \affiliation{%
  $^4$~Santa Fe Institute, Santa Fe, NM 87501, USA}
\date{\today}

\begin{abstract} 
Behavioral gender differences are known to exist for a wide range of human activities including the way people communicate, move, provision themselves, or organize leisure activities. Using mobile phone data from 1.2 million devices in Austria (15\% of the population) across the first phase of the COVID-19 crisis, we quantify  gender-specific patterns of communication intensity, mobility,  and circadian rhythms. We show the resilience of behavioral patterns with respect to the shock imposed by a strict nation-wide lock-down that Austria experienced in the beginning of the crisis with severe implications on public and private life. We find drastic differences in gender-specific responses during the different phases of the pandemic. After the lock-down gender differences in mobility and communication patterns increased massively, while sleeping patterns and circadian rhythms tend to synchronize. In particular, women had fewer but longer phone calls than men during the lock-down. Mobility declined massively for both genders, however, women tend to restrict their movement stronger than men. Women showed a stronger tendency to avoid shopping centers and more men frequented recreational areas. After the lock-down, males returned back to normal quicker than women; young age-cohorts return much quicker. Differences are driven by the young and adolescent population. An age stratification highlights the role of retirement on behavioral differences. We find that the length of a day of men and women is reduced by one hour. We discuss the findings in the light of gender-specific coping strategies in response to stress and crisis.
\end{abstract}

\date{Oct 8, 2020}
\keywords{mobile-phone-data $|$  communication patterns  $|$ mobility $|$ coping-strategies $|$ pandemic $|$ circadian rhythm } 

\maketitle

\section{Introduction}
Gender differences exist for a wide spectrum of human behavior. Behavioral differences are manifest in communication behavior, visible for example in the different investment in biological offspring across women and men's lifetimes \cite{palchykov2012sex}. 
Gender differences in mobility patterns do rise from a mix of cultural, infrastructure, resource, safety and socio-economic factors \cite{gauvin2020gender}. Psychological and cognitive and other non-reproductive differences have been studied for many decades maybe even centuries, see e.g. \cite{halpern2013sex}. Also differences in stress perception and respective coping mechanisms have been known to exist for a long time \cite{ben1996gender,matud2004gender}.
Non-reproductive biological differences include women having shorter circadian rhythms \cite{duffy2011sex} and showing different co-morbidity patterns than men across their lifetimes \cite{chmiel2014spreading}. 
Even in virtual societies of online game players strong behavioral gender differences were found. In particular, male and female players tend to behave differently in economic activities, their dealing with aggression and hostilities, and generally how they structure their social networks \cite{szell2013how}.

In the last two decades it became possible to collect data on human behavior on a population-wide scale, see e.g. \cite{watts2007twenty}.
Some of that data has been used to investigate human responses to crisis and emergency situations \cite{bagrow2011collective,lu2012predictability, wang2014quantifying, garcia2019collective}. Studying collective response to crisis is essential for catastrophy planning and coordination \cite{petrescu2005emergent,Gao2020} and policy makers in health and safety \cite{Oliver2020}. 
Response to crisis also reveals human qualities that only surface when facing different kinds of actual or perceived danger \cite{cohen2020risk,taylor2000biobehavioral,ben1996gender,garcia2019collective}. 

Times of stress may change social norms and typical behavior. It is {\em a priori} not clear if and how these changes increase or decrease behavioral gender differences. One the one hand one might speculate that stress leads to a more universal behavior, where gender differences become less important and thus less pronounced \cite{wang2020psychological}. 
On the other hand one might also find signs of some types of evolutionary benefit if gender differences are amplified to cope better with crisis \cite{ben1996gender,galea2002psychological,juul2020gender}.
Population-wide stress can be seen as a shock to the system and 
The COVID-19 crisis serves as a natural experiment to investigate the impact of population-wide stress and its consequences for 
gender-specific changes in behavior.

The natural experiment can be used to measure the {\em resilience} of behavioral changes, i.e. how long it takes after the onset of a well-defined shock to get back to pre-crisis patterns of behavior. This characteristic time might also be important for a better objective understanding of temporal psychological effects of emergencies, which are usually studied using self-reported data at few or single points in time \cite{ben1996gender,galea2002psychological,matud2004gender,cohen2020risk}.

At the end of 2019 the SARS-CoV2 virus emerged in China, causing an ongoing, world-wide pandemic. In response to sharply rising numbers during the ``first wave'', on March 15\textsuperscript{th} the Austrian government introduced a severe nation-wide lock-down. 
The implemented non-pharmaceutical interventions (NPIs) included: 
school closures, 
restaurant closures,
mandatory use of masks, 
incentives to use home-office, 
the complete prohibition of gatherings of any size,
closure of all non-essential shops,
and a general limitation of mobility. 
It was only possible to leave the house for one of four reasons: 
work that cannot be postponed, shopping for groceries, 
helping people and short walks \cite{Desvars-larrive2020}.
These measures led to a massive reduction of mobility as measured for example with cell-phone data \cite{Heiler2020}, or traffic counts \cite{asfinag2020verkehr}. 
The lock-down had severe consequences on public life: 
58\% of Austrians who were in employment or self-employed reported that they were employed in a company which introduced home-office to some extent\cite{market2020corona}, 
the number of people registered unemployed increased by 76\% \cite{ams2020arbeitsmarktdaten}, more than 1,300,000 persons were temporarily laid off \cite{hager2020kurzarbeit}, and public life, such as
theaters, cinemas, restaurants, bars, shopping-malls and even large parks, was shut down.

The uncertainty of the situation, especially the threat of job-loss and additional childcare duties caused a lot of stress and anxiety in the Austrian population \cite{pellert2020dashboard}. 
Right from the start this lead to the apprehension that women could be affected more by the lock-down due to additional childcare duties \cite{ifes2020homeoffice,Viglione2020,oecd2020women}, domestic violence \cite{bradbury2020pandemic}, employment in high exposure jobs and simultaneously higher unemployment \cite{oecd2020women}.
Austrian women were more affected by unemployment and partial layoffs \cite{ams2020arbeitsmarktdaten}, surveys registered an increase of domestic violence \cite{steiner2020domesticviolence} and female scientists posted less pre-prints and started less projects \cite{Viglione2020}. 
The fact that men and women react differently to stress and crises is not new. Women experience more stress \cite{Zeidner2006,ben1996gender} and employ more active and problem-focused coping strategies \cite{ben1996gender,matud2004gender}, while men tend to emotional avoidance and emotional coping mechanisms\cite{ben1996gender}. 

In this paper we want to understand the effects of the COVID-19 crisis on behavioral gender differences in five directions: 
Changes in communication patterns, 
changes in mobility, 
changes in food supply, 
changes in spending leisure time and
changes in circadian rhythms as seen in digital traces.
To this end we use longitudinal, nation-wide telecommunication data of 1.2 million cell-phones, covering about 15\% of the entire Austrian population. The anonymized data covers the time period across the government interventions from 
February 1\textsuperscript{st} to June 29\textsuperscript{th}. We split the observation time into 6 periods that characterize the different stages of the pandemic and the response to it. To relate behavioral changes to different phases of life, we stratify our results with respect to age. 
From this data we extract gender-specific features about communication patterns, such as the average interaction duration and the number of calls for all possible gender combinations of calling and being called.
The data further allows us to characterize mobility. From location data we estimate the number of people shopping for food and the usage of recreational areas. 
Finally, we estimate circadian activity of telecommunication and internet usage, from which we estimate e.g. gender differences in sleeping patterns. 

Telecommunication data has been used before to study the effect of crisis and emergencies \cite{candia2008uncovering,bagrow2011collective,Gao2020}. They were used to detect crisis \cite{candia2008uncovering}, study communication patterns subsequent to different emergencies  \cite{bagrow2011collective}, predict movement, e.g. subsequent to the Haiti earthquake 2010 \cite{lu2012predictability}, and to help explain the spread of SARS-CoV-2 \cite{Gao2020}.

\begin{figure}
	\centering
	\includegraphics[width=\linewidth ]{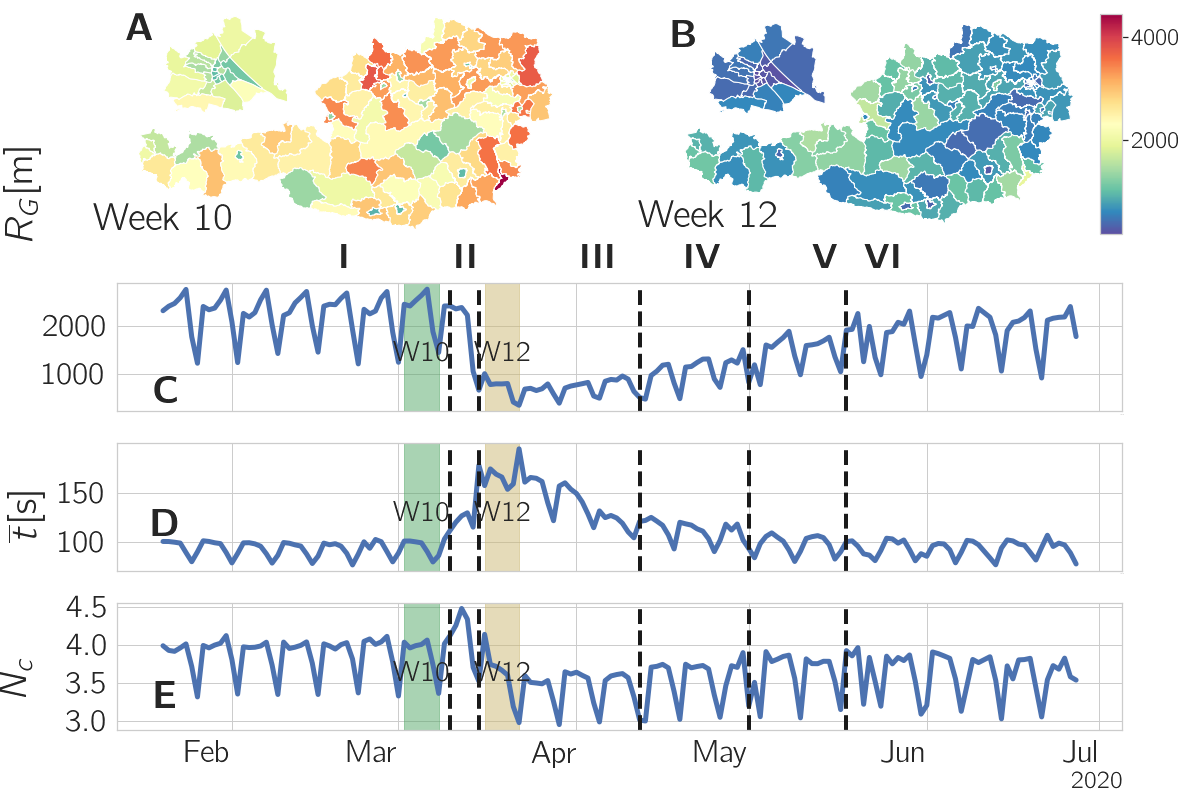}
	\caption{Population-wide response to the COVID-19 crisis. The maps show the  mobility (radius of gyration, $R_G$) for calendar week (\textbf{A}) 10  and (\textbf{B}) 12 for Austria. The timeseries below outline the changes in (\textbf{C}) $R_G$, (\textbf{D}) the call duration per call $\bar{t}$, and (\textbf{E}) the number of calls per device $N_c$. During the lock-down mobility was drastically reduced throughout Austria. The call duration per call $\bar{t}$ increased dramatically and the number of calls, after a brief increase around the beginning of the lock-down, dropped below the pre-lock-down level.}
	\label{fig:fig1_summary}
\end{figure}
Gender differences in human mobility and communication have been reported in \cite{palchykov2012sex,gauvin2020gender}. 
In \cite{palchykov2012sex} changes in communication behavior across age and gender were reported and how reproductive investments and preferred relationships of both sexes shift over a lifespan. 
It is well known  that males tend to have a workplace further away from home and thus generally move more, see e.g. \cite{prashker2008residential}.
Gender differences in mobility in Santiago de Chile are reported in  \cite{gauvin2020gender}. There, significantly different movement behavior is found which is interpreted as an interplay of socio-economic and urban factors.
The gender specific behavioral response to seven terrorist attacks in six cities is investigated in \cite{juul2020gender}. They compare mobile phone communication patterns in response to the attacks and report significant differences between the genders.

\section*{Results}

We partition the observation period from February 1\textsuperscript{st} to June 29\textsuperscript{th} 2020 into six periods:
\RNum{1}  Pre-awareness phase. The population is practically not yet aware of the presence of the disease in Austria. 
\RNum{2} Transition period from the announcement (March 12\textsuperscript{th}) to the actual lock-down on March 16\textsuperscript{th}.
\RNum{3} lock-down until first easing of NPIs (April 13\textsuperscript{th}).
\RNum{4} Period of some easing of NPIs. 
\RNum{5} Gatherings of more than 10 people are allowed, begins on May 1\textsuperscript{st}.
\RNum{6} Back to normal, restaurants and businesses re-open. 
For more details, see SI Text S1. 
We analyze 454,000 women and 452,000 men, for a description of the data see SI Text S2.

\subsection*{Overall behavioral changes during the lock-down}

Figure \ref{fig:fig1_summary} shows the effects of the lock-down. A reduction of mobility in the districts of Austria occurs from before the lock-down (panel A) to right after it (panel B). As a measure for mobility we use the median radius of gyration, $R_G$, see Methods in SI Text S3.
$R_G$ captures the time weighted, spatial extent of an individuals trajectory.
We observe a decrease of $R_G$ between 59\% and 14\%. Panel C shows the time evolution of $R_G$, averaged over all districts. After a sharp decline of almost 50\% in phase \RNum{3} a rebound to almost  pre-crisis levels is seen. 
In panel D we observe a more than 60\% increase of call duration per call, $\bar{t}$. For a definition, see Methods, SI Text S3. 
Panel E shows a brief increase of the number of calls per person, $N_{c}$, in the days just before the lock-down (phase \RNum{2}) followed by a 10\% decrease. 
We now stratify these changes with respect to gender and age.

\subsection*{Communication patterns}

\begin{figure}
	\centering 
	\includegraphics[width=0.95\linewidth]{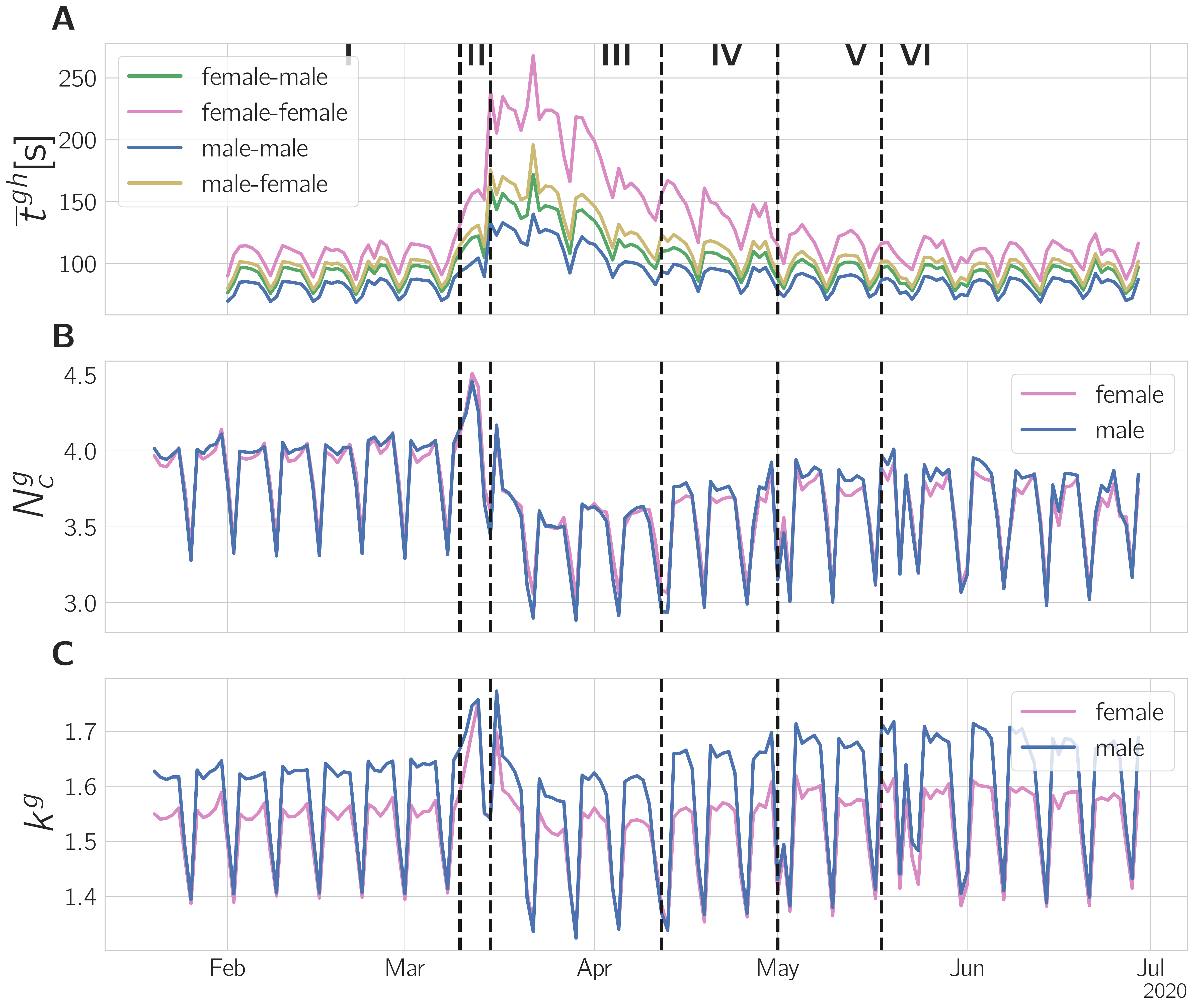} 
	\caption{Gender-specific changes in communication behavior. 
		(\textbf{A})  Median call duration of the four possible types of gender-specific calls, depending on who initiated the call and who received it. By mid-May pre-crisis levels are reached. Half-life times range from 17.3d in the female-female to 14.9 in the female-male case. 
		(\textbf{B}) Number of calls originating from  males (blue) and females (red). The median call duration peaks in phase III, particularly for female-female calls, whereas the number of calls assumes a minimum. Up to the end of the observation period, pre-crisis levels are not reached. 
		(\textbf{C}) The number of communication partners, the degree $k^g(t)$, rises briefly and then drops below pre-crisis levels.}
	\label{fig2_communication}
\end{figure}

As proxies for the strength of social interactions we first analyze the call duration per pair of interaction partners, $\bar{t}^{gh}(t)$, the number of calls, $N^{g}_c(t)$, and the number of calling partners per user, $k^g(t)$, see Methods in SI Text S3. The superscripts indicate  gender, $g$ represents the gender of the caller $h$ is the gender of the called.

\begin{figure}
	\centering 
	\includegraphics[width=0.9\linewidth]{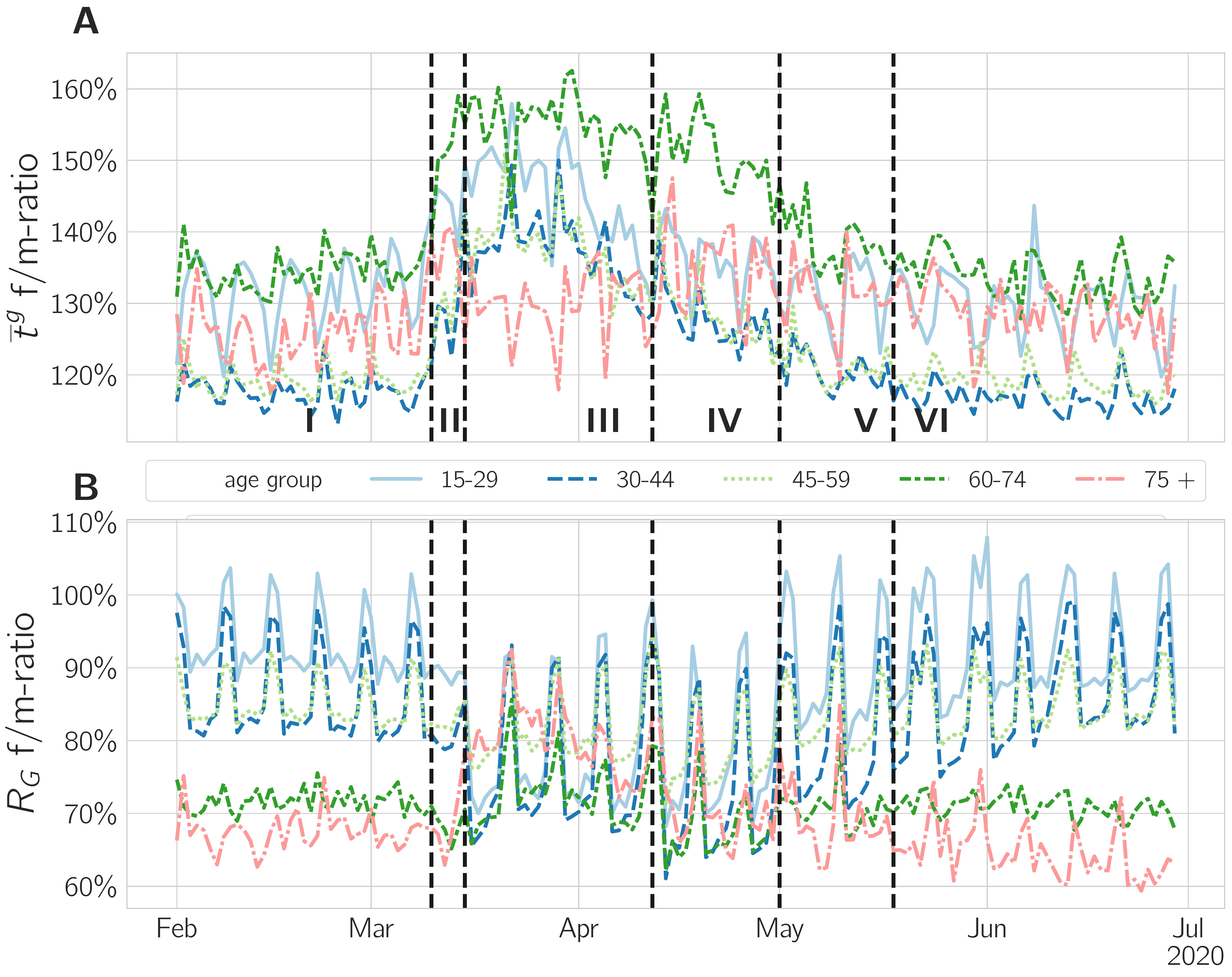} 
	\caption{Gender ratios of communication and mobility for different age cohorts. The gender ratio of (\textbf{A}) the median call duration $\bar{t}$ and (\textbf{B}) the radius of gyration, $R_G$, is seen. In \RNum{3} the $R_G$ gender ratio of young cohorts is shifted towards women moving significantly (p < 0.001) less, while for old cohorts it is shifted towards a more balanced value. In the same period, for all cohorts except 75+, the gender bias for the call duration increases towards women that have a higher call duration.}
	\label{fig:fig3_gender_age_ratios}
\end{figure}

Figure \ref{fig2_communication} depicts the situation over time. 
In panel A we see a massive increase of calling times for the different gender combinations in phase \RNum{2} and the beginning of \RNum{3}. For the female-female calls we observe an increase of up to 140\%, female-male and male-female rise by up to 81\% and 97\%, respectively, and male-male calls increase up to 66\%. 
We find that calls involving women are generally longer than those involving men. 
Moreover, the call time increase is larger when women are involved.

Calling times decrease gradually and reach pre-crisis levels in phase \RNum{6}. This decay can be fitted with an exponential function. The exponents of the fits translate into corresponding ``half-life'' times, which are  $t_{1/2, mm} = 15.9$d for male-male and $t_{1/2, ff} = 17.3$d for female-female interactions, the mixed interactions have half-life times of $t_{1/2, mf} = 15.5$d and $t_{1/2, fm} = 14.5$d for male-female and female-male interactions, respectively. For details, see SI Text S4.
Call times show a pronounced bias towards female initiated calls being longer. In phase \RNum{1}, female originated calls were 10\% longer than male originated, and up to 30\% longer on weekdays in phase \RNum{3}. From its maximum in phase \RNum{3}, the gender ratio continuously declines to normal levels in phase \RNum{5}, see Supplementary Fig. \ref{fig:final_female_inititated_unique_src_dst_count_sum} A. 

The age profile for the median 
call duration is relatively flat for the adult and senior age cohorts and has very low values for the youngest cohort. The call duration increases slightly for the two youngest, but strongly for the two oldest cohorts. For a visualization, see Supplementary Fig. \ref{fig:age_profile_call_time_sum}.
The gender ratio in 
call duration is biased towards women for all ages during the crisis, as seen in Fig. \ref{fig:fig3_gender_age_ratios} A. Notably, the age cohort 15-29 is the only cohort having a more balanced call duration on weekends. For all other  cohorts gender differences are increased on weekends. Around the beginning of phase \RNum{3}, the ratios for all except the 75+ cohort reach a maximum. The 75+ cohort reaches a maximum of the gender imbalance in phase \RNum{4}.

In Fig. \ref{fig2_communication} B we show the number of calls, $N^g_c$, for male and female generated calls. After a short increase in calls in phase \RNum{2} (female: +13\%, male +6\%) we see a significant drop in calls in phase \RNum{3} (both -9\%), which never reaches pre-lock-down levels in the observation period. It stabilizes at a level of  -5\% and -4\% of the previous level for women and men, respectively.
There are only small gender difference in the number of calls. 
For a discussion see SI Text S5.

In Fig. \ref{fig2_communication} C we show the timeseries for the number of different communication partners, $k^g$, i.e. the degree of men and women in their communication networks.
After a brief rise (up to 8\% and 13\% for men and women, respectively) in phase \RNum{2}, $k^g$ falls below its pre-crisis level (-3\% and -2\%). In phases \RNum{4} and \RNum{5} $k^g$ rises to values higher than the initial values in phase \RNum{1}. In phase \RNum{6} $k_g$ is about 4\% higher for men and 2.5\% higher for women.

During normal times (phase \RNum{1}) we find that men have a slightly higher average degree (communication partners) on weekdays (f/m ratio 95\%, men 1.6, women 1.55 unique contacts per day), while on weekends it is more or less balanced (women and men 1.4).
In phase \RNum{2}, $k_g$ is increased for both genders to a maximum around 1.73, with an increasingly smaller gender bias. In phase \RNum{3} the degree drops below pre-crisis levels, but men reduce $k^g$ stronger, resulting in a smaller gender divide in phase \RNum{3} (96\%). From phase \RNum{4} onward, the degree slightly increases (even above pre-crisis levels: men 1.7 and women 1.6), even stronger for men, hence resulting in an increased gender divide (less than 94\%).
Supplementary Fig. \ref{fig:communication_age_ratios} C shows the age dependence of the gender ratio for the degree. Again, there is a weekend trend towards women. They have more communication partners on weekends, except for the 15-29 age cohort. The gender ratio increases in phase \RNum{3} for all age cohorts.

Call duration increases much more than the number of calls decreases, regardless of gender. This is visible in Fig.  \ref{fig:fig1_summary} D and E. Just in phase \RNum{2} there is a drastic rise in both, call time per call, and the number of calls. 
The concentration of communication partners is higher for females and increases during crisis. The bias is also shifted towards men having more communication partners in phase \RNum{6}.
All proxies indicate a strengthening of individual contacts and a focus on important contacts. 

Gender ratios of different phases are compared with a two-sided Mann-Whitney-U test and reject the null hypothesis that they are from the same distribution. The results of the significance tests are presented in SI Text S6 for all age groups, separated into weekdays and weekends.

\subsection*{Mobility}

\begin{figure}
	\centering 
	\includegraphics[width=0.9\linewidth]{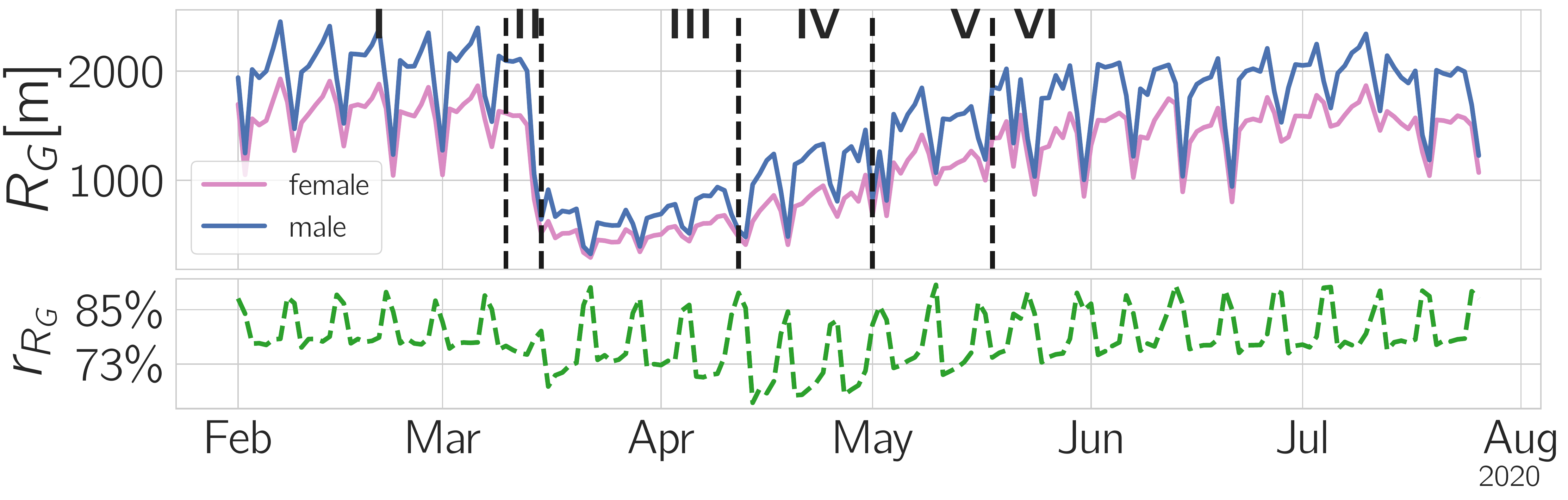} 
	\caption{Mobility quantified by $R_G$. The upper panel shows $R_G$ for men (blue) and women (pink). The lower panel depicts the gender ratio, $r_{R_G}$, over time.
		We observe a large drop in $R_G$ for both genders in phase \RNum{3} and a drop in gender ratio in phases \RNum{3} (lock-down), \RNum{4}, and \RNum{5} (lock-down eased).
	}
	\label{fig:fig4_rog}
\end{figure}

In Fig. \ref{fig:fig1_summary} C we see the overall decline of mobility in terms of the radius of gyration, $R_G$.
Austrians move drastically less during the lock-down, start to move again when the first easing occurs in mid April, and return back to normal in phases \RNum{5} and \RNum{6}.
Figure \ref{fig:fig4_rog} A shows $R_G$
for the two genders, $R^f_G$ (red) and $R^m_G$ (blue). The gender-ratio, defined as $r_{R_G} = R^f_G / R^m_G$ is depicted in panel B. 
The female population is moving less than males in pre-crisis times (phase \RNum{1}), as seen in the ratio $r_{R_G}$ of 78\% on weekdays and 88\% on weekends.
After a brief transition period \RNum{2} the  Weekday ratio drops to around 73\% during the lock-down phase \RNum{3}, while on weekends the ratio remains at initial levels.
In phase \RNum{4}, once restrictions were lifted, $R_G$ for males returns back to normal more quickly than for females, hence decreasing the gender ratio further down to 67\%. The ratio starts to recover towards pre-crisis levels starting from phase \RNum{5} onward, once the main restrictions were lifted. 
When fitting the $R_G$ curves as they converge to pre-crisis levels after the lock-down, we report a \emph{half-life time} for men of $t^m_{1/2} = 34.8$d, and $t^f_{1/2} = 36.0$d for women. For details of the fitting, see SI Text S2. 

The changes in gender ratios of $R_G$ are significant between the phases. For the significance tests, see Supplementary Tab. \ref{t:significance} in SI Text S6. 
Especially the changes from phase \RNum{1} to the subsequent phases and from \RNum{3} to phase \RNum{4} are indeed highly significantly.
We find similar results if we replace the radius of gyration by an alternative measure for mobility that is inspired by entropy, $S^{f/m}_i$. It is presented and discussed in SI Text S7. 

In Fig. \ref{fig:fig3_gender_age_ratios} B we show the age-stratification of the  gender-ratios. 
Before the crisis we observe very different gender ratios for different ages. 
Generally the ratio decreases with increasing age. 
For the young cohort of 15-29 years, the weekday-ratio is above 90\%. 
For the two age cohorts above the average age of first childbirth\footnote{women 26.3 and men 28.7 \cite{statistikaustria2020lebensformen}},
30-44 and 45-59, the ratio is reduced to about 83\%. 
For the age cohorts of retirement, 60-74 and 75+, gender disparity becomes even more biased towards men with a ratio of about 70\%. 
In phase \RNum{3}, the three younger cohorts show an overall trend of increasing gender biases. 
For the age cohort 45-59, this trend is much less pronounced.
Strikingly, the effect is reversed for the retirement cohorts where the gender ratio changes from around 70\% to more than 80\%, which again decreases towards pre-crisis levels in phase \RNum{4}.
The ratio for the old cohorts returns much more quickly to pre-crisis values  than all the younger ones, 
which do not return to the previous values until the end of the observation period.
We do not observe large differences in half-life times across gender, but $t_{1/2}$ is much smaller for older cohorts. For all 
cohorts we find values between $t_{1/2} = 38.8$d for 15-29 year old women to $t_{1/2} = 28.8$d  for 75+ year old men. For more details, see SI Text S2.
For the corresponding statistical tests, see SI Text S6.

For the radius of gyration, we can compare it with data of the previous year (2019) along the same time period. We find that in 2020, during the lock-down phase, there is less than 40\% of the movement than in 2019, see SI Text S8.

\subsection*{Basic provisioning}

In SI Fig. \ref{fig:shopping1_2020} A we show the number of unique devices as a proxy for the number of people at a shopping center across the lock-down. We count the number of unique subscribers in a specifically defined area, see SI Text S3 for the exact definition.
In panel B the corresponding gender ratio is shown.  
The shopping center is the largest of its kind in Austria and one of the largest in Europe. It is a cluster of 359 shops spread over an area of 670,000 m$^2$. Shops sell a wide range of products, including sports equipment, garments, furniture and electronics.
It is visited by more than 20 million visitors each year from Vienna and its hinterland, especially in the south, as well as from Hungary and Slovakia.  
There are also 14 shops, including supermarkets, drug stores and pharmacies that were not affected by the lock-down. 

The visiting patterns of the shopping center in phase \RNum{1} show a pronounced weekly periodicity with a maximum on Saturdays and very few visitors on Sundays, when all stores except cinemas and restaurants are closed. The gender ratio in phase \RNum{1} is close to one, indicating gender balance. In phase \RNum{3} the shopping complex was shut down to a large extent. No  businesses other than stores for basic provisioning were allowed to open. Nevertheless we find a small number of visitors that we account mainly to persons shopping for food and drugs. 
The gender ratio in phases \RNum{3} and \RNum{4} is clearly male-dominated (see Fig. \ref{fig:shopping1_2020} B and for p-values, see SI Tab. \ref{t:significance pi} in SI Text S10). In phase \RNum{5}, when shops were allowed to re-open, visitor numbers rose to pre-crisis levels at the beginning of the week, however without the strong peaks on Saturdays. The gender ratio returns to a balanced situation; compare with SI Tab.  \ref{t:significance pi}. For a comparison with the same period in 2019 we refer to SI Text S9.

\subsection*{Leisure activities}

In SI Fig. \ref{fig:kahlenberg2020} A we count the numbers in a popular recreational area nearby Vienna, the Kahlenberg, frequented mainly for walks, and easy hikes. The number of visitors does not drop in phases \RNum{2}--\RNum{5}, but increases with the usual seasonal trend from March to June. For a comparison with the situation in the year 2019, see SI Text S11.
We find more visitors on weekends and on days with good weather, explaining the high variance in numbers.
The gender ratio is biased towards women during phase \RNum{1}, which changes in phase \RNum{3}, where we find a more balanced gender ratio. Interestingly, the gender ratio does not return to pre-crisis values after the lock-down, 
see SI Fig. \ref{fig:kahlenberg2020} B. The corresponding statistical tests are found in SI Tab.  \ref{t:significance pi} in SI Text S10.

\subsection*{Circadian rhythms}

\begin{figure}[tb]
	\centering 
	\includegraphics[width=0.39\textwidth]{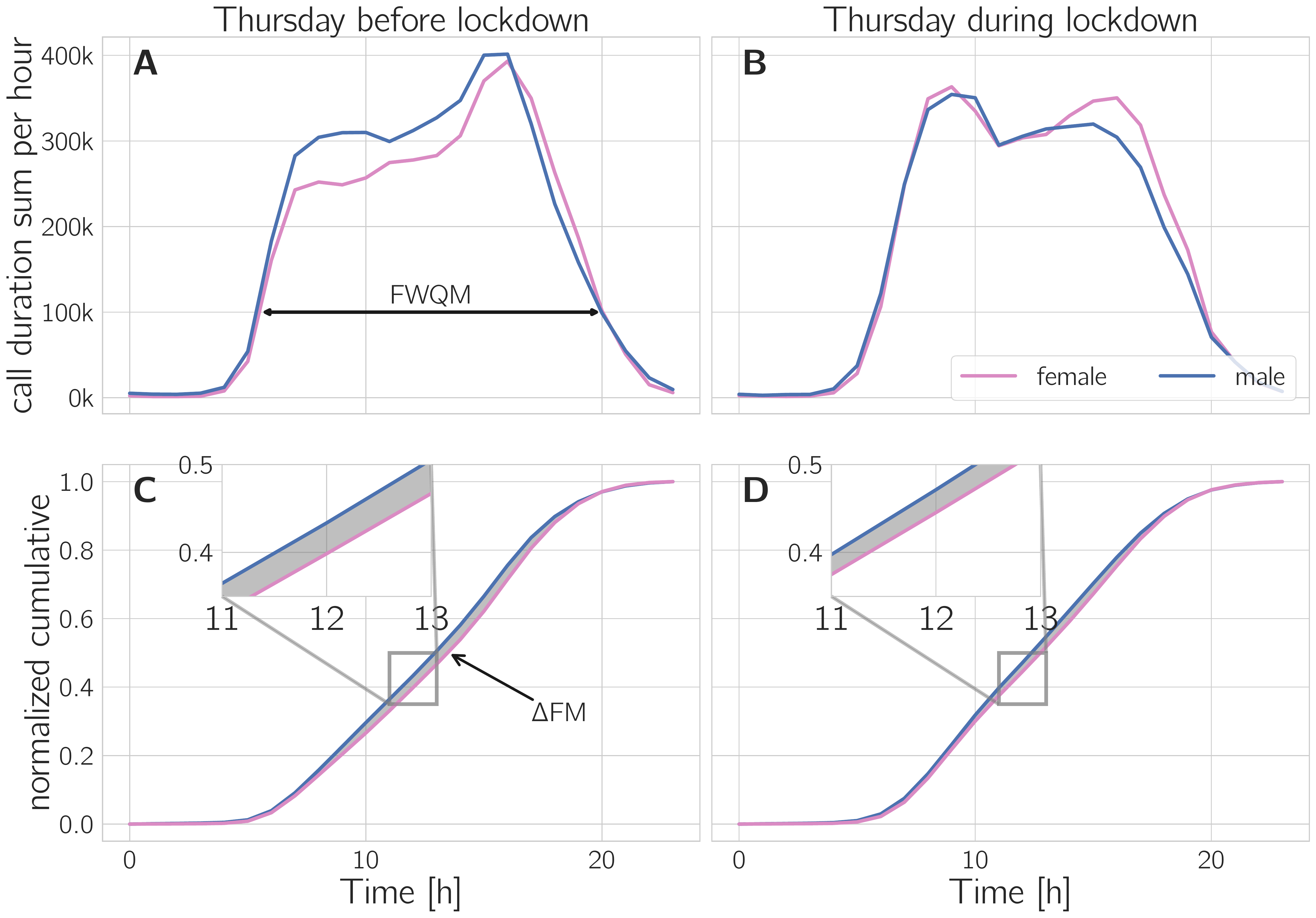} \\ 
	\includegraphics[width=0.39\textwidth]{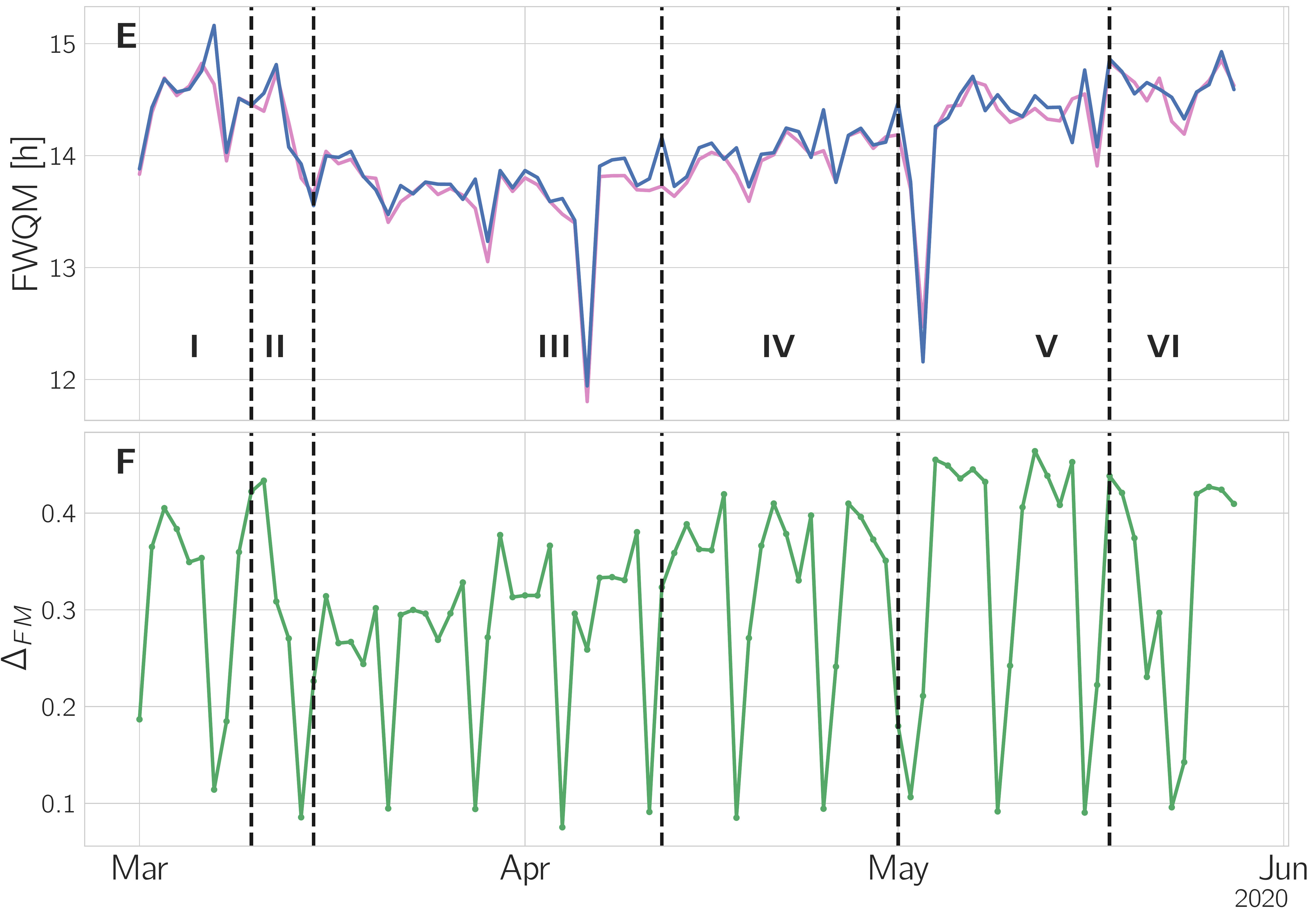}
	\caption{Changes in circadian rhythms during the lock-down measured by the call duration (in seconds) in the whole network. 
		(\textbf{A}) Phone network traffic measured by call duration per hour on the last Thursday in phase \RNum{1}, March 4\textsuperscript{th}. The horizontal arrow marks the full-width-quarter-maximum length (FWQM). 
		(\textbf{B}) Call duration per hour on Thursday, March 18\textsuperscript{th}, the first Thursday in phase \RNum{3}.
		(\textbf{C}) Normalized cumulative activity for the day shown in panel A. The inset highlights the difference of the male and female curve. The gray shaded area marks the difference between the circadian rhythm of men and women, denoted by $\Delta_{FM}$. 
		(\textbf{D}) Same as in C, but for the curve in panel B. 
		(\textbf{E}) FWQM for men and women over time. 
		(\textbf{F}) As the gender ratio of FWQM does not change significantly, we show the gender difference in circadian rhythm, $\Delta_{FM}$, over time. For both genders, the activity maximum shifts from late afternoon to morning and the length of the activity period is approximately 45 minutes shorter during the lock-down. A reduction in $\Delta_{FM}$ means that  circadian rhythms of men and women become more synchronized.
	}
	\label{fig:fig5_sleep_call_duration}
\end{figure}

We compare aggregated phone network traffic across the 24 hours of a day for women and men, to estimate gender differences in circadian rhythms. For  definitions, see SI Text S3. 
Figure \ref{fig:fig5_sleep_call_duration} A and B show 
the call time per hour for the last Wednesday in phase \RNum{1} and the first Wednesday during the lock-down. The maximum activity shifts from the late afternoon to the morning hours.
The average full-width-quarter-maximum (FWQM) captures the length of the daily activity period.
We find that the FWQM is reduced by approximately $53$ minutes from $14 \textrm{h} 33 \textrm{min}$ in phase \RNum{1} to $13 \textrm{h} 40 \textrm{min}$ in phase \RNum{3}. The results are displayed in panel C.
We do not find a significant change in the gender ratio of the activity FWQM 
(Mann-Whitney U, $p>0.05$). 

To capture the shift of the activity of men and women to different times of the day, we calculate the normalized cumulative functions of the call duration, as shown in Fig. \ref{fig:fig5_sleep_call_duration} C and D, thereby correcting for different total activity. Following \cite{juul2020gender} we compare gender differences by calculating the area between the curves $\Delta_{FM}$, see Fig. \ref{fig:fig5_sleep_call_duration} C by the gray shaded area, see Methods in SI Text S3. A large (small) value indicates that the activity of men and women takes place at different (the same) times of day. We find that $\Delta_{FM}$ reduces by 25\% from phase \RNum{1} to phase \RNum{3}.
The values for FWQM and $\Delta_{FM}$ across the crisis are shown in panels E and F. 
The significance of these findings is again shown with a two-sided Mann-Whitney U test that rejects the null-hypothesis that the values are drawn from the same distribution with $p < 3 \times 10^{-5}$. It confirms that $\Delta_{FM}$ is indeed lower in phase \RNum{3}.

Qualitatively we find  the same behavior for the sum of gigabytes up- and downloaded and the number of calls. The reductions in lengths of day range from 40 minutes for the downloaded gigabytes to 60 minutes for the number of calls. 
For the corresponding analyses, see SI Text S12

\section*{Discussion}

The COVID-19 pandemic represents a unique natural experiment to understand individual and collective coping mechanisms with respect to stress and crisis.
Telecommunication data reveals almost real-time insights into many aspects of daily life without interfering with the subjects' actions and interactions. 
Using anonymized mobile phone data of a large fraction of the Austrian population, we find that gender differences that exist in communication patterns, mobility and spending leisure time are amplified during the crisis, imposed by a severe lock-down in the first phase of the COVID-19 crisis. 
In the context of basic provisioning, we find that during the crisis there is a bias toward men doing the shopping for food that is absent in normal times. 
For circadian rhythms we see that the daily activity is concentrated on a shorter period of the day for men and women equally and that the circadian rhythm of men and women becomes more synchronized during the lock-down.

For both genders we observe an increase in total call duration, 
which is due to an increase of the call time per call and interestingly a decrease in the number of calls. This is a clear sign that conversation focuses on the core communication partners. This result is in line with a general decline of the number of communication partners during the lock-down. 
The reduction in communication partners could also result from the loss of conversation partners from work, but we also observe a reduction on the weekends, where we would not expect the effect of work to dominate.
The degree distribution before the crisis is in line with earlier work on mobile phone data \cite{onnela2007structure}.
While they find a mean degree 2.34 (averaged over 18 months), we get a smaller value of 1.53 because we average over 24 hours. However, we find the same power-law exponent $\sim-8$ for the degree distribution.
In these quantities we see a clear increase and amplification of the gender-biases. 

Women show a smaller decrease in the number of calls and a stronger increase in call time per call. As a consequence, the gender ratios of both quantities shift towards females. 
Women have been reported to have more tightly knit (online) networks than men \cite{szell2013how,igarashi2005gender}. We interpret our findings as a signal that this behavior intensifies during crisis.
Additionally, our findings agree with the well known \emph{tend and befriend reaction}, which says that facing (real or perceived) danger, females are more likely to defend offspring (tend) and turn to their social group for protection (befriend) \cite{taylor2000biobehavioral}. The tightening of the social network can also be attributed to social carework, such as calling lonely elderly, which is performed more often by women during the lock-down \cite{prainsack2020solpan}. 
Women were reported to employ more active, problem-oriented coping strategies such as emotional and social support, while men show rational and detachment strategies \cite{ben1996gender,matud2004gender}, again supporting the expectation, that women seem to tighten their social networks more than men.

We find that the recovery time to women's total call time initially is as fast as for men, but later on clearly slows down. 
The higher, longer increase in demand for communication can be interpreted in the context with higher needs for communication as a coping strategy in an ongoing crisis \cite{ben1996gender,matud2004gender}. It also aligns well with the fact that women experience more stress than men \cite{michael2009gender}, have higher levels of post traumatic stress disorder \cite{stein2000gender} and have a higher prevalence to depression, partly due to `stress responsiveness' \cite{parker2010gender}. 
Our result could be confounded by gender differences introduced by work environments. However, increasing gender-ratios in the call times per call and the number of calls on weekends are a strong indicator that the confounder indeed weakens the effect on weekdays.

The age profiles for call times and the number of calls seemingly suggest that younger cohorts communicate less than older ones.
We attribute this to the increased use of instant messaging services  \cite{torok2016big} and other modern channels of communication by the younger cohorts.
Here a channel selection bias towards younger cohorts using web-based communication services more actively acts as a severe confounding factor.

The female population is moving less over the entire period, confirming earlier work in different countries and situations \cite{gauvin2020gender, prashker2008residential}.
The decrease in mobility following the lock-down is stronger for women. In addition, men recover their mobility behavior much quicker after the measures are lifted. 
This effect depends on age. For the young and adolescent population the existing gender-bias in mobility is enhanced, while for those above retirement age the bias reduces.
We relate this to childcare duties during the reproductive age and gender specific differences in occupation.
This confirms the apprehensions that females would be required to restrict their mobility stronger due to childcare duties and higher unemployment \cite{ifes2020homeoffice,Viglione2020,oecd2020women}. 

In addition to care-taking duties, we expect more effects. 
Women were shown to exhibit more ethical behavior, at least where it is socially desirable, while men often behave less community-aware \cite{betz1989,dalton2011}. For women, it has been shown that they are 50\% more likely to adopt  non-pharmaceutical interventions in response to a respiratiory epidemic \cite{moran2016meta}.
In this context, the reduction of mobility in women could be partly attributed to responsible behavior in staying at home to protect vulnerable parts of the population. This argument is supported by a qualitative panel survey, that reports women taking the COVID-19 pandemic more seriously in Austria  \cite{eberl2020die,kittel2020austrian}. 

Since it seems that men go out for work more and are more often responsible for gathering basic provisions during the lock-down, they are more exposed to the perceived danger of catching SARS-CoV-2. This could be interpreted as increased risk-taking behavior in men, in line with  \cite{gustafsod1998gender,byrnes1999gender,szell2013how}. 

Generally, gender differences in mobility decrease on week-ends. We confirmed that the radius of gyration is larger for men because they commute more/farther \cite{gauvin2020gender}. This suggests that a main factor for our observed behavioral changes is indeed driven by work.
Further evidence for this hypothesis is found in the fact that only for the 60+ age cohort the gender-ratio does not change between weekends and weekdays.
Nevertheless, the effects discussed above persist on weekends and our conclusions stay valid.

We proxy the activity of people by analyzing their Internet traffic loads across a day. On average, the daily activity period, as observed by several quantities, is reduced by around 40 to 60 minutes by the lock-down. 
The gender-ratio of the length of a day does not change significantly. 
However, mobile phone usage of men and women takes place at different times of the day, with the maximum shifted from the late afternoon to the morning during the lock-down. 
Network traffic starts to increase later in the day and ends earlier. This can be explained by commutes becoming obsolete because of home office and the rise in unemployment \cite{ifes2020homeoffice,Viglione2020,oecd2020women}.
We believe that the shift of the maximum activity from evening to morning is on one hand caused by different activity patterns in home office and by different spare time activities during the lock-down.

The gender-difference in cumulative activity, correcting for different activity levels, drops by 25\%, hence showing that the circadian rhythms of women and men become more synchronized during the lock-down. We interpret this as a consequence that people stay at home much more than usual, where often an opposite-gender partner is present \cite{statistikaustria2020lebensformen}. 
It would be interesting to understand if the synchronization is stronger for couples spending much time together. More detailed studies of the effects of the pandemic on the circadian rhythm, with respect to age and personal attributes, such as morningness or eveningness \cite{aledavood2015digital}, could give more detailed insights in e.g. disturbed sleep patterns.

We have shown that massive collective crisis results in tighter social networks with a focus on a social core environment.  
Women seem to focus more on this tightening, indicating stronger, more active coping strategies, a different perception of the dangers of COVID-19 and stronger pro-social behavior.
We see that mobility is reduced much more in females, and the time to recovery is much slower for females. This is partly connected to work-related influences and maybe a stronger community-aware behavior in response to public mobility restrictions.
We might see a slight indication of increased risk taking in males when it comes to basic provisioning. 
Finally, we report synchronization effects of (electronic) activity behavior during the day between males and females during crisis.

\section*{Methods}
For a description of the data set see SI Text S2 and for definitions of the used metrics see SI Text S3.

\begin{acknowledgments}
\subsection{ACKNOWLEDGEMENTS}
We thank Eva Bauer, Barbara Prainsack, Christian Diem, and Hannah Metzler for helpful discussions. This work was supported in part by Austrian Science Fund FWF under projects P29252 and I3073, the Austrian Research Promotion Agency (FFG) under projects 857136 and 873927, the WWTF under projects COV 20-017 and COV 20-035, and the Medizinisch-Wissenschaftlicher Fonds des 
B\"urgermeisters der Bundeshauptstadt Wien under project CoVid004.  
\end{acknowledgments}

 



\newpage 
.
\pagebreak 
 
\onecolumngrid
\appendix

\section*{Supplementary Information}

\subsection*{SI Text S1: Partitioning of observation interval}
The observation period ranges from Feb 1 to June 29 2020.
Our analyses compare various periods of time before, during and after the lock-down. Each period is based on the introduction or easing of one or more non pharmaceutical interventions (NPIs) and identified by roman numerals. Period \RNum{1} refers to the phase before the Austrian population was widely aware of the COVID-19 disease in Austria, and ends with the first press conference announcing public restrictions such as school closings and a call for home office on March 12\textsuperscript{th}.This was followed by a transition period \RNum{2} until Sunday, March 15, 2020, when an extensive lock-down was announced Including all shops closing and restrictions on public movement and gatherings, such as public transport only being allowed to be used for commuting, a ban on meetings with non-household persons and leaving the home only permitted for work that cannot be postponed, grocery-shopping, helping people and short walks. The following lock-down \RNum{3} lasted until Easter (April 13, 2020) which marks the first easing of the measures  with small shops and construction stores being allowed to re-open. Phase \RNum{4} continues until the beginning of phase \RNum{5} on May 1, where the lock-down was eased further and gatherings of up to 10 people were allowed. May 15\textsuperscript{th} marks the beginning of phase \RNum{6}, with restaurants and bars re-opening. The dates and description of the measures is based on \cite{Desvars-larrive2020}.

\subsection*{SI Text S2: Data}

We partnered with a large Austrian internet service provider (ISP) to get access to data from mobile phones.
We use a combination of classical Call Data Records for the voice domain as well as X Data Records for the data domain.
Thus we do not only register an event when a call is performed, but rather perceive additional events when data packages are transferred.
Various network interfaces are connected via probes so we get data points from a multitude of network technologies for mobile data usage (2G, 3G, 4G), calls, text messages as well as Voice over LTE, from both user- as well as control plane.
On average we observe approximately 1 Billion events per day, 4.5 Million devices per day and for 80\% of the devices the next event is received in 1.7 minutes, on average 4 minutes.
When evaluating gender differences we need to filter the data to approx 1.2 Million devices per day where demographic details are defined.
Demographic information is not available for roamers or virtual mobile operators (MVNO) and thus they are excluded from this analysis.
Furthermore, only devices with a radius of gyration $R_G$ (see eq. \eqref{eq:rogtw} below) larger than 0m and lower than 300km are considered to exclude resting internet of things-devices and devices above the theoretical maximum of $R_G$ in Austria.
Calls are filtered to a length of at least 25 seconds prior to aggregation to exclude calls that were not picked up, which form a distinct peak just below 25 seconds, shown in Supplementary Fig. \ref{fig:histogram_mo_call_duration_dist}.

As the gender attribute in the data set is self-reported and we discuss psychological, social and biological reasons for the observed behavioral changes, we will use the term `gender', referring to the psycho-social construct.

\begin{figure}[htb]
	\centering
	\includegraphics[width=0.4\linewidth]{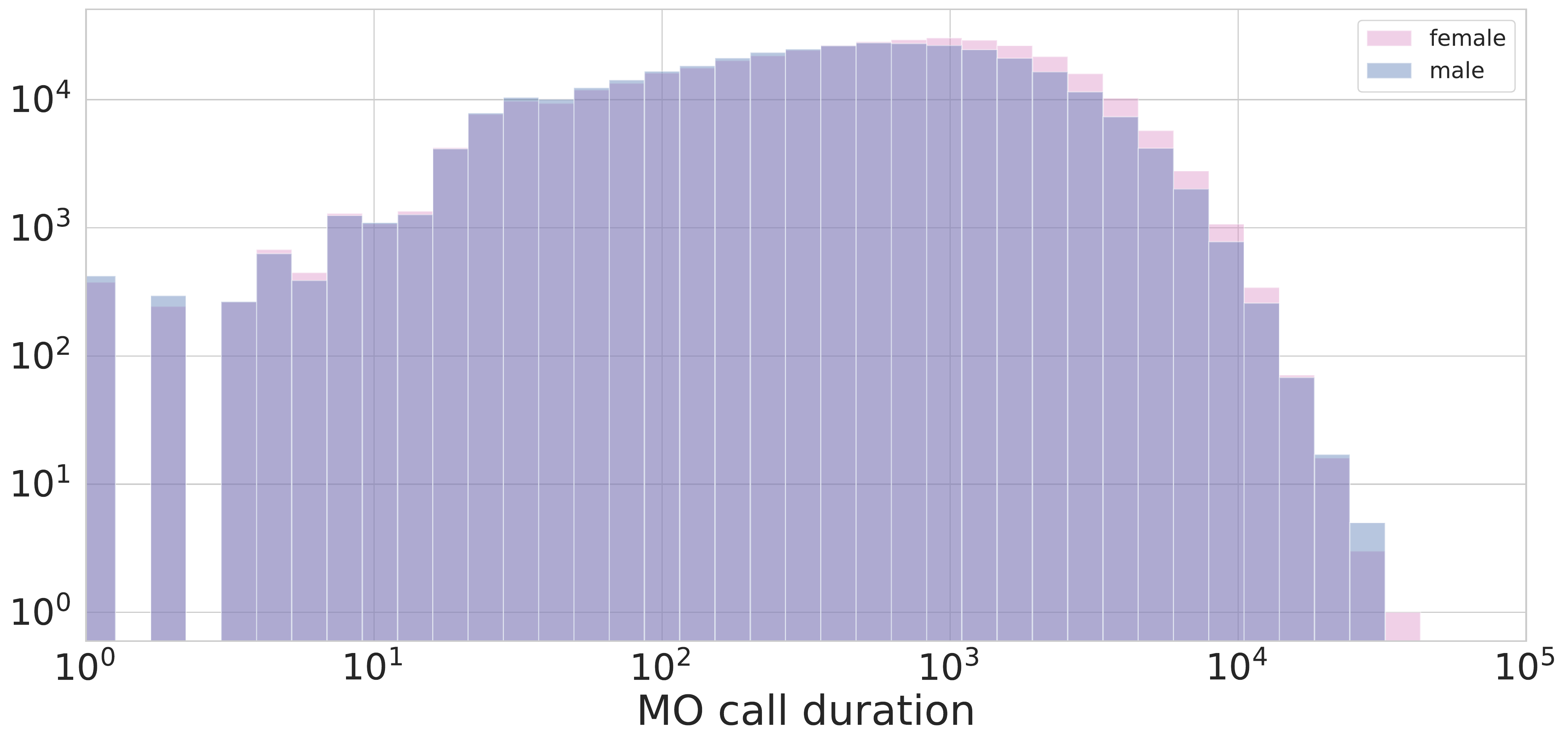}
	\caption{Histogram of the call duration. There is a local maximum around 25 seconds, corresponding to calls not answered, which we get exclude by introducing a cutoff at 25s. Note the double logarithmic axes.}
	\label{fig:histogram_mo_call_duration_dist}
\end{figure}

Table \ref{t:cohort_overview} outlines the distribution of the devices per cohort. As not all devices are active every day we give the mean and standard deviation as an approximation for the overall counts.
As we cannot analyze a device for more than 24 hours (see below), we need to calculate aggregate statistics over the analyzed time period.

\begin{table}[hb]
	\centering
	\caption{
		Sample sizes of the gendered demographic cohorts.
		\label{t:cohort_overview}
	}
\begin{tabular}{llll}
\hline
 gender & age group &    mean &    std \\
\hline
 female &     15-29 &   44318 &   3038 \\
 female &     30-44 &  130137 &   9349 \\
 female &     45-59 &  151173 &  11576 \\
 female &     60-74 &   80479 &   6269 \\
 female &      75 + &   18890 &   1726 \\
   male &     15-29 &   42023 &   2553 \\
   male &     30-44 &  116230 &   7493 \\
   male &     45-59 &  158313 &  12686 \\
   male &     60-74 &   84800 &   7368 \\
   male &      75 + &   19085 &   1915 \\
\hline
\end{tabular}
\end{table}

\paragraph{Localization}
Our localization methodology is based on the topology of the network, namely the observed cell-id.
This means that the accuracy is limited, and much less accurate than GPS based localization or the result of custom apps combining Bluetooth, WiFi and GPS.
However, the data is available for a large quantity of devices.
The ISP provides us with the localization information for each cell-id, which is based on the centroid of the network coverage simulation.

\paragraph{Privacy}
\label{l:privacy}
The data is anonymized, any identifiers are hashed every 24 hours with a changing key by the ISP prior to making the data available for the researchers.
Only cell-id based localization is used to enhance the privacy of the subscribers and
only aggregate and k-anonymized statistics are reported.
With this procedure we adhere to the recommendations of the GSMA \cite{gsmacovid} with regards to data privacy handling as well as the law of the local jurisdiction.

\subsection*{SI Text S3: Metrics}

\paragraph{Communication}

By analyzing calls, social interactions can be modeled.
This part of the data consists of a list of outgoing (MO) and incoming (MT) calls, each associated with a source and destination. We filter to calls with a duration of at least 25 seconds to adjust for a shift in the distribution corresponding to calls that were not answered (see Fig \ref{fig:histogram_mo_call_duration_dist}).

For each device we find $N_c^{\mathrm{MO}}$ outgoing and $N_c^{\mathrm{MT}}$ incoming calls with $k^{\mathrm{MO}}$ and  $k^{\mathrm{MT}}$ other individuals, respectively (in- and out-degree). The call duration is denoted by $\bar{t}$.
Additionally, as described earlier for the mobility dimension, for each device, age group and gender are specified. 

For all of these device-level metrics we report the median of the whole population, or for cohorts specified by age groups or gender. We will add superscripts $g$ and $h$ to indicate gender.

\paragraph{Mobility}

We obtain mobility data as a stream of spatially localized network signaling events.
It is transformed into a list of locations $\vec{x}_{i\mu} = (x_{i\mu}, y_{i\mu})$, with associated stay duration $t_{i\mu}$ for every individual $i = 1 ... N_{indiv}$ at location index $\mu = 1 ... N_{locations}$, where $x$ and $y$ represent longitude and latitude, respectively.
Due to the anonymization procedure the location index $\mu$ is reset every day and the individual index $i$ is reshuffled accordingly.
For the individuals, metadata is collected in a vector $m_i = (g_i, a_i)$, containing gender $g_i \in {female, male}$ and age $a_i$ aggregated into cohorts of 15 years. 

\begin{figure}[ht]
	\centering
	\includegraphics[width=0.4\linewidth]{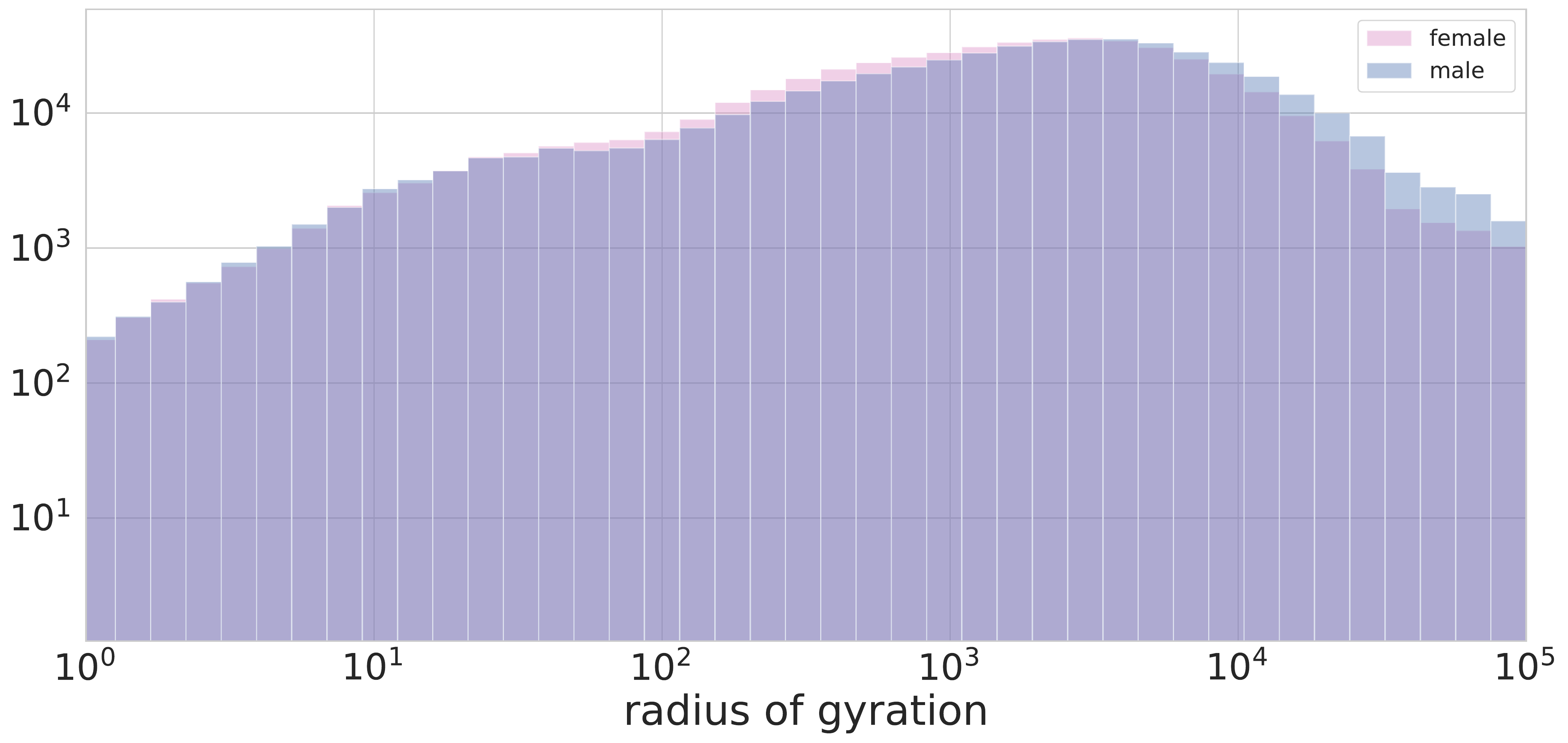}
	\caption{Histogram of the radius of gyration $R_G$. Note the double logarithmic axes and the truncated fat tail.}
	\label{fig:histogram_rog_dist}
\end{figure}

The \emph{radius of gyration} $R_G$ is calculated as the square root of the time-weighted mean of the squared distances\footnote{Calculated as the Haversine distance which calculates a distance in meters from latitude and longitude coordinates given in degrees.} $d$
of the locations $\vec{x}_{i\mu}$ to the daily centroid $\overline{x}_i = \frac{\sum_\mu \vec{x}_{i\mu} t_{i\mu}}{\sum_\mu t_{i\mu}}$:
\begin{equation}
	R_{\mathrm{G},i} = \sqrt{\frac{\sum_\tau d(\overline{x}_i, \vec{x}_{i\tau})^2}{\sum_\tau t_{i\tau}}} \label{eq:rogtw}
\end{equation}
It captures the amount of movement in a time weighted manner and has the dimension of a length in meters. The distribution of $R_{\mathrm{G},i}$ is fat tailed, see Fig. \ref{fig:histogram_rog_dist}. In the main paper we report the median because it robust to heavy tails. 

For our second mobility measure  \emph{entropy} the locations $\vec{x}_{i\mu}$ are binned into a hexagonal raster using Uber's H3  \cite{uber_h3}.
The chosen resolution level for the raster yields hexagons with an area of approximately $800m^2$\footnote{This is H3's resolution level 8.}.
For each hexagon $\Tilde{x}_\nu$ ($\nu = 1...N_{hex}$), the stay duration of the locations in each hexagon are aggregated to $\Tilde{t}_{i\nu}$
\begin{equation}
	\Tilde{t}_{i\nu} = \sum_{\nu \forall \vec{x}_{i\nu} \in \Tilde{x}_\nu} t_{i\nu} 
\end{equation}
The stay time distribution of an individual $i$ is then defined as the share of its time spent in a given hexagon $\Tilde{x}_\nu$
\begin{align}
	p(\Tilde{x}_{i\nu}) = \frac{\Tilde{t}_{i\nu}}{\sum_{\nu}\Tilde{t}_{i\nu}}   \label{eqn:sproba} 
\end{align}
The entropy of an individual's stay time distribution, $S_i$, is defined, using the standard formulation of Shannon Entropy, as:
\begin{align}
	S_i = - \sum_\nu p(\Tilde{x}_{i\nu})\log_{2}(p(\Tilde{x}_{i\nu}))   \label{eq:entropy}
\end{align}

%
%

\paragraph{Points of interest (Shopping, Leisure)}
Specific points of interest reflecting shopping and leisure zones in Vienna were analyzed in more detail. We first used H3 by Uber \cite{uber_h3} to create a discrete raster for the whole country to speed up the analysis of specific locations afterwards.
Then we count the number of unique subscribers in a set of manually defined hexagons.
We limit our investigations to stays longer than 10 minutes and shorter than 4 hours. We assume this eliminates devices passing the shopping complex on the nearby highway, as well as persons working there, because these activities take much shorter or longer, respectively.

\paragraph{Circadian rhythm}

We investigate the circadian rhythm using network traffic measures $A(t)$ aggregated by gender and ranging from the sum of call duration per hour to downloaded gigabytes per hour. Irrespective of the quantity, we observe a broadened, peak-like structure with a rise in the morning and a drop in the evening. We quantify the duration by the \emph{full-width-quarter-maximum} distance (FWQM). It denotes the time span between point where the activity is larger than the quarter of the maximum activity in the morning and the point where the activity drops below the same value in the evening. We choose the threshold relative to the maximum, so we are independent of the total activity; its value is set to a quarter without loss of generality.

\begin{equation}
	A(t)- max(A(t))/4=0 \quad \forall t_1, t_2
\end{equation}

\begin{equation}
	\mathrm{FWQM} = t_2 - t_1
\end{equation}

Male and female activity, corrected for the difference in total activity, is not distributed across the 24 hours of a day in the same way. Inspired by \cite{juul2020gender} we apply the following procedure. We correct for the difference in total activity by calculating and normalizing the cumulative activity
\begin{equation}
	C(t) = \frac{\int_0^t A(\tau) d\tau}{\int_0^{24} A(\tau) d\tau} \quad\textrm{,}
\end{equation}
where 0 and 24 are the time at the beginning and end of the chosen 24 hour period.
Now we calculate the gender difference $\Delta_{FM}$ in circadian rhythm by calculating the absolute area between the cumulative activity functions for men and women 
\begin{equation}
	\Delta_{FM} = \int_0^{t_{max}} | C_m(t) - C_f(t) | dt 
\end{equation}

\paragraph{Gender differences}
To investigate gender differences we calculate the gender ratio $r_x$ for the various aggregations $x$ presented in this work. The ratio $r_x$ is calculated as the quotient of the aggregate for the female cohort divided by the aggregate for the male cohort $r_x = x_\mathrm{female} / x_\mathrm{male}$ ($x$ represents the aggregation, e.g. median $R_G$ or median call duration $\bar{t}$). A gender ratio $r_x$ close to 1 (or 100\%) indicates that the quantity is of similar size for both genders, less (more) than 100\% indicates smaller (larger) values for females.

\subsection*{SI Text S4: Time-back-to-normal: Fitting Half-life times}

For many of the investigated quantities, the COVID-19 lock-down represents a perturbation from an equilibrium with a subsequent, smooth return to the previous value. 
We fit an exponential function $$f(t; a_0, a_1 ,b) = a_0 + a_1 b^t \mathrm{,}$$ where $a_0$ is fixed to the mean value in phase \RNum{1}, $a_1$ is the additional offset at the beginning of the decay and $b < 1$ is the average daily reduction factor.  
From $b$ we can calculate the half-life time $$t_{1/2} = \frac{ln(1/2)}{ln(b)}$$ in days, which quantifies the time it takes the fitted quantity to return halfway back to ``normal''.
The non-linear least squares fit is performed to the 7-day moving average of the investigated quantity, starting from the beginning of phase \RNum{3} to the end of the observation period.
We report standard deviations $\sigma$ calculated as the square root of the diagonal elements of the covariance matrix. 

\begin{figure}
	\centering
	\includegraphics[width=0.4\linewidth]{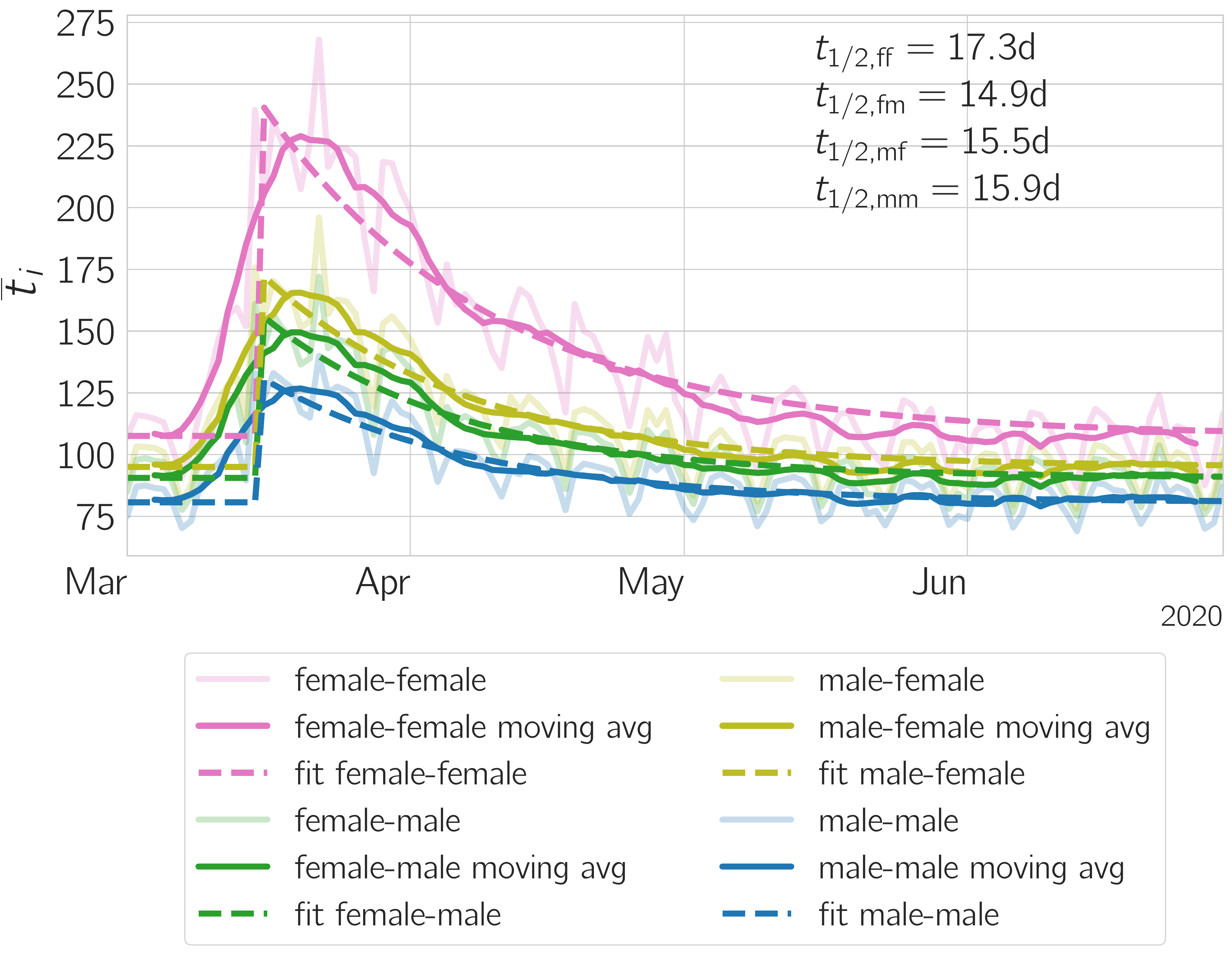}
	\caption{Decay parameters for the call duration $\bar{t}^{gh}$. We show $\bar{t}^{gh}$ (transparent line), the seven day moving average (solid line) and the fitted curve (broken line). The half-life times for the return from perturbed state to normal are 17.3d for male-male, 14.9d for female-male, 15.5d for male-female and 15.9d for male-male interactions. Detailed results of the fitted parameters are in Tab.\ref{tab:exp_fit_results}.}
	\label{fig:expfit_avgt}
\end{figure}

Figure \ref{fig:expfit_avgt} shows the average interaction time $\bar{t}^{gh}$, its 7-day moving average and the fitted exponential. We find half-life times between 14.9d and 17.3d for female-male and female-female interactions, respectively. The detailed results are given in Tab. \ref{tab:exp_fit_results}. The female-female interaction $t_{1/2}$ is more than one standard deviation larger than for the interactions where males are involved.

\begin{figure}
	\centering
	\includegraphics[width=0.4\linewidth]{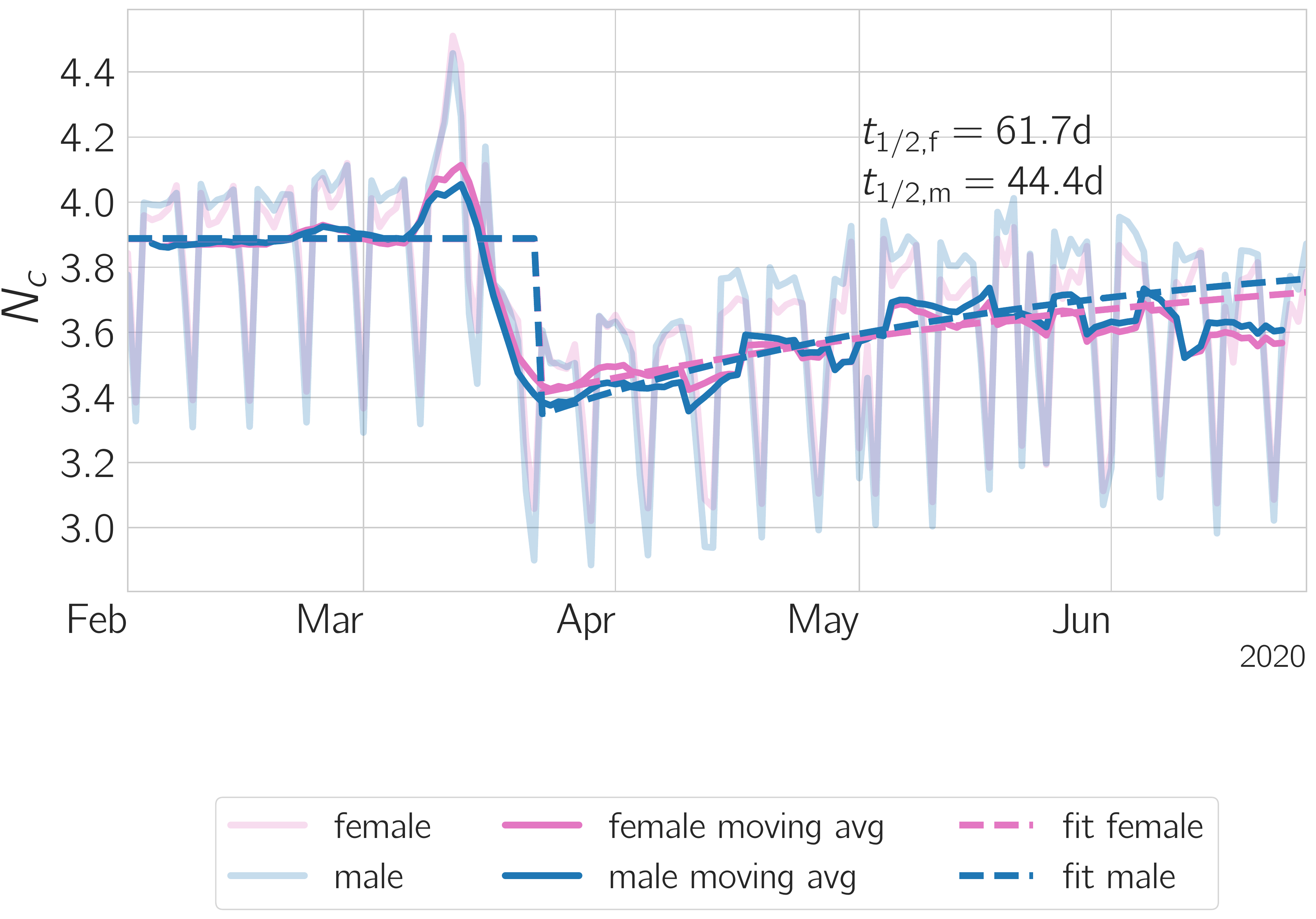}
	\caption{Decay parameters for the number of calls $N_c$. We show $N_c$ (transparent line), the seven day moving average (solid line) and the fitted curve (broken line). The half-life times for the return from perturbed state to normal are 61.7d for women and 44.4d for men. Detailed results of the fitted parameters are in Tab. \ref{tab:exp_fit_results}.}
	\label{fig:expfit_nc}
\end{figure}

\begin{figure}
	\centering
	\includegraphics[width=0.4\linewidth]{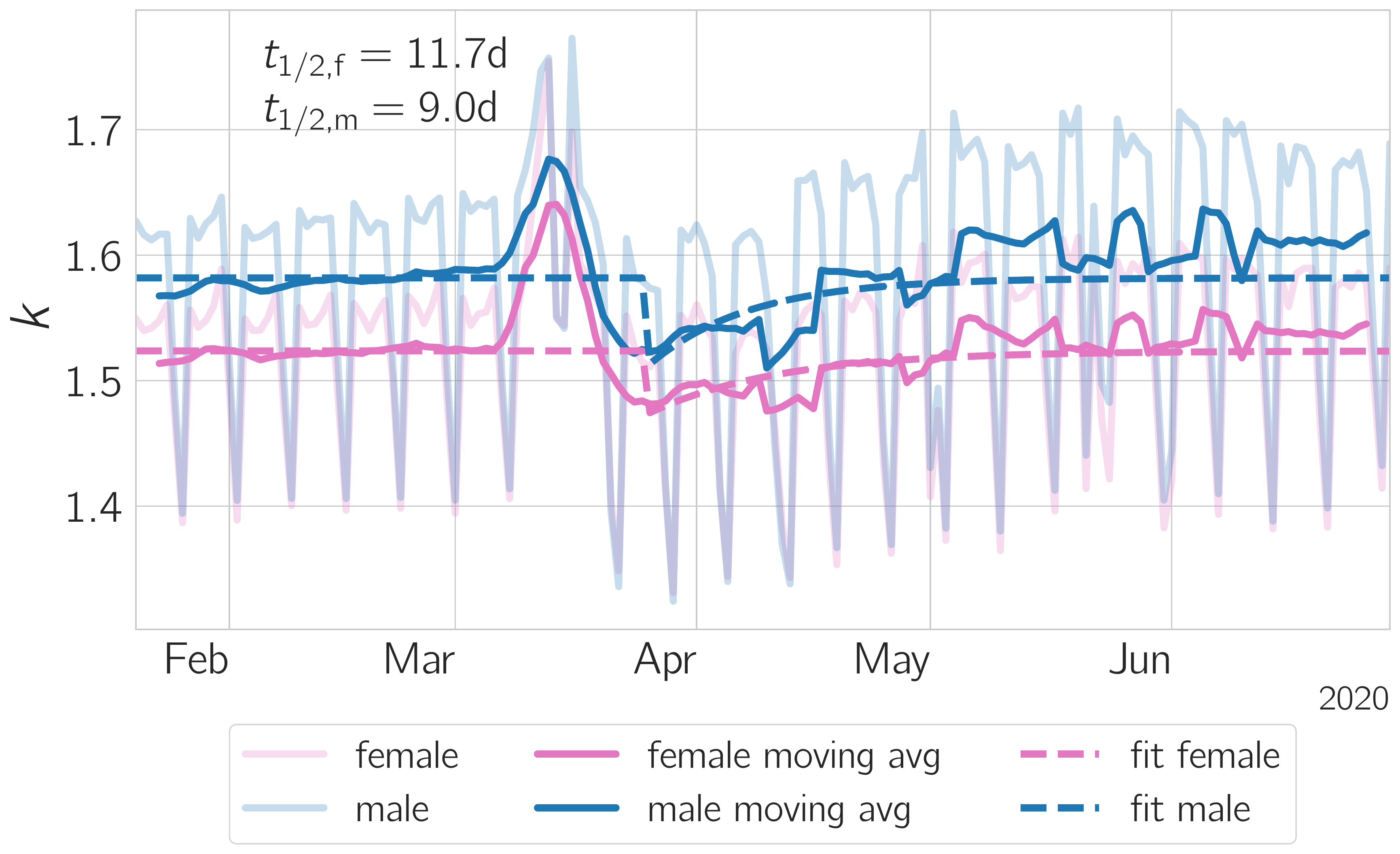}
	\caption{Decay parameters for the number of interaction partners $k$. We show $k$ (transparent line), the seven day moving average (solid line) and the fitted curve (broken line). The half-life times for the return from perturbed state to normal are 11.7d for women and 9.0d for men. Detailed results of the fitted parameters are in Tab. \ref{tab:exp_fit_results}.}
	\label{fig:expfit_degree}
\end{figure}

We also fit for the number of calls $N_c$ and the number of unique contacts $k$, see Figs. \ref{fig:expfit_nc} and \ref{fig:expfit_degree}, and Tab. \ref{tab:exp_fit_results} for details. Both times, the half-life times for females are more than one standard deviation larger. For $N_c$ the half-life times are large compared to the other quantities, and in fact it has not reached the pre-crisis levels by the end of the study period.

\begin{table}
	\centering
	\caption{Fitted parameters for the exponential decay. The confidence intervals are one standard deviation.}
	\begin{tabular}{c c c c c c}
		\hline
		quantity & gender & $a_0$ & $a_1$ & $b$ & $t_{1/2}$  \\
		\hline
		$\bar{t}_i$ & female-female & $107 $ & $138 \pm 4$ & $0.9607 \pm 0.0015$  & $17.3 \pm 0.7$d \\
		$\bar{t}_i$ & female-male & $91$ & $69 \pm 2$ & $0.9546 \pm 0.0017$  & $14.9 \pm 0.6$d \\ 
		$\bar{t}_i$ & male-female & 95 & $81 \pm 2$ & $0.9562\pm 0.0017$ & $15.5 \pm 0.6$d \\ 
		$\bar{t}_i$ & male-male & 81 &  $52 \pm 1$ & $0.9573\pm 0.0015$ & $15.9 \pm 0.6$d \\ 
		
		$N_c$ & female & 3.89 & $-0.48 \pm 0.01$ & $0.9888 \pm 0.0010$ & $61.7 \pm 5.6$d \\ 
		$N_c$ & male & 3.89 & $-0.55 \pm 0.02$ & $0.9845\pm 0.0012$ & $44.4 \pm 3.5$d \\ 
		
		$k$ & female & 1.524 &  $-0.052 \pm 0.007$ & $0.9426\pm 0.0115$ & $11.7 \pm 2.4$d \\ 
		$k$ & male &  1.582 &   $-0.074 \pm 0.014$ & $0.9260 \pm 0.0198$ & $9.0 \pm 2.5$d \\ 
		
		$R_G$ & female & 1564 & $-1331 \pm 41$& $0.9809 \pm 0.0010$ & $36.0 \pm 1.8$d \\ 
		$R_G$ & male & 1959 & $-1606 \pm 48$ & $0.9803 \pm 0.0009$ & $34.8 \pm 1.7$d \\ 
		
		$S_i$ & female & 0.73 & $-0.53 \pm 0.02$ & $0.9761 \pm 0.0015$ & $28.7 \pm 1.9$d \\ 
		$S_i$ & male & 0.78  & $-0.52 \pm 0.02$ & $0.9753 \pm 0.0016$ & $27.6 \pm 1.8$d \\ 
		\hline
	\end{tabular}
	\label{tab:exp_fit_results}
\end{table}

\begin{figure}
	\centering
	\includegraphics[width=0.4\linewidth]{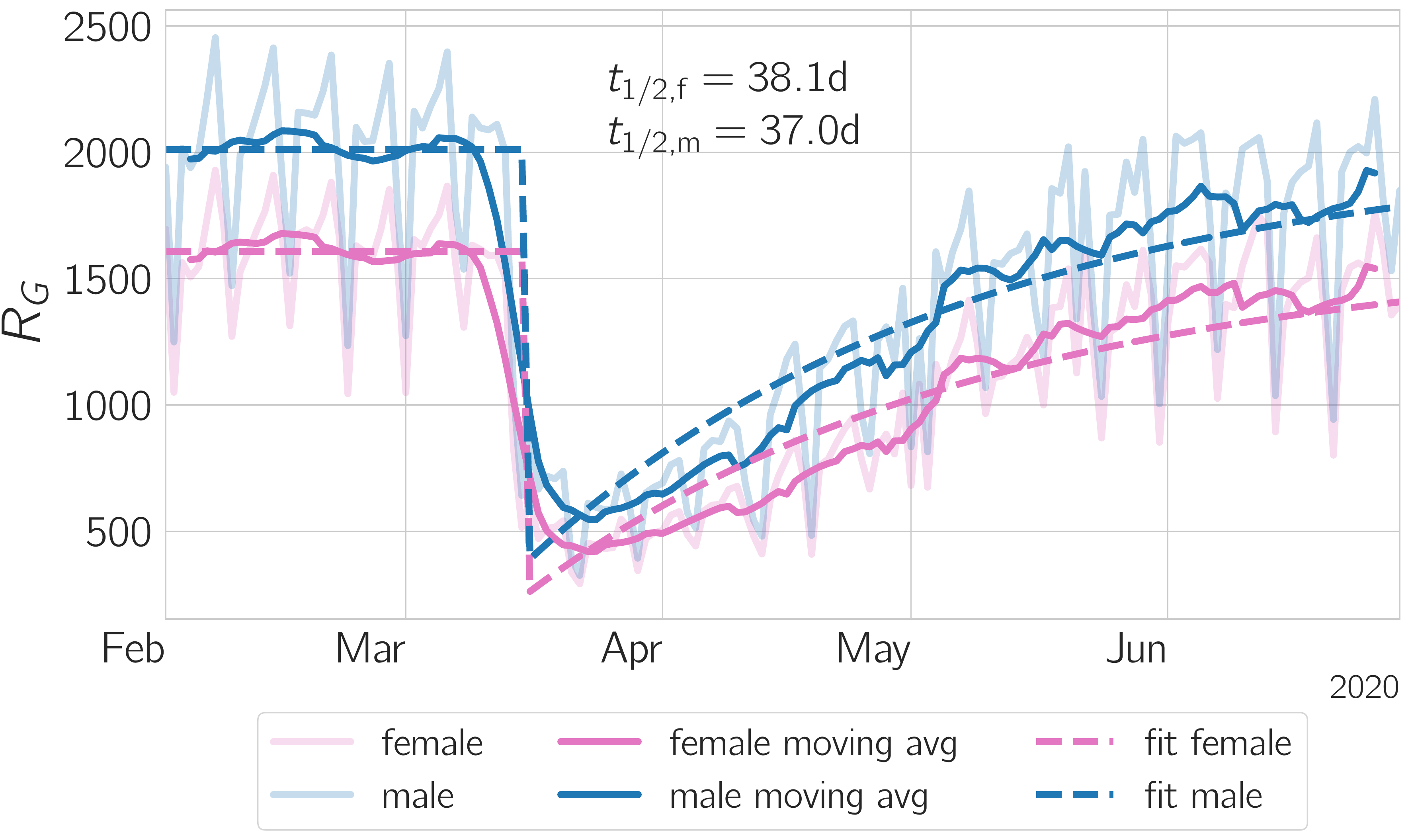}
	\caption{Decay parameters for the radius of gyration $R_G$. We show  $R_G$ (transparent line), the seven day moving average (solid line) and the fitted curve (broken line). The half-life times for the return from perturbed state to normal are 36.0d for women and 34.8d for men. Detailed results of the fitted parameters are in Tab. \ref{tab:exp_fit_results}.}
	\label{fig:expfit_rog}
\end{figure}

\begin{figure}
	\centering
	\includegraphics[width=0.4\linewidth]{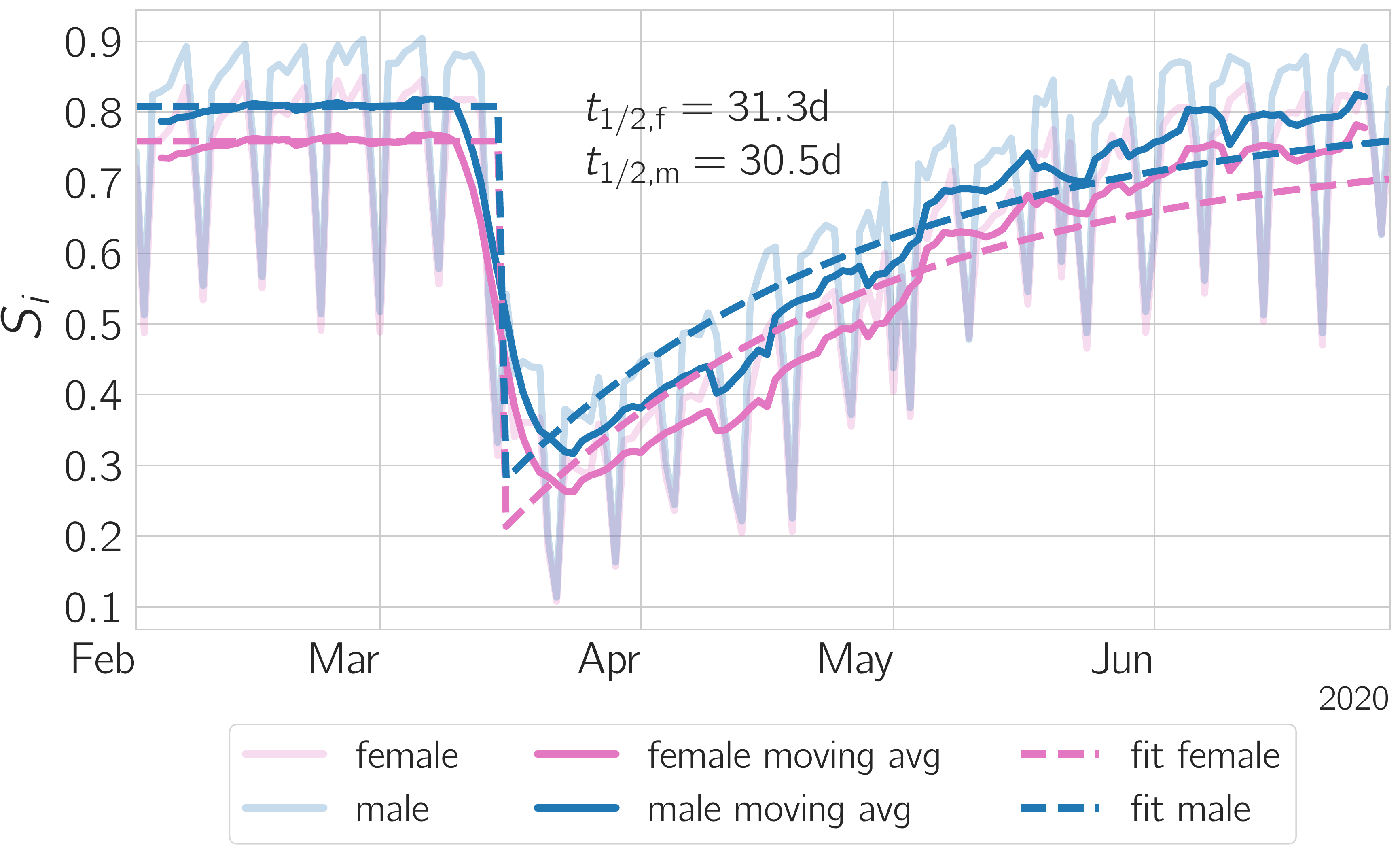}
	\caption{Decay parameters for the stay time entropy $S_i$. We show  $S_i$ (transparent line), the seven day moving average (solid line) and the fitted curve (broken line). The half-life times for the return from perturbed state to normal are 27.6d for women and 28.7d for men. Detailed results of the fitted parameters are in Tab. \ref{tab:exp_fit_results}.}
	\label{fig:expfit_entropy}
\end{figure}

We perform the same procedure for the mobility-related quantities radius of gyration $R_G$ and stay-time distribution entropy $S_i$ and show the fits in Figs. \ref{fig:expfit_rog} and \ref{fig:expfit_entropy}. The results for the fits are reported in Tab. \ref{tab:exp_fit_results}. The half-life times of females are slightly, but not significantly larger.

For $R_G$ we also stratify for age, see Tab. \ref{tab:exp_fit_results_rg_age}.

\begin{table}
	\centering
	\caption{Stratifying the fitted parameters of the exponential function for age. The confidence intervals are one standard deviation.}
	\begin{tabular}{ccccccccc}
		\hline
		gender & age &      $a_0$ &       $a_1$ &    stdev $a_1$ &        b &  stdev b &   thalbe & thalbe stdev \\
		\hline
		female & 15.0 &  2318.96 &  0.9822 &  0.0009 & -1997 &  63 &  38.6 &    2.0 \\
		female & 30.0 &  1809.65 &  0.9810 &  0.0009 & -1564 &  48 &  36.2 &    1.8 \\
		female & 45.0 &  1761.09 &  0.9810 &  0.0010 & -1446 &  45 &  36.2 &    1.8 \\
		female & 60.0 &   1038.4 &  0.9799 &  0.0010 &  -860 &  27 &  34.2 &    1.7 \\
		female & 75.0 &  526.376 &  0.9788 &  0.0011 &  -346 &  12 &  32.3 &    1.7 \\
		\hline
		male & 15.0   &  2498.53 &  0.9819 &  0.0009 & -1948 &  60 &  38.0 &      2.0 \\
		male & 30.0   &   2126.9 &   0.9802 &  0.0010 & -1696 &  51 &  34.7 &     1.7 \\
		male & 45.0   &  2089.17 &   0.9808 &  0.0009 & -1663 &  51 &  35.7 &     1.8 \\
		male & 60.0   &  1446.69 &  0.9791 &   0.0010 & -1195 &   37 &  32.8 &    1.6 \\
		male & 75.0   &  786.449 &  0.9762 &   0.0014 & -610 &  24 &  28.8 &     1.7 \\
		\hline
	\end{tabular}
	\label{tab:exp_fit_results_rg_age}
\end{table}

Through visual inspection it is clear that the quality of the fits of the mobility quantities are not good because the exponential function does not represent the functional form of the return to normal well. The derivative of $R_G$ and $S_i$ is small during the lock-down in phase \RNum{3} and gets larger in subsequent phases when the restrictions are lifted. We fit a logistic function of the form $$f(t;\alpha, t_0) = \frac{1}{1+e^{-\alpha (t-t_0)}} \quad \mathrm{.}$$ The turning point, with the largest derivative, is at $t_0$ and $\alpha$ controls the time it takes the logistic function to transition from one niveau to the other.

\begin{figure}
	\centering
	\includegraphics[width=0.4\linewidth]{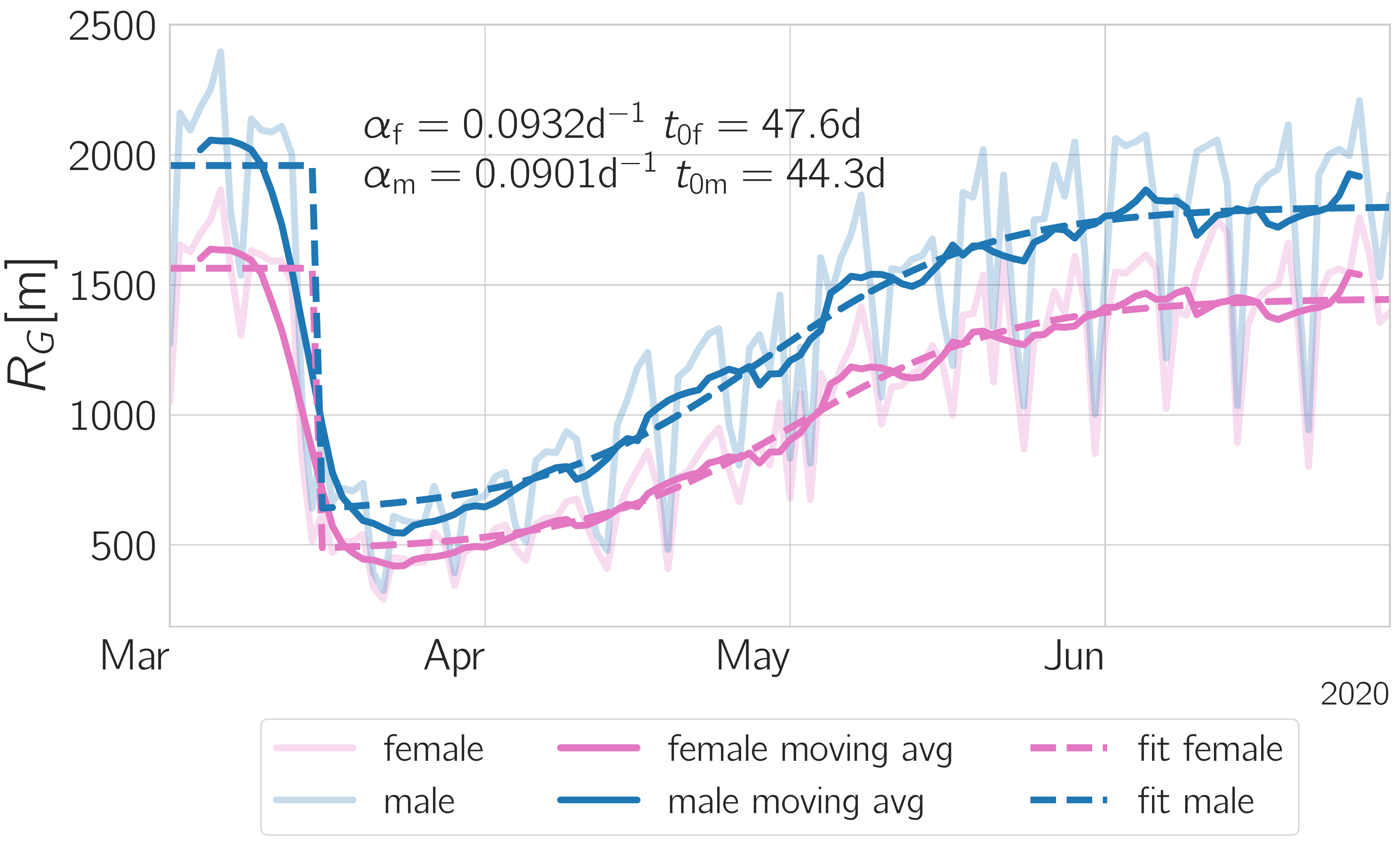}
	\caption{Logistic fit for the radius of gyration $R_G$. We show  $R_G$ (transparent line), the seven day moving average (solid line) and the fitted curve (broken line). The decay rate parameter $\alpha$ is similar for both genders, but the turning point $t_0$ for men is earlier than for women with 44.3d and 47.6d after beginning of the lock-down. Detailed results of the fitted parameters are in Tab. \ref{tab:step_fit_results}.}
	\label{fig:stepfit_rog}
\end{figure}

In Fig. \ref{fig:stepfit_rog} we show the results of the results of fitting the step function, the values for $t_0$ and $\alpha$ are reported in Tab.  \ref{tab:step_fit_results}. The $\alpha$ parameter is not significantly different, but the turning point $t_0$ is significantly larger for females.

\begin{figure}
	\centering
	\includegraphics[width=0.4\linewidth]{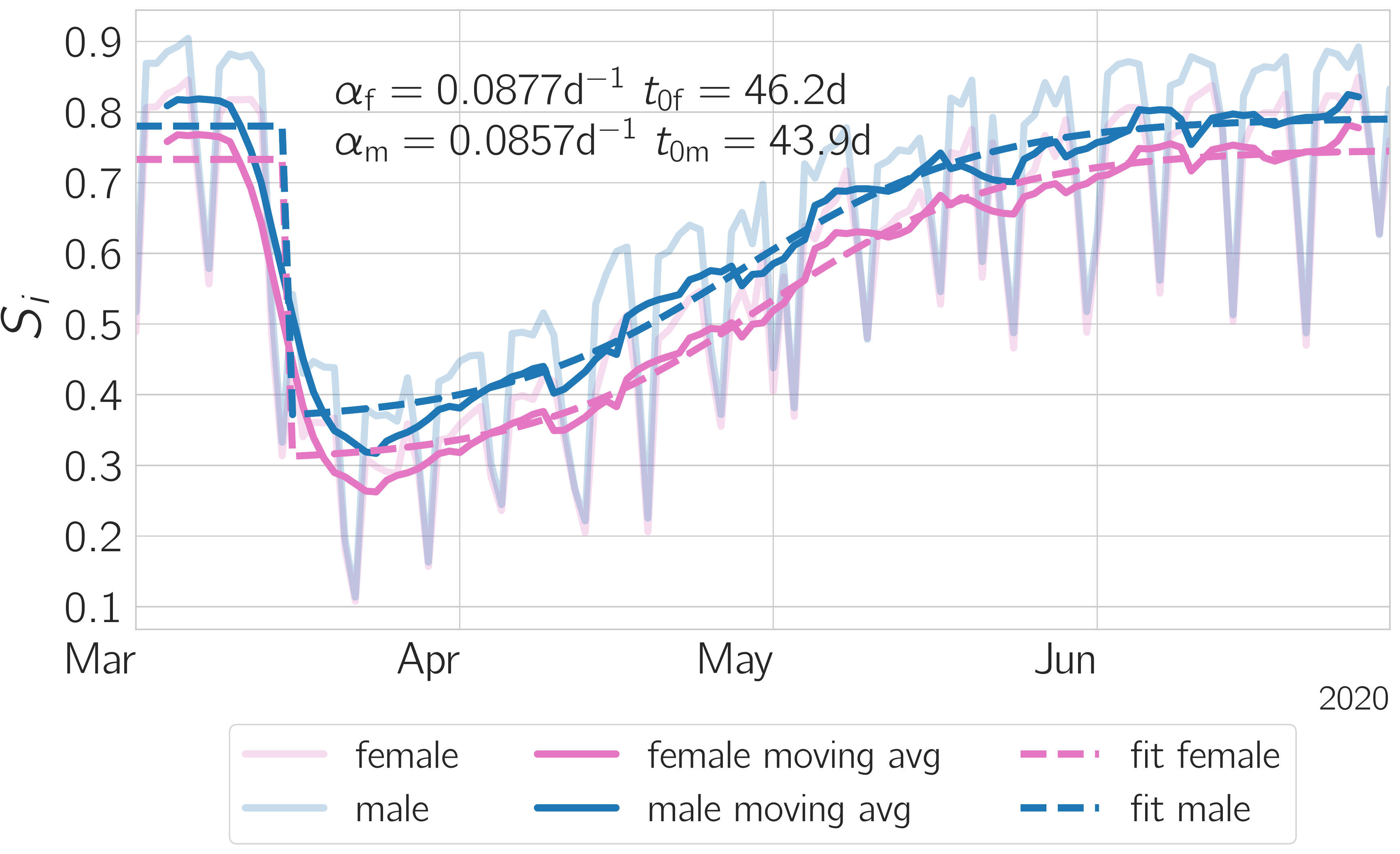}
	\caption{Logistic fit for the stay time entropy $S_i$. We show  $S_i$ (transparent line), the seven day moving average (solid line) and the fitted curve (broken line). The decay rate parameter $\alpha$ is similar for both genders, but as for $R_0$ the turning point $t_0$ for men is earlier than for women with 43.9d and 46.2d after beginning of the lock-down. Detailed results of the fitted parameters are in Tab. \ref{tab:step_fit_results}.}
	\label{fig:stepfit_entropy}
\end{figure}

The same holds for our second measure of mobility, the entropy $S_i$. The fits shown in Fig. \ref{fig:stepfit_entropy} and values reported in Tab.  \ref{tab:step_fit_results} have not significantly different values for $\alpha$, but the turning point $t_0$ for females is significantly later. The extended flat period in the beginning of the logistic function accounts for the extended duration of the lock-down.

\begin{table}
	\centering
	\caption{Fitted parameters for the logistic function. The confidence intervals are one standard deviation.}
	\begin{tabular}{c c c c}
		\hline
		quantity & gender & $\alpha$ & $t_0$ \\
		\hline
		$R_G$ & female & $0.0932 \pm 0.0070 \mathrm{d}^{-1}$ & $47.6 \pm 0.8$d \\ 
		$R_G$ & male & $0.0901 \pm 0.0078 \mathrm{d}^{-1}$ & $44.3 \pm 1.0$d\\ 
		
		$S_i$ & female & $0.0877 \pm 0.0078 \mathrm{d}^{-1}$ & $46.2 \pm 1.0$d \\ 
		$S_i$ & male & $0.0857 \pm 0.0087 \mathrm{d}^{-1}$ & $43.9 \pm 1.2$d\\
		\hline
	\end{tabular}
	\label{tab:step_fit_results}
\end{table}


\subsection*{SI Text S5: Further discussion of communication quantities}
In this SI Text we provide additional results concerning communication patterns.

\begin{figure}[htb]
	\centering
	\includegraphics[width=0.4\linewidth]{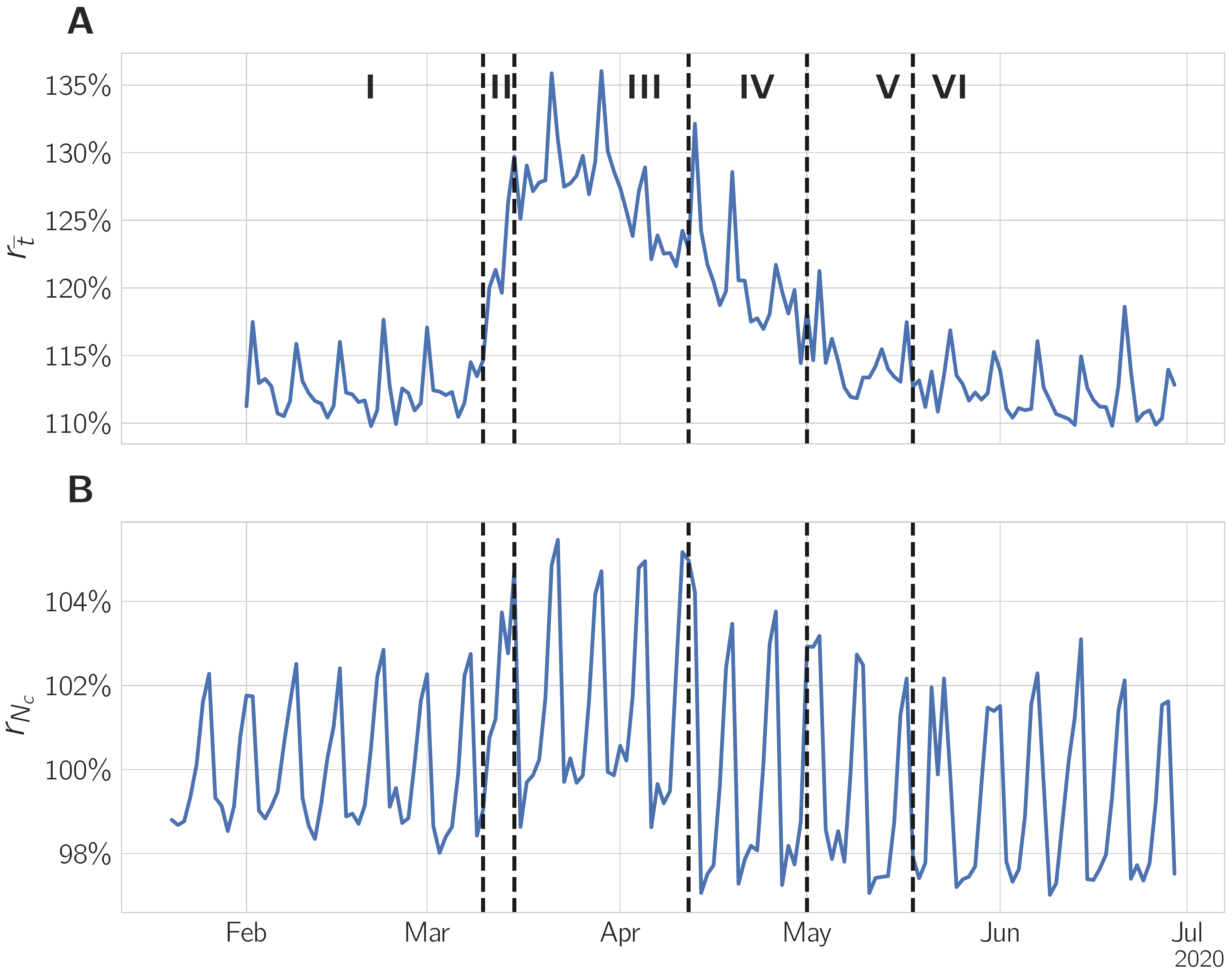}
	\caption{Gender ratios of communication quantities.
		(\textbf{A}) The ratio $r_{\bar{t}}$ of the call time of female over male initiated interactions starts to spike in \RNum{2}, increases in \RNum{3} and recovers subsequently. (\textbf{B})
		The ratio  $r_{N_c}$ of the number of female over male initiated calls is increased during \RNum{3}, but recovers subsequently.
	}
	\label{fig:final_female_inititated_unique_src_dst_count_sum}
\end{figure}

The absolute values of the interaction time $\bar{t}^{gh}$ are given in the main text Fig. \ref{fig2_communication}. Here we discuss the gender ratio and stratify for age. 
Supplementary Fig. \ref{fig:final_female_inititated_unique_src_dst_count_sum} A shows the call time female initiated over male initiated ratio $r_{\bar{t}}$ across the six phases. It is typically shifted towards women (+12\%) and even more so on weekends (+16\%). In the transition phase \RNum{2} the relative gender gap approximately doubles to +27\% on weekdays and up to +35\% on weekends. Subsequently we observe a smooth return to pre-lock-down values.

\begin{figure}
	\centering
	\includegraphics[width=0.4\linewidth]{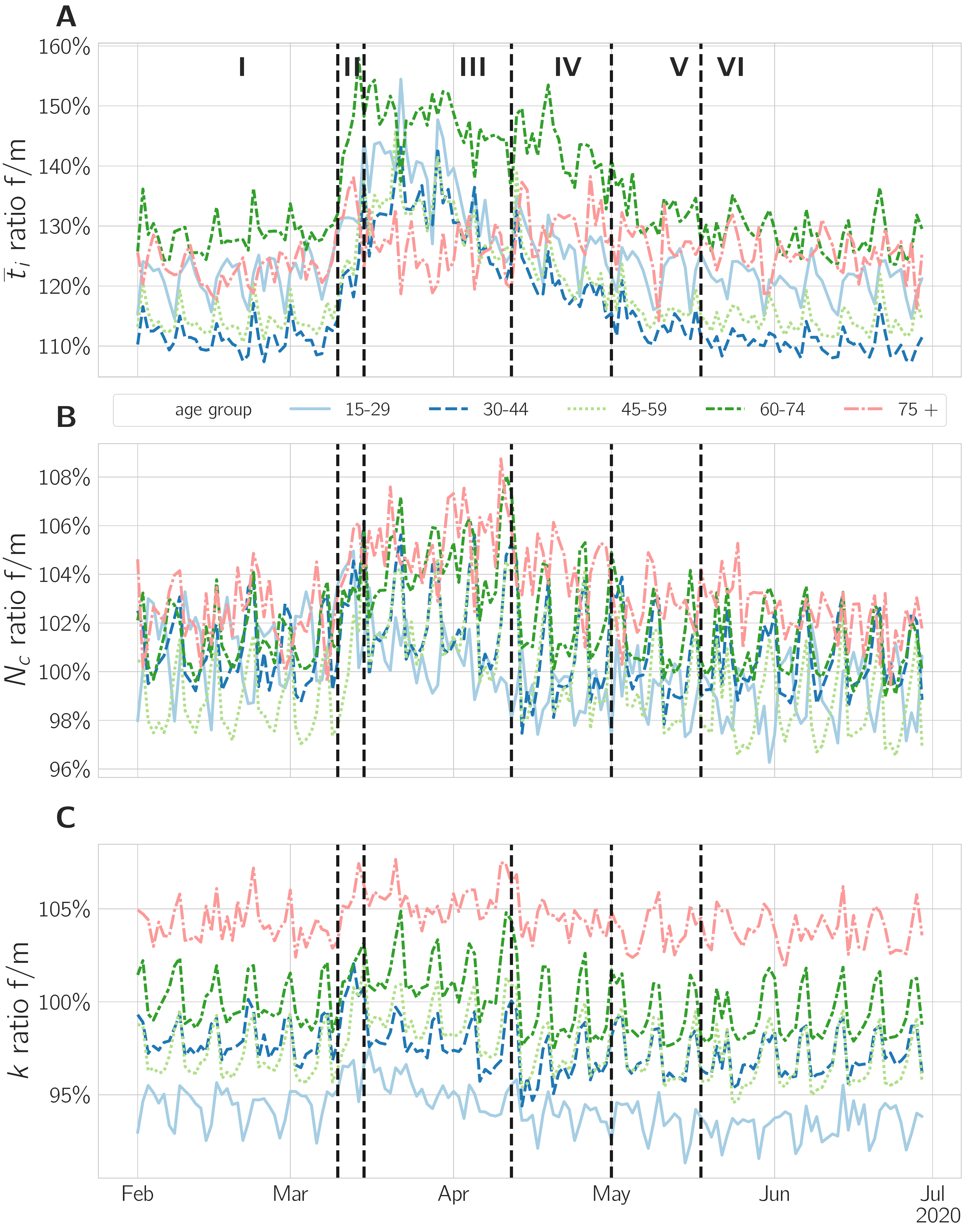}
	\caption{Age dependence of communication strength gender ratios. (\textbf{A}) Gender ratio for call time $\bar{t}$ , (\textbf{B}) gender ratio for number of calls (\textbf{C}) gender ratio for degree. The gender ratios overwhelmingly shift towards women communicating more. The effect is the weakest for the oldest cohort. The youngest cohort 15-29 has a different weekday - weekend pattern than the rest of the age cohorts.}
	\label{fig:communication_age_ratios}
\end{figure}

\begin{figure*}
	\centering
	\includegraphics[width=0.4\linewidth]{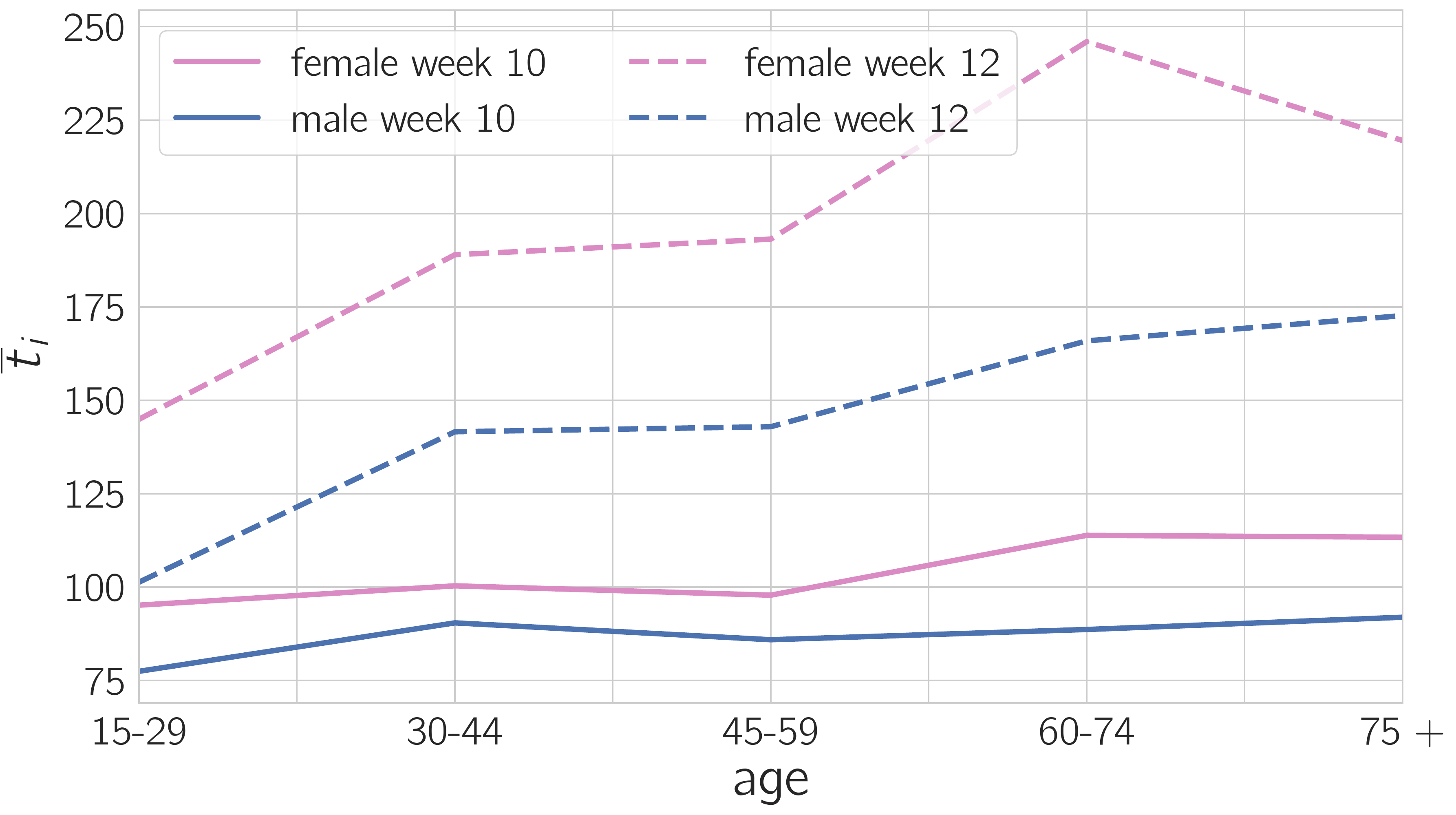}  \includegraphics[width=0.4\linewidth]{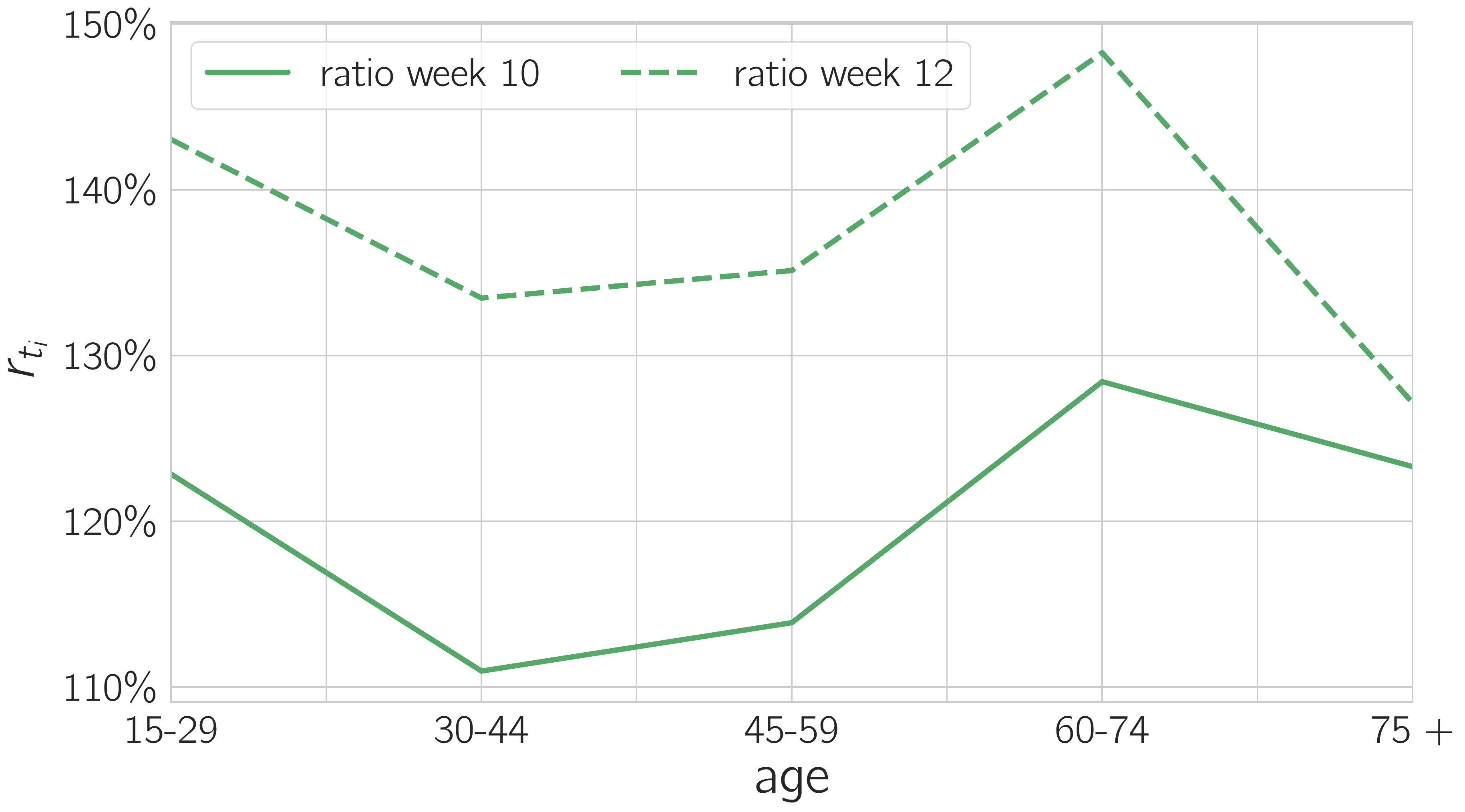}
	\caption{Age profile call duration $\bar{t}$. (\textbf{A}) Weekly average in $\bar{t}$ of week 10 (solid line) and week 12 (broken line). (\textbf{B}) Gender ratio of the weekly average in $\bar{t}$ of week 10 (solid line) and week 12 (broken line). All age cohorts have an absolute increase in $\bar{t}$, but its smallest for the youngest age cohort. The ratio changes only little for the oldest cohort. }
	\label{fig:age_profile_call_time_sum}
\end{figure*}

We stratify for age in Fig. \ref{fig:age_profile_call_time_sum} A, where we show the average $\bar{t}^{g}$ over calendar week 10 in phase \RNum{1} and over calendar week 12 in phase \RNum{3} for all age cohorts. The phase \RNum{1} profile is flat with values slightly above 300s for women and slightly below 300s for men. In the lock-down phase \RNum{3} the call time increases for both genders and all ages, but stronger for women  and stronger for older age cohorts. The 15-29 cohort increases their communication behavior the least. 
The gender ratio for $\bar{t}^g$ is shown in Fig. \ref{fig:age_profile_call_time_sum} B. It is shifted towards women for all age cohorts and the lowest for ages 30-59. We see a strong increase in week 12 for all ages, except the 75+ cohort.
In Fig.  \ref{fig:communication_age_ratios} A we show the time series of the age ratios, already discussed in the main text.

Next, we discuss the the number of calls as shown in main text Fig.  \ref{fig2_communication} B.
As shown in Fig. \ref{fig:final_female_inititated_unique_src_dst_count_sum} B, before the crisis women tended to have slightly fewer calls than men on weekdays (3\% less) and more on weekends (2\% more). This changes in phase \RNum{3}, where the bias shifts towards women having equally as many  calls on weekdays as well as an increased bias on weekends (4\% more). The bias returns to normal levels in phase \RNum{4}, however, the weekday ratio is shifted to be more male-biased. 
We report significance tests in SI Text S6. 

The gender ratio of $N_c$, shown in SI Fig. \ref{fig:communication_age_ratios} B, deviates only weakly from equality before the lock-down. All cohorts, except 15-29, are biased towards females having longer calls on weekends, except for the 75+ which does not exhibit a weekly rhythm.
In phase \RNum{3} the gender ratio for all age cohorts shifts towards women having more calls. 

\begin{figure*}
	\centering
	\includegraphics[width=0.4\linewidth]{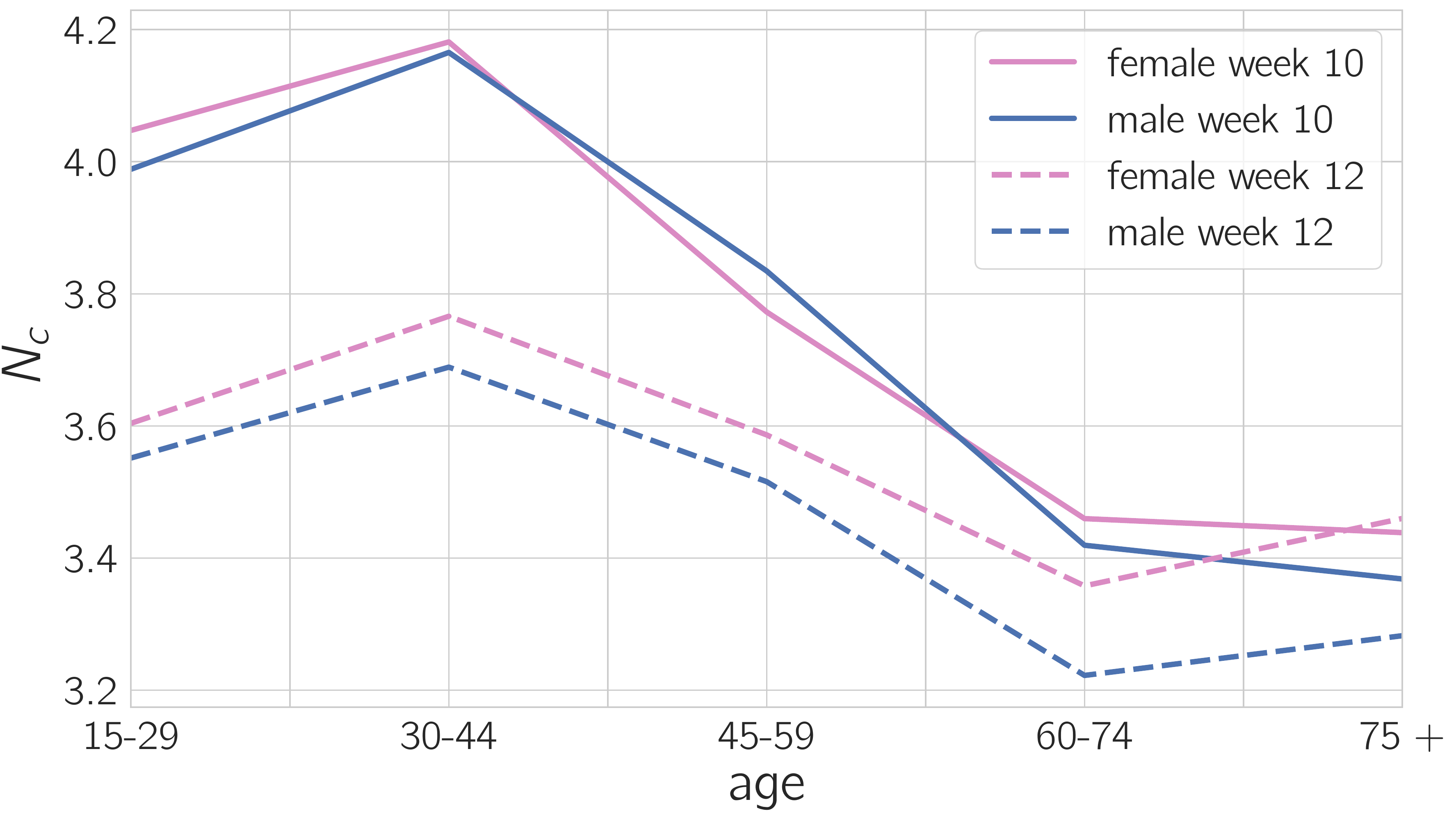}
	\includegraphics[width=0.4\linewidth]{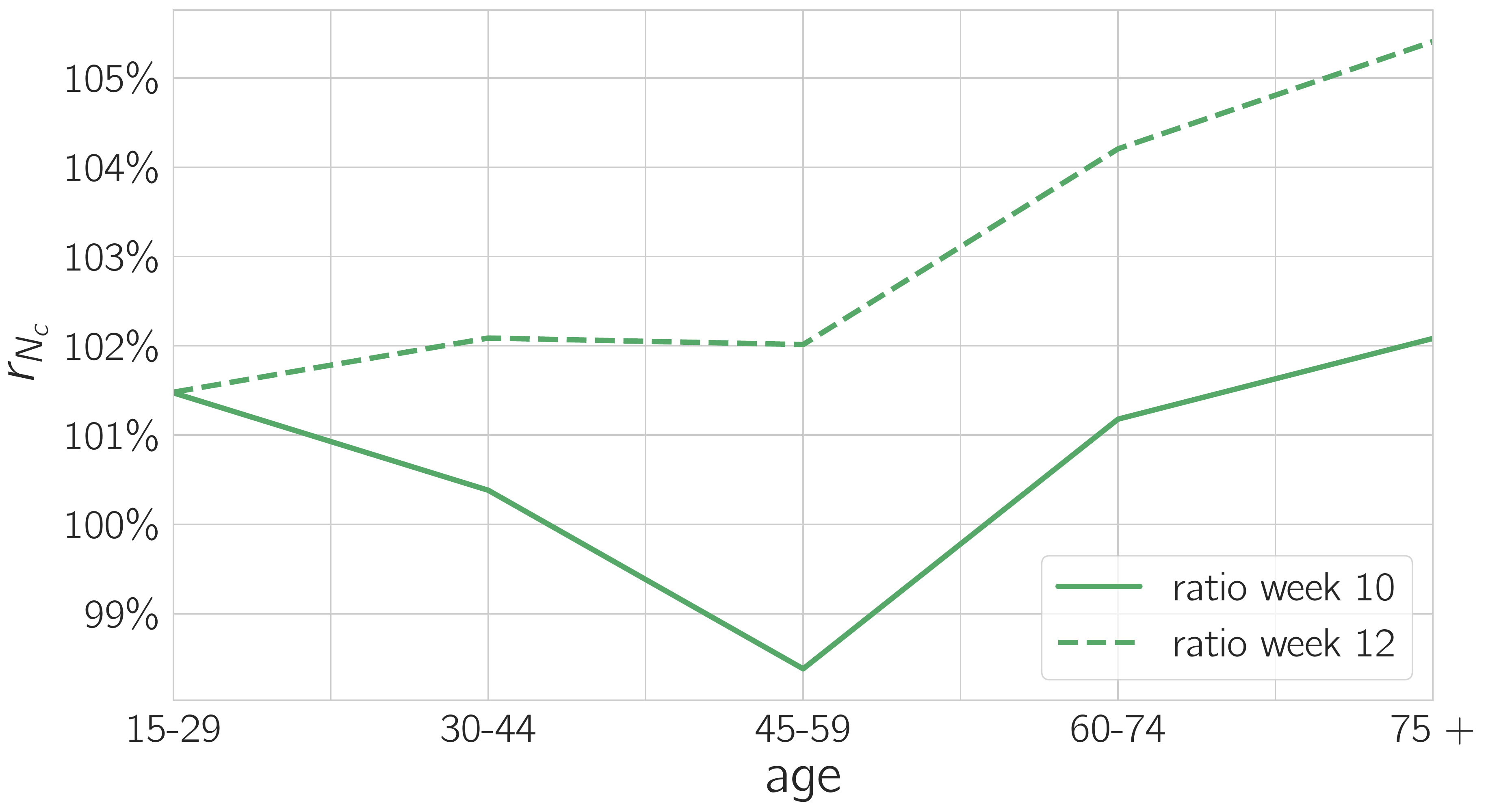}
	\caption{Age profile number of calls $N_c$. (\textbf{A}) Weekly average in $N_c$ of week 10 (solid line) and week 12 (broken line). (\textbf{B}) Gender ratio of the weekly average in $N_c$ of week 10 (solid line) and week 12 (broken line). All age cohorts have an absolute increase in $N_c$, but its smallest for the oldest age cohort. However, the ratio changes only little for the youngest and much for the oldest cohort. }
	\label{fig:age_profile_Nc}
\end{figure*}

The age profiles in Fig.  \ref{fig:age_profile_Nc} elucidate this a bit more, by showing that from week 10 to week 12 the number of calls decreased for all age cohorts and especially for young cohorts, which have a higher initial level in week 10. The gender ratio increases from week 10 to week 12 for all age cohorts, except for the cohort 15-29, which remains at the same value.

\begin{figure*}
	\centering
	\includegraphics[width=0.4\linewidth]{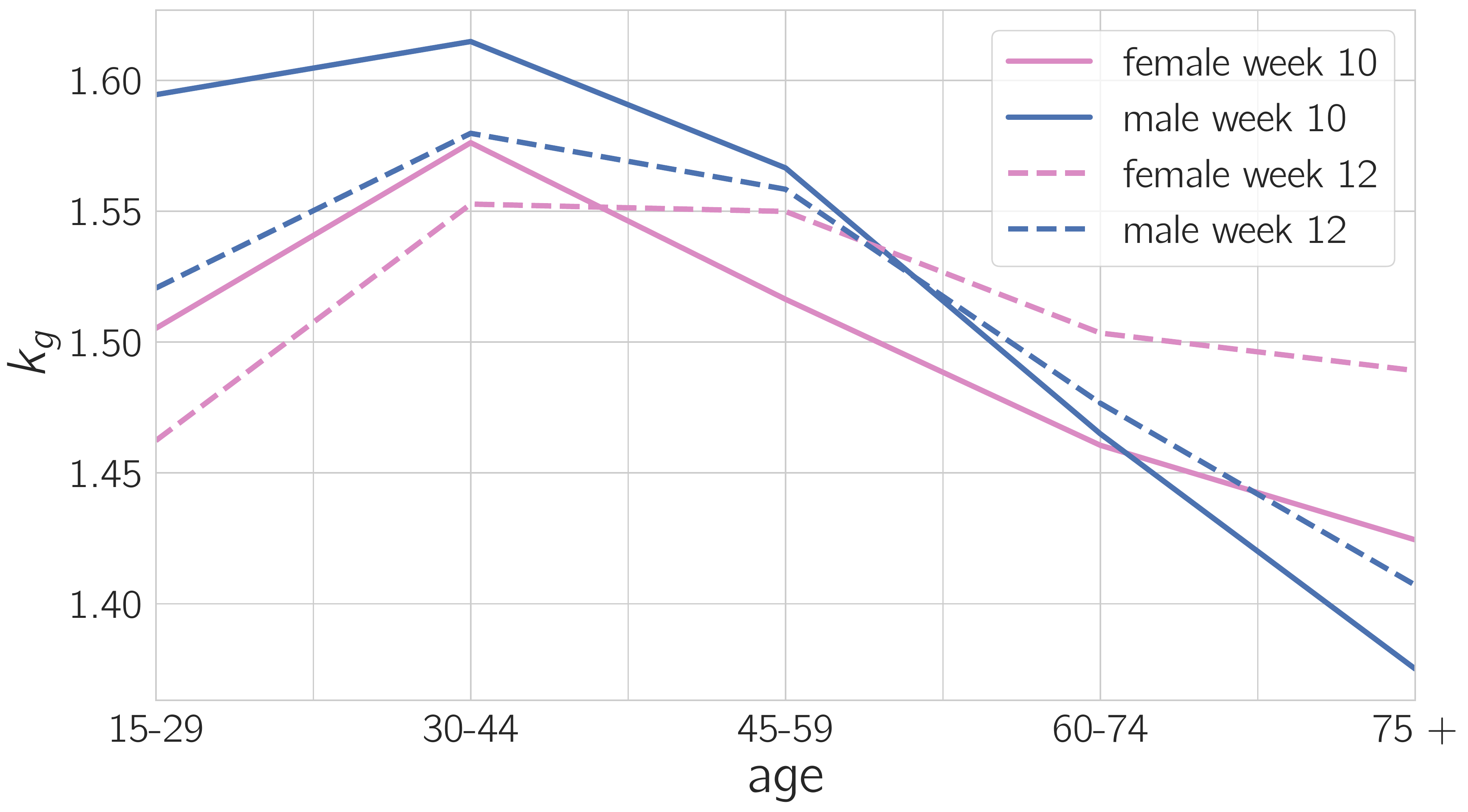}  
	\includegraphics[width=0.4\linewidth]{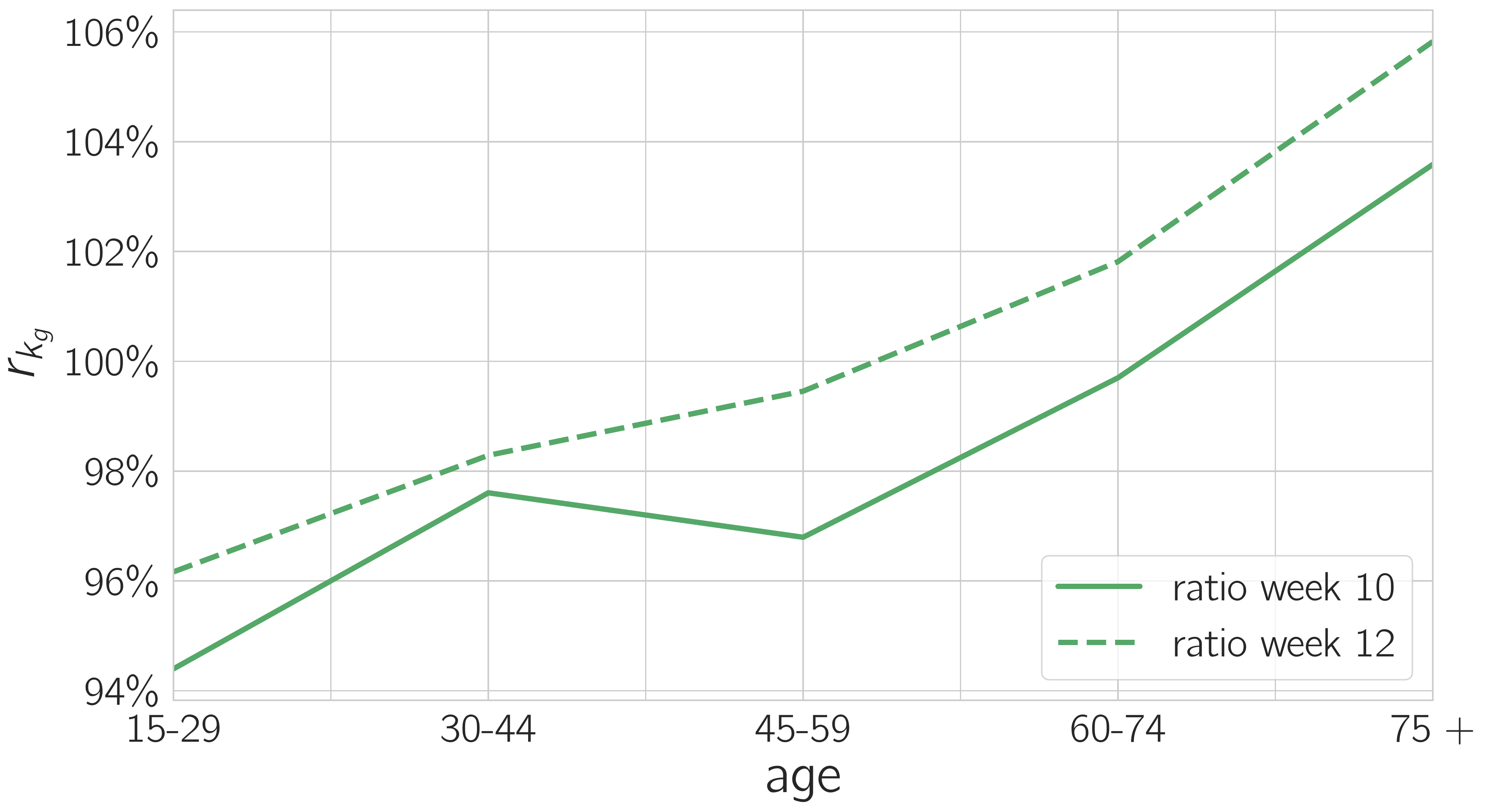}
	\caption{Age profile degree $k_i$. (\textbf{A}) Weekly average in $k_i$ of week 10 (solid line) and week 12 (broken line). (\textbf{B}) Gender ratio of the weekly average in $k_i$ of week 10 (solid line) and week 12 (broken line). This curve is interestingly different from the number of calls in Fig. \ref{fig:age_profile_Nc}. The absolute values of unique communication partners only increase for younger cohorts and decrease for senior cohorts. Nevertheless, gender ratio rises for all age cohorts.}
	\label{fig:age_profile_degree}
\end{figure*}

Mean degree varies around 1.4 and 1.7, so the range is relatively small. The age stratification in Fig. \ref{fig:age_profile_degree} A shows that degree is highest for the age cohort 30-44. We see that for the age cohorts younger than 45 there is a reduction in degree, for men 45-59 there is a very small reduction and for women older than 45 and men 60+ there is an increase in degree.

\subsection*{SI Text S6: Statistical significance of gender-ratio-changes across phases}

We perform significance tests to ensure that the changes gender ratio in the observed quantities are indeed significant.
Gender ratios of different samples of observables (e.g. phases) are compared with a two-sided Mann-Whitney-U test to reject the null hypothesis that they are from the same distribution. The results are shown in Tab. \ref{t:significance} for the overall quantities and in Tab. \ref{t:age-group-sign} we report age stratified tests.

For $R_G$ we can show that the reduction from phase \RNum{1} to phases \RNum{2}, \RNum{3}, \RNum{4}, \RNum{5} and \RNum{6} is highly significant on weekdays, while the change on weekends is less significant, except for phase \RNum{4}. The change from before to after Easter, phase \RNum{3} vs. \RNum{4} is significant as well on weekends and weekends.

\begin{table}
	\centering
	\caption{
		A Mann Whitney U test was applied to test for significance.
		We compare the ratio female/male for various metrics of one period against another period, to reject the hypothesis that they are drawn from the same distribution.
		The following rules were used when assigning the significance stars: $* < 0.05, ** < 0.01, *** < 0.001$
		\label{t:significance}
	}
\begin{tabular}{llll}
\hline
                               metric & compared periods & \multicolumn{2}{l}{Mann-Whitney} \\
                                      &                  &         Weekday &        Weekend \\
\hline
                            $R_G$ [m] &        I vs. III &  3.94e-10, *** &    2.69e-01, - \\
                            $R_G$ [m] &         I vs. IV &  6.19e-08, *** &   6.29e-03, ** \\
                            $R_G$ [m] &          I vs. V &  3.07e-05, *** &    3.95e-01, - \\
                            $R_G$ [m] &         I vs. VI &   5.02e-03, ** &    2.08e-02, * \\
                            $R_G$ [m] &       III vs. IV &  1.45e-05, *** &    2.02e-02, * \\
                                  $S$ &        I vs. III &   2.88e-03, ** &  7.94e-05, *** \\
                                  $S$ &         I vs. IV &  6.19e-08, *** &    4.37e-01, - \\
                                  $S$ &          I vs. V &  3.31e-04, *** &    3.16e-02, * \\
                                  $S$ &         I vs. VI &  4.19e-04, *** &  6.51e-04, *** \\
                                  $S$ &       III vs. IV &  9.15e-07, *** &    2.02e-02, * \\
                                $S_i$ &        I vs. III &  3.94e-10, *** &    2.17e-02, * \\
                                $S_i$ &         I vs. IV &  6.19e-08, *** &   6.29e-03, ** \\
                                $S_i$ &          I vs. V &  7.49e-07, *** &   4.64e-03, ** \\
                                $S_i$ &         I vs. VI &    3.27e-01, - &  5.78e-04, *** \\
                                $S_i$ &       III vs. IV &    1.31e-01, - &    8.96e-02, - \\
 total call duration (MO) [s] &        I vs. III &  3.93e-10, *** &  2.35e-05, *** \\
 total call duration (MO) [s] &         I vs. IV &  2.63e-07, *** &   7.24e-03, ** \\
 total call duration (MO) [s] &          I vs. V &    3.77e-01, - &    3.55e-01, - \\
 total call duration (MO) [s] &         I vs. VI &  2.13e-11, *** &    5.12e-02, - \\
 total call duration (MO) [s] &       III vs. IV &  9.15e-07, *** &   9.81e-03, ** \\
 total call duration (MT) [s] &        I vs. III &  3.94e-10, *** &  2.35e-05, *** \\
 total call duration (MT) [s] &         I vs. IV &  6.19e-08, *** &    1.91e-02, * \\
 total call duration (MT) [s] &          I vs. V &  1.28e-04, *** &    3.16e-01, - \\
 total call duration (MT) [s] &         I vs. VI &  3.55e-07, *** &    2.35e-02, * \\
 total call duration (MT) [s] &       III vs. IV &  1.10e-06, *** &   4.44e-03, ** \\
                               degree &        I vs. III &  5.97e-04, *** &  2.36e-05, *** \\
                               degree &         I vs. IV &  5.88e-06, *** &    3.55e-01, - \\
                               degree &          I vs. V &  5.21e-04, *** &    3.55e-01, - \\
                               degree &         I vs. VI &  1.34e-07, *** &   2.22e-03, ** \\
                               degree &       III vs. IV &  3.79e-06, *** &   2.91e-03, ** \\
\hline
\end{tabular}\end{table}

\begin{table*}
	\centering
	\caption{
		Significance testing results per gendered age group for selected metrics. We apply a Mann Whitney U test, comparing the ratio female/male for various metrics of one period against another period, to reject the hypothesis that they are drawn from the same distribution.
		The following rules were used when assigning the significance stars: $* < 0.05, ** < 0.01, *** < 0.001$
		\label{t:age-group-sign}
	}
\begin{tabular}{lllll}
\hline
                               metric & compared periods & age group & \multicolumn{2}{l}{Mann-Whitney} \\
                                      &        Weekday &        Weekend \\
\hline
                            $R_G$ [m] &        I vs. III &     15-29 &  1.18e-09, *** &  1.26e-04, *** \\
                            $R_G$ [m] &        I vs. III &     30-44 &  1.18e-09, *** &  3.42e-04, *** \\
                            $R_G$ [m] &        I vs. III &     45-59 &  1.68e-09, *** &    3.83e-01, - \\
                            $R_G$ [m] &        I vs. III &     60-74 &    2.29e-02, * &  8.68e-04, *** \\
                            $R_G$ [m] &        I vs. III &      75 + &  1.18e-09, *** &  4.37e-05, *** \\
                            $R_G$ [m] &         I vs. IV &     15-29 &  1.17e-07, *** &   2.19e-03, ** \\
                            $R_G$ [m] &         I vs. IV &     30-44 &  1.17e-07, *** &   2.19e-03, ** \\
                            $R_G$ [m] &         I vs. IV &     45-59 &  1.17e-07, *** &    1.24e-02, * \\
                            $R_G$ [m] &         I vs. IV &     60-74 &  1.34e-07, *** &    4.28e-01, - \\
                            $R_G$ [m] &         I vs. IV &      75 + &    5.54e-02, - &    3.01e-02, * \\
                            $R_G$ [m] &          I vs. V &     15-29 &  3.94e-05, *** &    1.82e-01, - \\
                            $R_G$ [m] &          I vs. V &     30-44 &  3.50e-05, *** &    1.51e-01, - \\
                            $R_G$ [m] &          I vs. V &     45-59 &  7.96e-05, *** &    4.76e-01, - \\
                            $R_G$ [m] &          I vs. V &     60-74 &   3.29e-03, ** &    4.76e-01, - \\
                            $R_G$ [m] &          I vs. V &      75 + &    4.21e-01, - &    2.27e-02, * \\
                            $R_G$ [m] &         I vs. VI &     15-29 &  7.57e-07, *** &   2.65e-03, ** \\
                            $R_G$ [m] &         I vs. VI &     30-44 &    1.69e-02, * &    6.02e-02, - \\
                            $R_G$ [m] &         I vs. VI &     45-59 &    1.03e-01, - &    1.12e-01, - \\
                            $R_G$ [m] &         I vs. VI &     60-74 &   1.73e-03, ** &    3.48e-02, * \\
                            $R_G$ [m] &         I vs. VI &      75 + &  1.68e-10, *** &    2.11e-01, - \\
                            $R_G$ [m] &       III vs. IV &     15-29 &    1.93e-01, - &    4.72e-01, - \\
                            $R_G$ [m] &       III vs. IV &     30-44 &  5.56e-04, *** &    2.81e-02, * \\
                            $R_G$ [m] &       III vs. IV &     45-59 &  1.23e-04, *** &   9.81e-03, ** \\
                            $R_G$ [m] &       III vs. IV &     60-74 &  6.33e-06, *** &    3.85e-02, * \\
                            $R_G$ [m] &       III vs. IV &      75 + &  1.58e-06, *** &    3.85e-02, * \\
 total call duration (MO) [s] &        I vs. III &     15-29 &  7.41e-07, *** &  7.49e-05, *** \\
 total call duration (MO) [s] &        I vs. III &     30-44 &  1.18e-09, *** &  4.37e-05, *** \\
 total call duration (MO) [s] &        I vs. III &     45-59 &  1.18e-09, *** &  4.37e-05, *** \\
 total call duration (MO) [s] &        I vs. III &     60-74 &  1.18e-09, *** &  4.37e-05, *** \\
 total call duration (MO) [s] &        I vs. III &      75 + &  1.17e-08, *** &    2.44e-01, - \\
 total call duration (MO) [s] &         I vs. IV &     15-29 &    2.60e-01, - &    2.93e-01, - \\
 total call duration (MO) [s] &         I vs. IV &     30-44 &  4.48e-07, *** &    6.48e-02, - \\
 total call duration (MO) [s] &         I vs. IV &     45-59 &  1.17e-07, *** &   4.56e-03, ** \\
 total call duration (MO) [s] &         I vs. IV &     60-74 &  1.17e-07, *** &   2.19e-03, ** \\
 total call duration (MO) [s] &         I vs. IV &      75 + &  2.31e-07, *** &   2.19e-03, ** \\
 total call duration (MO) [s] &          I vs. V &     15-29 &    1.02e-02, * &    3.93e-02, * \\
 total call duration (MO) [s] &          I vs. V &     30-44 &    4.09e-01, - &    1.82e-01, - \\
 total call duration (MO) [s] &          I vs. V &     45-59 &  1.40e-04, *** &    3.36e-01, - \\
 total call duration (MO) [s] &          I vs. V &     60-74 &  7.48e-04, *** &    2.52e-01, - \\
 total call duration (MO) [s] &          I vs. V &      75 + &  6.78e-04, *** &    4.28e-01, - \\
 total call duration (MO) [s] &         I vs. VI &     15-29 &  1.03e-07, *** &   4.90e-03, ** \\
 total call duration (MO) [s] &         I vs. VI &     30-44 &  1.02e-12, *** &  4.05e-04, *** \\
 total call duration (MO) [s] &         I vs. VI &     45-59 &   1.43e-03, ** &    2.81e-01, - \\
 total call duration (MO) [s] &         I vs. VI &     60-74 &    3.78e-01, - &    3.75e-01, - \\
 total call duration (MO) [s] &         I vs. VI &      75 + &   3.89e-03, ** &    2.70e-02, * \\
 total call duration (MO) [s] &       III vs. IV &     15-29 &  2.16e-05, *** &    2.02e-02, * \\
 total call duration (MO) [s] &       III vs. IV &     30-44 &  9.15e-07, *** &   4.44e-03, ** \\
 total call duration (MO) [s] &       III vs. IV &     45-59 &  1.32e-06, *** &   6.66e-03, ** \\
 total call duration (MO) [s] &       III vs. IV &     60-74 &  1.57e-06, *** &    4.72e-01, - \\
 total call duration (MO) [s] &       III vs. IV &      75 + &    4.49e-01, - &    5.19e-02, - \\
\hline
\end{tabular}
\end{table*}

\subsection*{SI Text S7: Alternative measure for mobility: entropy}
We calculate an alternative measure for mobility, the entropy $S_i$ of the stay time distribution of device i, for a definition see SI Text S3 Methods. A high (low) value of entropy denotes a spread out (concentrated) stay time distribution, i.e. the device was moving a lot (little).

\begin{figure*}[ht]
	\centering
	\includegraphics[width=0.8\linewidth]{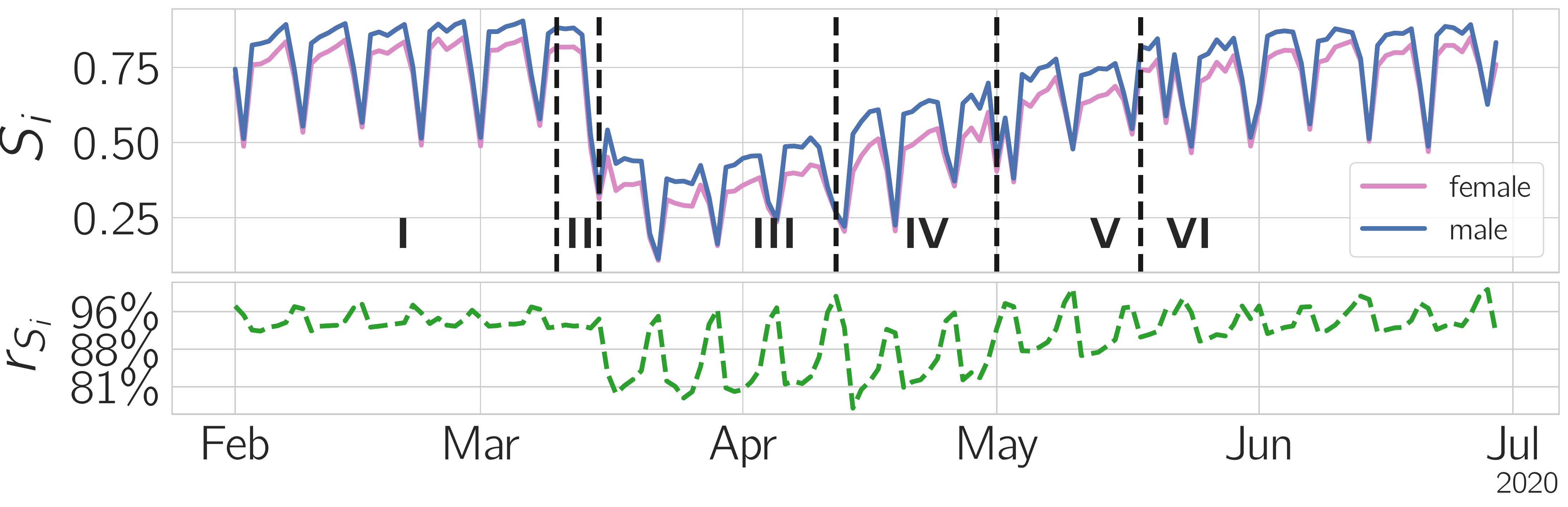}
	\caption{Mobility in Austria, measured by entropy. The upper panel shows entropy $S_i$ for men and women, the lower the gender ratio for entropy $r_{S_i}$. The drop at the beginning of phase \RNum{3} is very pronounced and so is the change in gender ratio. The weekday gender ratio is shifted towards men moving more and slowly returns to pre-crisis levels.}
	\label{fig:entropy_Si}
\end{figure*}

Figure \ref{fig:entropy_Si} A shows $S_i$ for both genders across the six phases. We observe a very similar pattern as for the radius of gyration. During the week $S_i$ is high and on weekends it is significantly less. In phase \RNum{3} $S_i$ reaches a minimum and only slowly recovers over the subsequent phases.

The gender ratio of $S_i$, shown in Fig.  \ref{fig:entropy_Si} B also corroborates the radius of gyration. In phase \RNum{1} the gender ratio is shifted towards men moving more, an effect which is smaller on weekends. During the lock-down in phase \RNum{3} the ratio is shifted strongly towards men and changed little to nothing on weekends. In phase \RNum{5} the ratio recovers to pre-crisis values.

\subsection*{SI Text S8: Long term analysis}
\begin{figure}[h]
	\centering
	\includegraphics[width=0.5\linewidth]{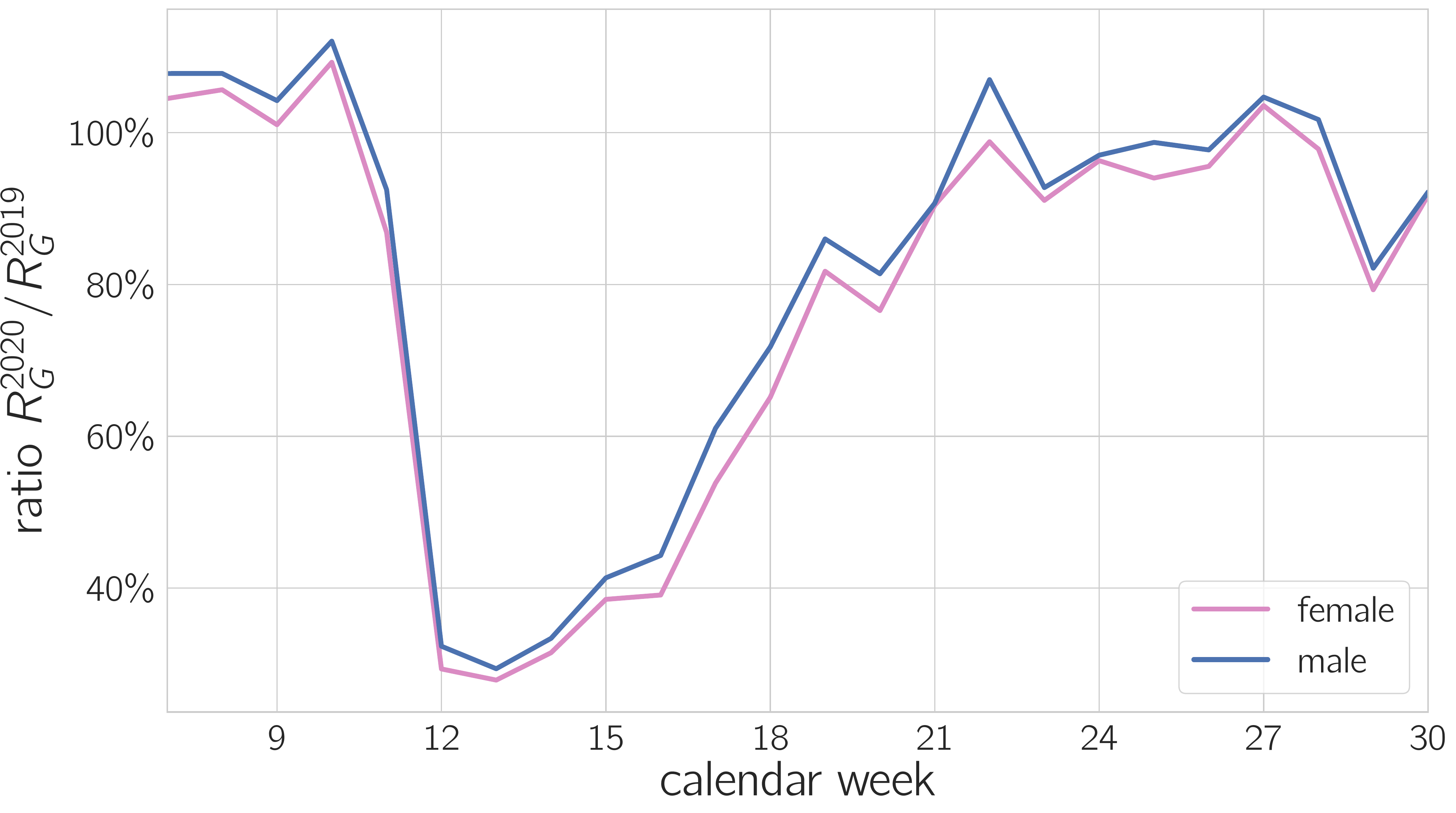}
	\caption{Comparison of the gendered $R_G$ with the same time period of the previous year. We plot the quotient of 2019's $R_G$ with 2020's. During the lock-down the ratio drops below 40\%.}
	\label{fig:last_year_comparison}
\end{figure}

A comparison with the data from the same time period of the previous year is available in Fig. \ref{fig:last_year_comparison}. The figure shows the ratio of radius of gyration from 2020 and 2019. If $R_G$ was smaller (larger) in 2020 than in 2019, the ratio is less (more) than 100\%.

The drastic reduction of movement to approximately 30\% compared to the previous year shows the impact of the measures. We also observe that women have a stronger reduction from week 16 to 20.

These results should be interpreted with care, as the accuracy of the localization in the GSM network was improved from 2019 to 2020, potentially resulting in different absolute values of $R_G$.

\subsection*{SI Text S9: Shopping centers}
\begin{figure}[h]
	\centering
	\includegraphics[width=0.5\linewidth]{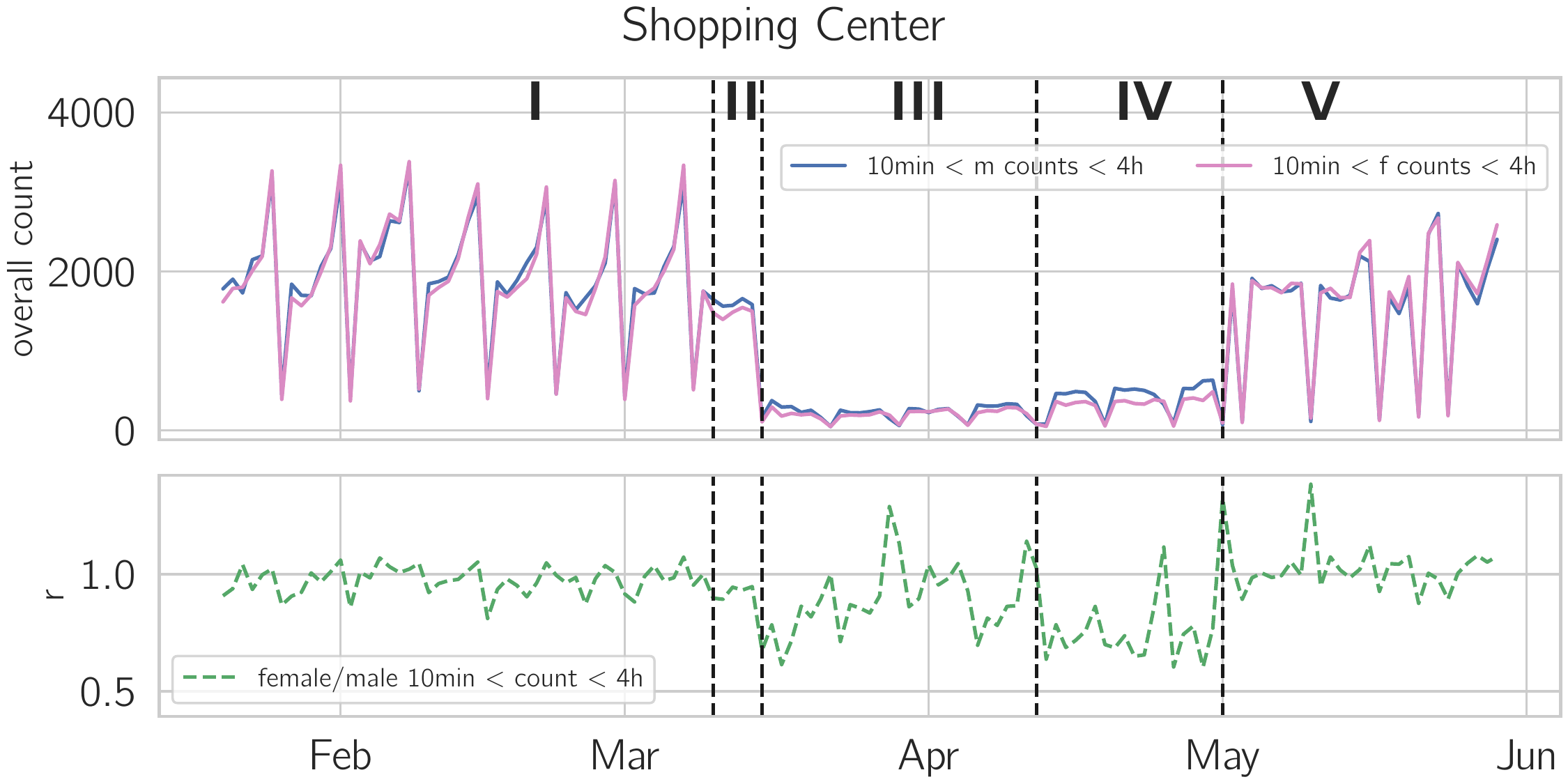}
	\caption{Shopping center in 2020. The upper panel shows the visitor count for a large shopping center in the south of Vienna. In the lower panel the gender ratio for the visitor count in panel A is shown. The lock-down is prominently visible as a drop in the visitor count and a deviation from equality in the ratio.}
	\label{fig:shopping1_2020}
\end{figure}

Here we present the visiting data for a large shopping center in the south of Vienna, as described in the main text. Figure \ref{fig:shopping1_2020} depicts data from the lock-down and is discussed in the main text. 

\begin{figure}[htbp]
	\centering
	\includegraphics[width=0.5\linewidth]{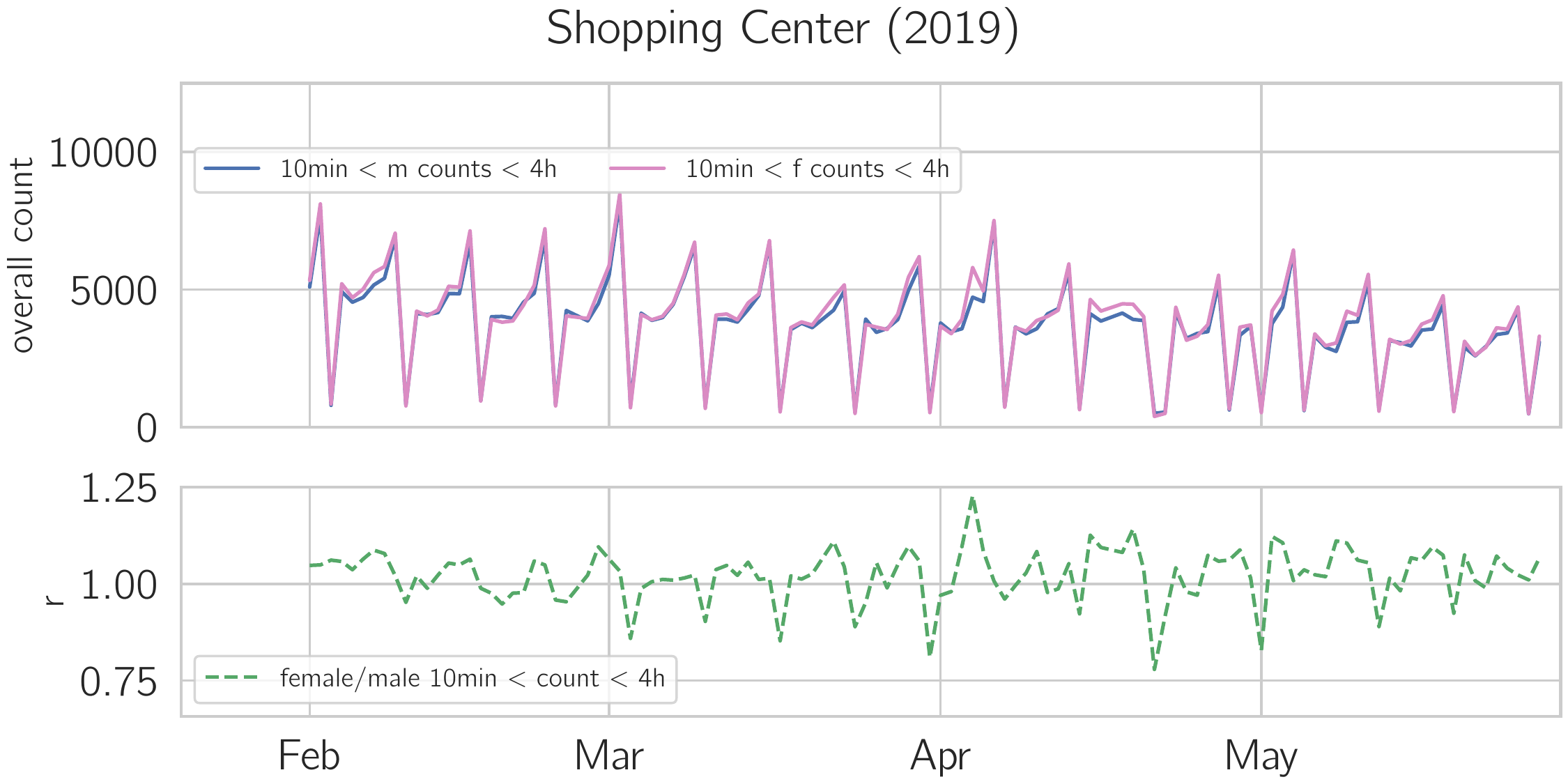}
	\caption{Shopping center in 2019. The upper panel shows the visitor count for the same period as in Fig. \ref{fig:shopping1_2020} in 2019 for a large shopping center in the south of Vienna. The lower panel displays the gender ratio for the visitor count in panel A. The changes from 2020 are not visible in this figure, however a declining trend from February towards summer is present.}
	\label{fig:shopping1_2019}
\end{figure}

For comparison we analyze the same time from the previous year, 2019, see Fig. \ref{fig:shopping1_2019}. In the visitor numbers we find the same weekly pattern as in phase \RNum{1} and a declining trend towards the summer months, which is, nevertheless, much weaker than the reduction in phases \RNum{3} and \RNum{4}. The gender ratio is quite stable around (or slightly above) 1.

\subsection*{SI Text S10: Statistical significance of gender ratio shopping center and recreational area}

We perform statistical tests to check if the observed changes in gender ratio from one phase to another are significant. A two sided Mann-Whitney U test is performed to reject the hypothesis that the values in both phases are drawn from the same distribution.

The changes in gender ratio from phase \RNum{1} to phases \RNum{3}, \RNum{4} and \RNum{5} are are highly significant on weekdays and not significant on weekends.

For the recreational area we find that changes from phase \RNum{1} to phases \RNum{3}, \RNum{4} and \RNum{5} are highly significant on weekdays and for phases \RNum{3} and \RNum{4} significant on weekends with a p<0.01 and p<0.05 significance level. For the second presented recreational area only the weekday change from phase \RNum{1} to \RNum{3} is significant.
\begin{table}[h]
	\centering
	\caption{A Mann Whitney U test is applied to test for significance. We compare the ratio of female over male count in an area of one phase against another. The following rules  were  used when assigning the significance stars: *$<$0.05, **$<$0.01, ***$<$0.001.}    
	
\begin{tabular}{llll}
\hline
                & \hphantom{end} &       Weekday &      Weekend \\
Point of Interest & compared time periods &               &              \\
\hline
Recreational Area 1 & I vs. III &  9.60e-08 *** &  7.66e-03 ** \\
                & I vs. IV &  6.85e-05 *** &   1.92e-02 * \\
                & I vs. V &  6.63e-08 *** &    2.81e-01  \\
                & III vs. IV &     8.90e-02  &    3.18e-01  \\
Recreational Area 2 & I vs. III &   1.68e-03 ** &    4.85e-01  \\
                & I vs. IV &     7.24e-02  &    4.79e-01  \\
                & I vs. V &     2.03e-01  &    4.86e-01  \\
                & III vs. IV &     1.59e-01  &    2.54e-01  \\
Shopping Center & I vs. III &  4.04e-07 *** &    1.40e-01  \\
                & I vs. IV &  2.75e-08 *** &    1.11e-01  \\
                & I vs. V &  3.23e-04 *** &    4.59e-01  \\
                & III vs. IV &  4.72e-04 *** &   3.63e-02 * \\
\hline
\end{tabular}	
	\label{t:significance pi}
\end{table}

\subsection*{SI Text S11: Recreational areas} 

\begin{figure}[htbp]
	\centering
	\includegraphics[width=0.6\linewidth]{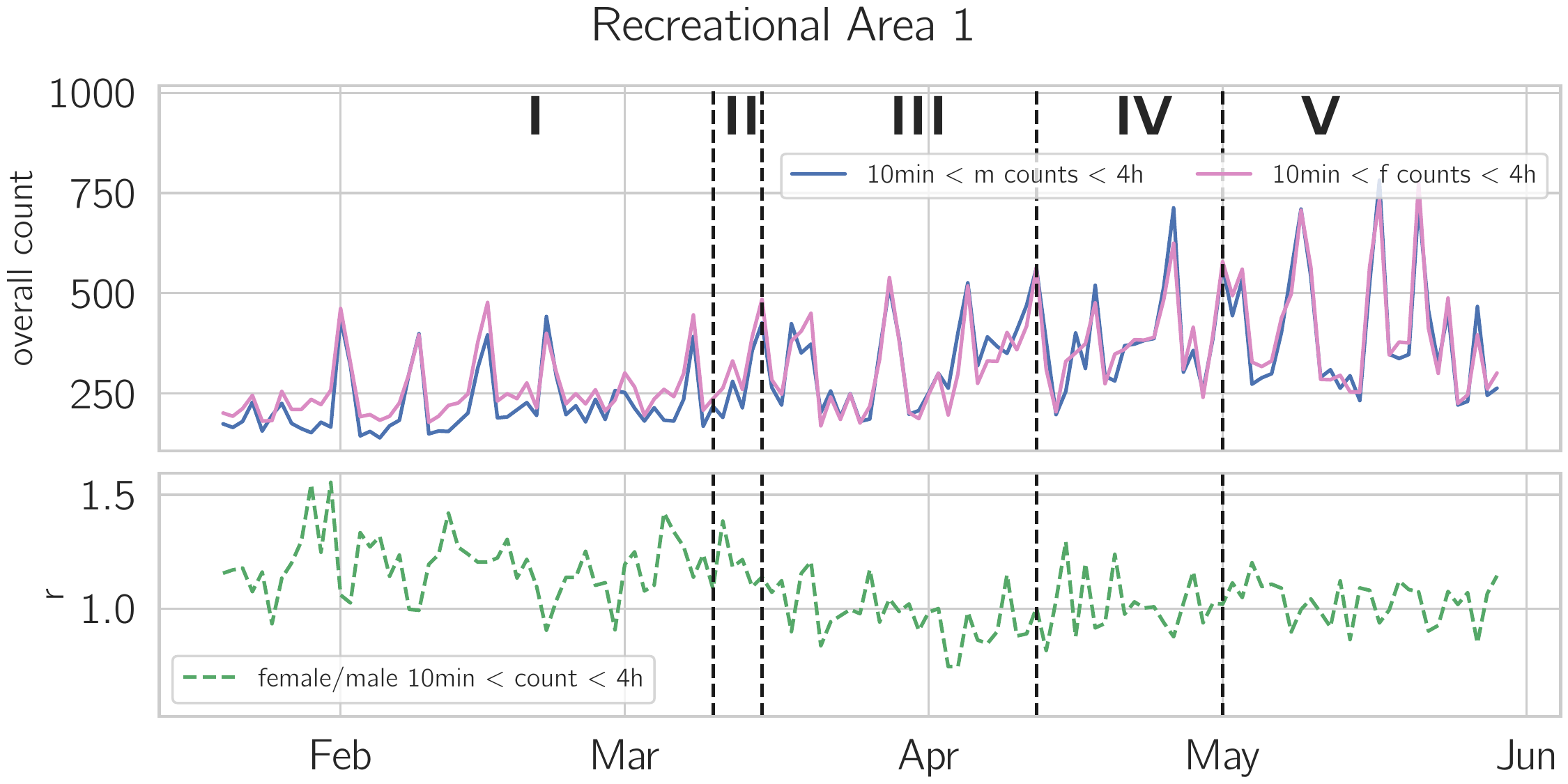}
	\caption{Visitors in a leisure area outside of Vienna, \emph{Kahlenberg}, during the Covid-19 crisis. The upper panel shows the counts of men and women present in the defined area. The lower panel shows the gender ratio of the counts. The overall counts are unaffected from the lock-down, but the gender ratio changes from being from female-biased to equality.}
	\label{fig:kahlenberg2020}
\end{figure}

\begin{figure}[htbp]
	\centering
	\includegraphics[width=0.6\linewidth]{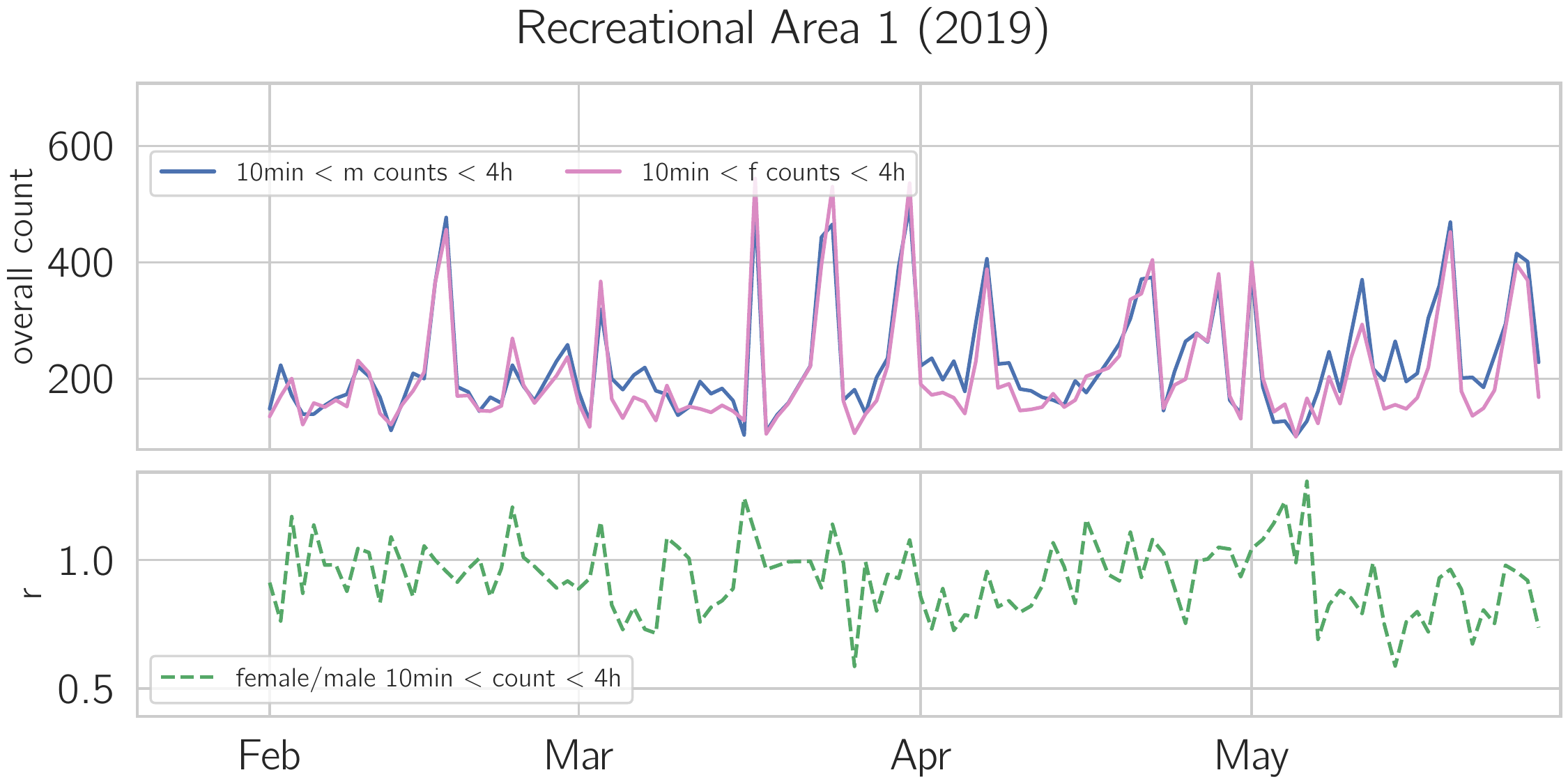}
	\caption{Visitiors in the same leisure area as in Fig. \ref{fig:kahlenberg2020}, but for the same time period in 2019. The upper panel shows the number of men and women present, the lower panel the gender ratio women over men $r$. We find a weakly increasing trend towards summer, but the gender ratio stays constant throughout the study period. Note that in contrast to 2020 the gender ratio balanced in February and March.
	}
	\label{fig:kahlenberg2019}
\end{figure}

Figure \ref{fig:kahlenberg2020} shows the recreational area described in the main text. For comparison we also show the relevant time period in the previous year, 2019, in Fig. \ref{fig:kahlenberg2019}. The limited comparability due to improvements in the network localization can be seen in the lower number of visitors, compared to 2020. However, the increasing trend towards summer is clearly present. The gender ratio is different to 2020, as it is more shifted towards men, however there is no stark change around march 15\textsuperscript{th}, supporting the claim that the change from phase \RNum{1} to  \RNum{3} is unique to 2020.

\begin{figure}[htbp]
	\centering
	\includegraphics[width=0.6\linewidth]{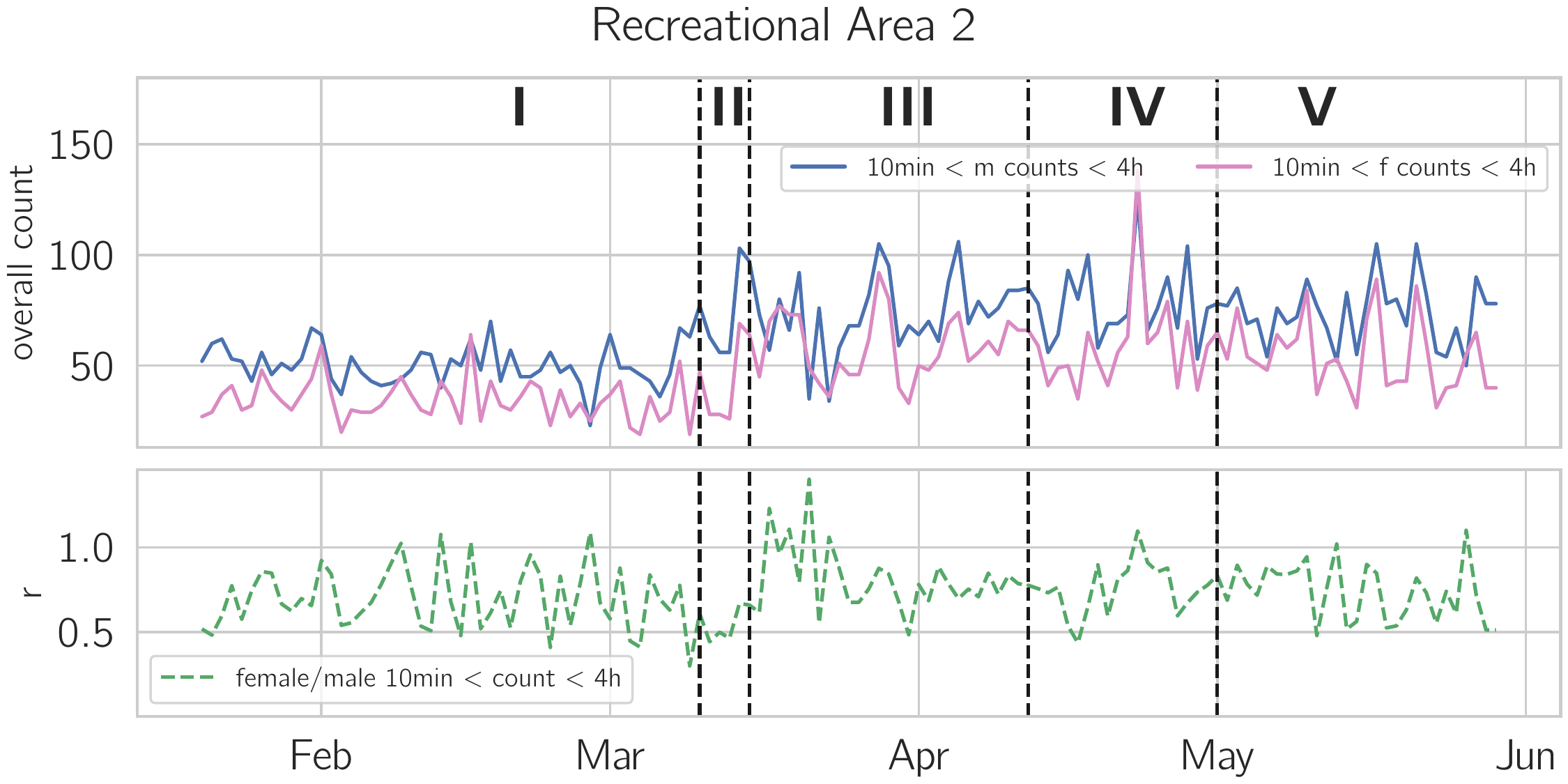}
	\caption{Visitors in another leisure area outside of Vienna, \emph{Lobau}, during the Covid-19 crisis. The upper panel shows the counts of men and women present in the defined area. The lower panel shows the gender ratio of the counts. The overall counts increase during phase \RNum{2} and stays high until the end of the study period. The gender ratio changes from being from male-biased to being more balanced in phase \RNum{3}.
	}
	\label{fig:lobau2020}
\end{figure}

\begin{figure}[htbp]
	\centering
	\includegraphics[width=0.6\linewidth]{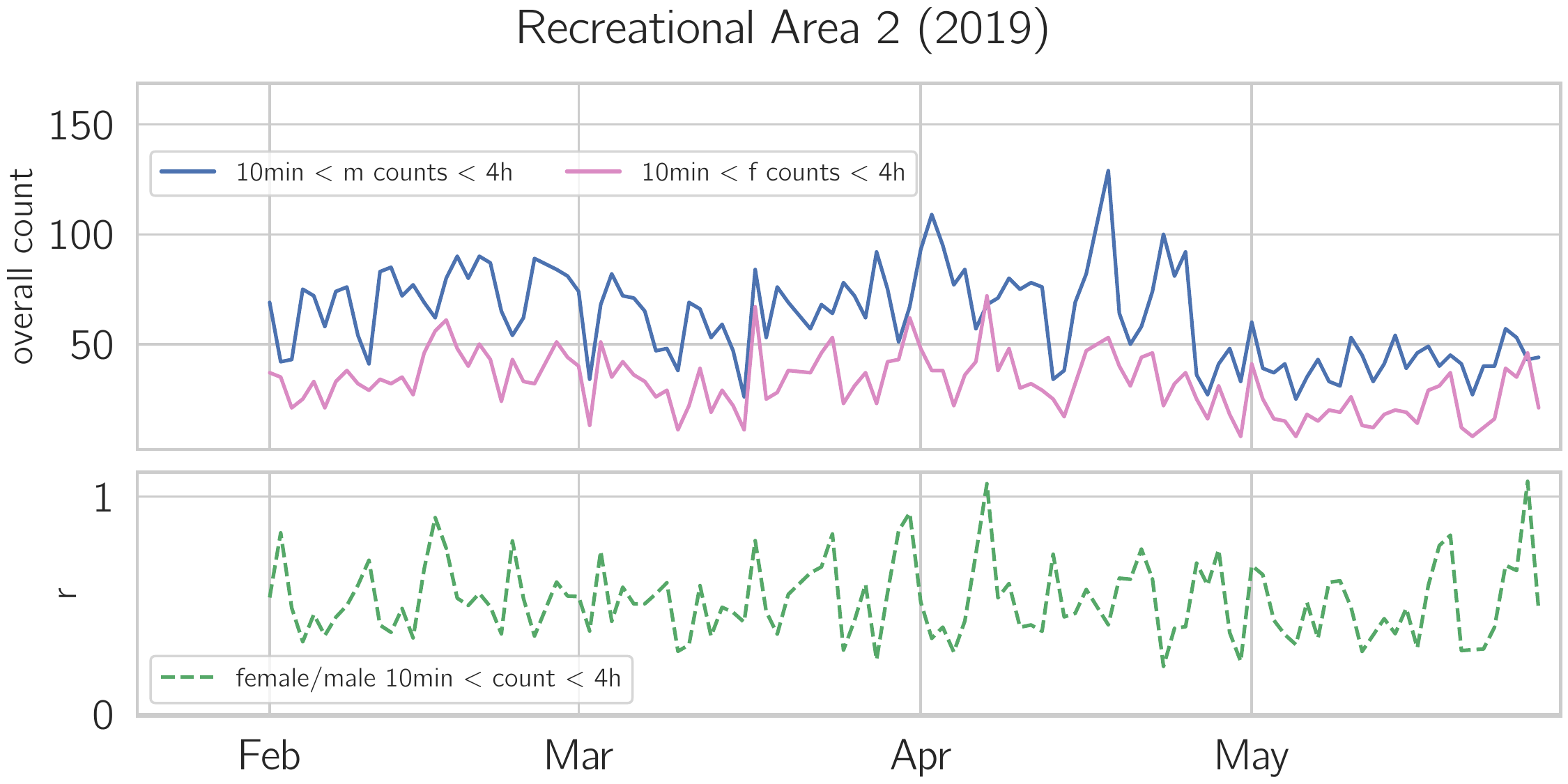}
	\caption{
		Visitiors in the same leisure area as in Fig. \ref{fig:lobau2020}, but for the same time period in 2019. The upper panel shows the number of men and women present, the lower panel the gender ratio women over men $r$. We do not find an increase in the middle of March, and also gender ratio stays the same throughout the study period.}
	\label{fig:lobau2019}
\end{figure}

To corroborate these findings we analyze a second recreational area on the other side of Vienna, presented in Fig.  \ref{fig:lobau2020}. There the gender ratio in phase \RNum{1} is typically shifted towards more men present. In the beginning of the lock-down, the gender ratio shifts towards equality, while at the same time more devices are present. A comparison with the previous year, 2019, in Fig.  \ref{fig:lobau2019} shows that these changes are specific to 2020.

\subsection*{SI Text S12: Circadian rhythm as observed by other quantities}

We investigate the number of calls, down- and uploaded gigabytes per hour as network traffic measures to corroborate the results presented for the call duration in the main text. 
All investigated quantities show similar patterns, the maximum of the daily activity shifts to the morning, the FWQM gets shorter and the days get more synchronized.

\begin{figure*}[h]
	\centering
	\includegraphics[width=0.39\textwidth]{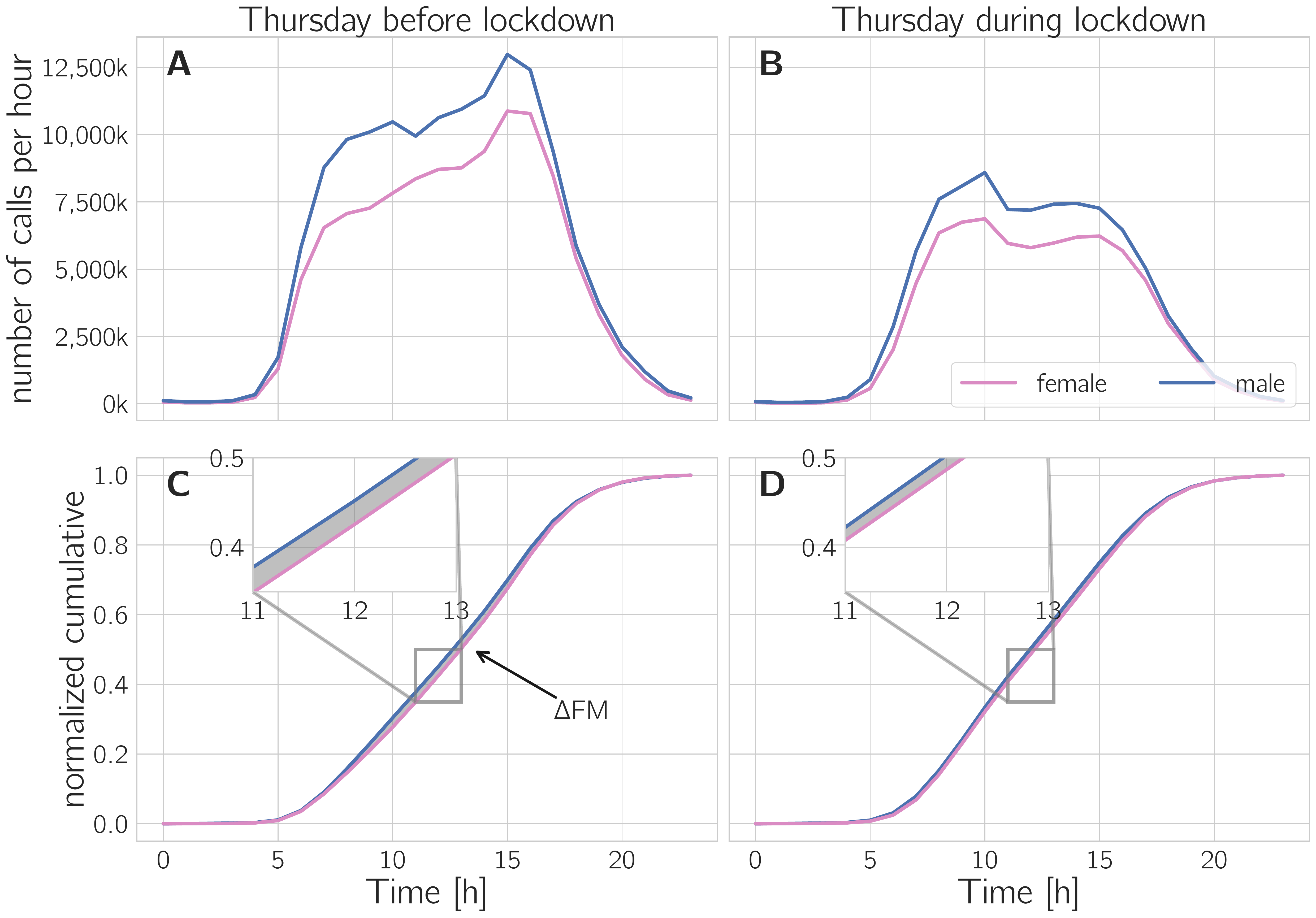} 
	\includegraphics[width=0.39\textwidth]{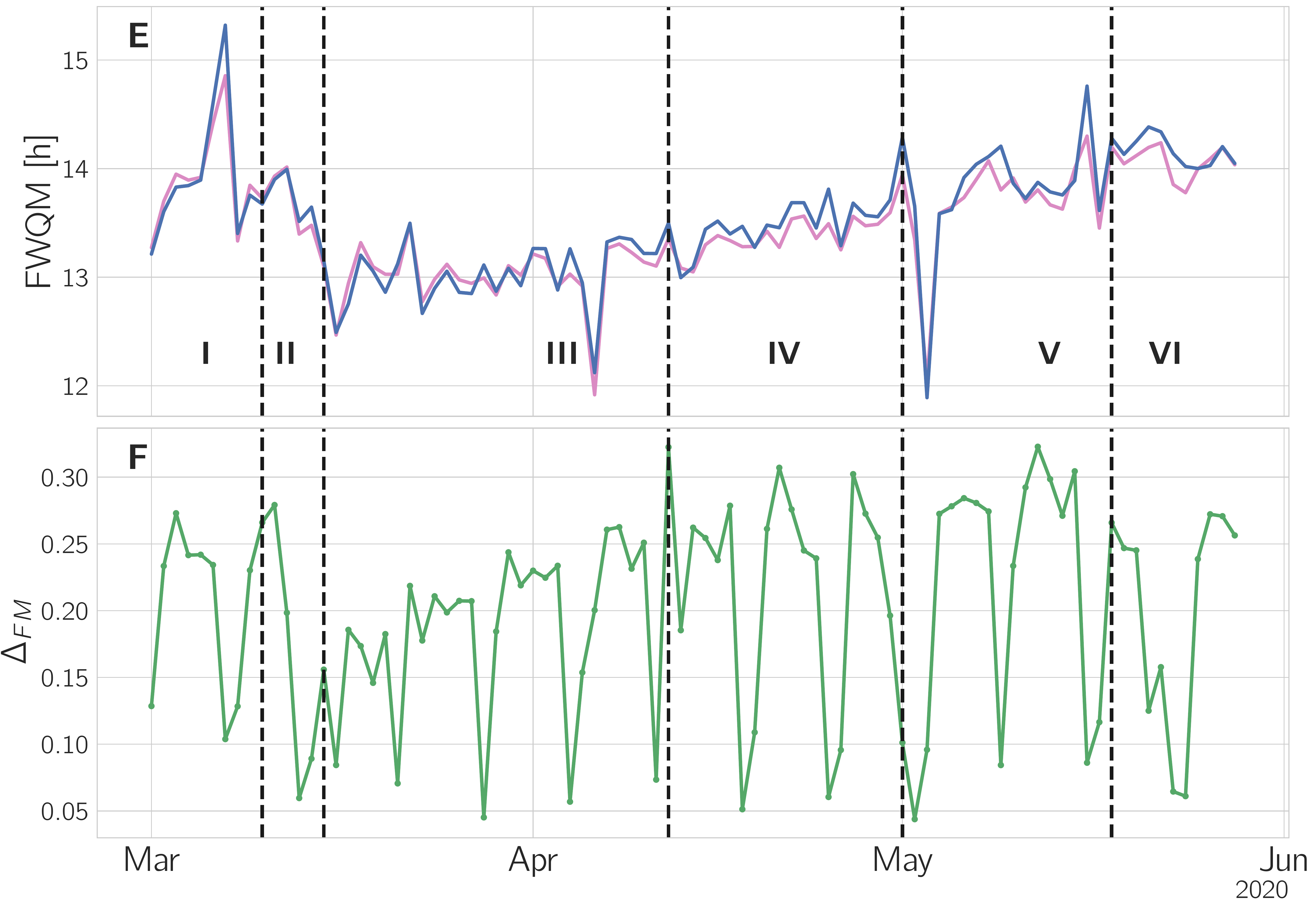}
	\caption{Changes in circadian rhythms during the lock-down measured by number of calls per hour. (\textbf{A}) Network traffic on the last Thursday in phase \RNum{1}, March 4\textsuperscript{th}. The horizontal arrow marks the full-width-quarter-maximum length (FWQM). 
		(\textbf{B}) Calls per hour on Thursday, March 18\textsuperscript{th}, the first Thursday in phase \RNum{3}.
		(\textbf{C}) Normalized cumulative activity for the day shown in panel A. The inset highlights the difference of the male and female curve. The grey shaded area marks the difference between the circadian rhythm of men and women, denoted $\Delta_{FM}$. (\textbf{D}) Same as in C, but for the curve in panel B. 
		(\textbf{E}) FWQM for men and women over time. 
		(\textbf{F}) The gender ratio $r_{FWQM}$ does not deviate strongly from equality, hence, we show
		$\Delta_{FM}$ over time. For both genders the activity maximum shifts from late afternoon to morning and the length of the activity period is approximately 60 min shorter during the lock-down. The reduction in $\Delta_{FM}$ indicates that the circadian rhythms of men and women become more synchronized.
	}
	\label{fig:sleep_calls}
\end{figure*}

We present the results for the call count per hour in Fig.  \ref{fig:sleep_calls}.
Panel E epicts how the length of day as measured by FWQM reduces by one hour from 14h before to 13h during the lock-down. 
Again, the gender ratio of FWMQ does not show any significant differences.
The difference in daily circadian rhythm, $\Delta_{FM}$, reduces by about one quarter from phase \RNum{1} to phase \RNum{3} and smoothly recovers until phase \RNum{4}, see panel F.

\begin{figure*}[h]
	\centering
	\includegraphics[width=0.39\textwidth]{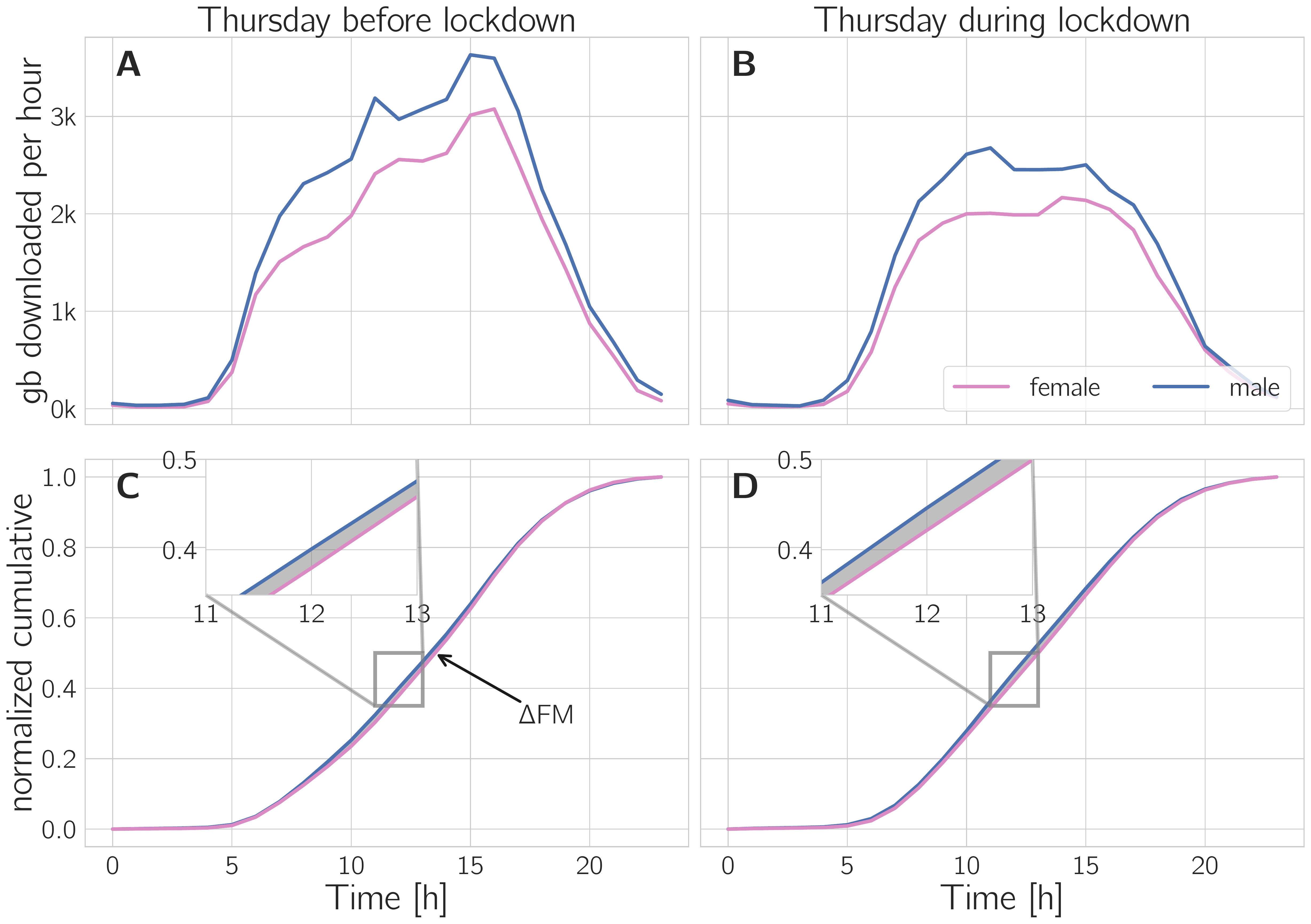} 
	\includegraphics[width=0.39\textwidth]{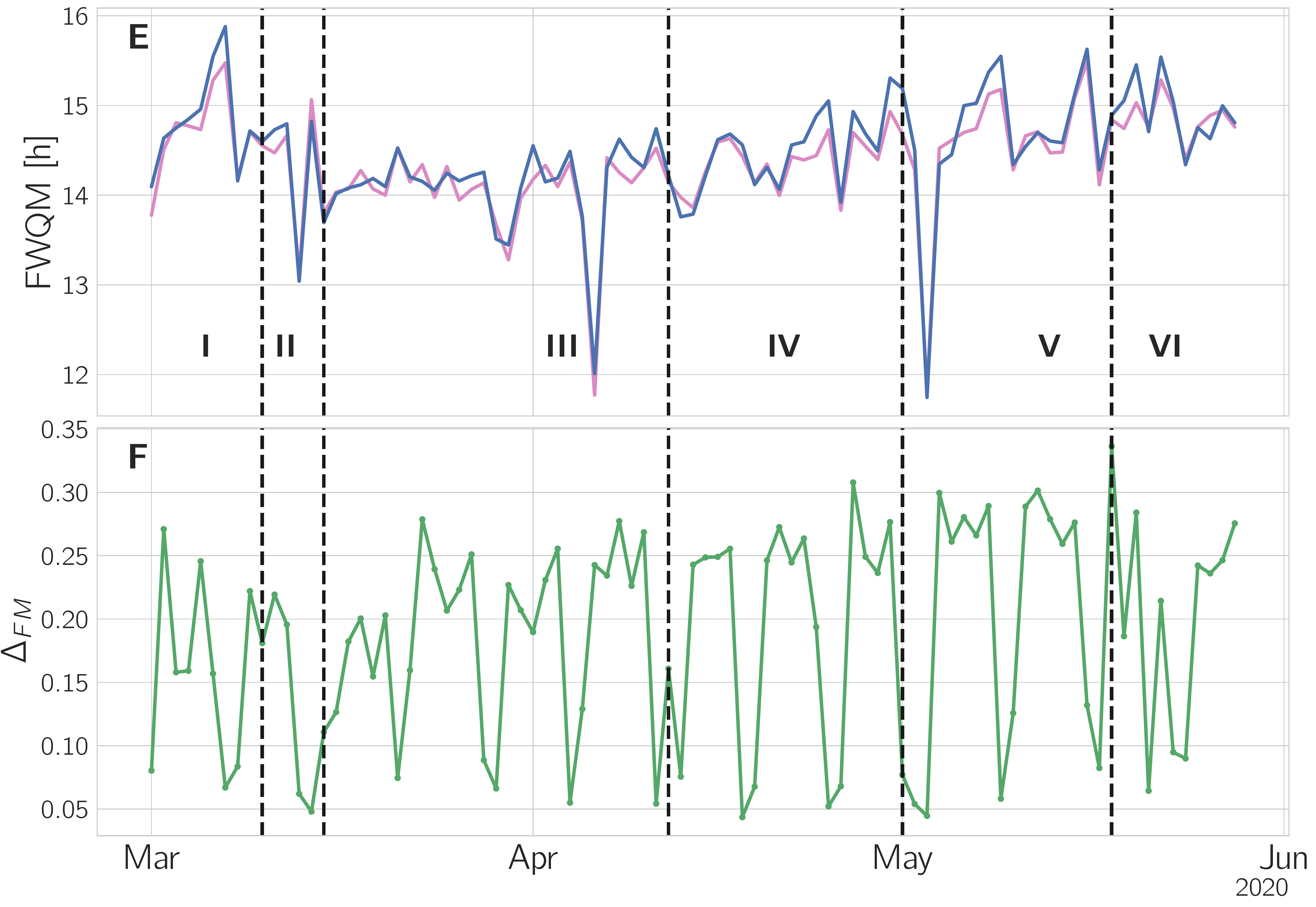}
	\caption{Changes in circadian rhythms during the lock-down measured by downloaded gigabytes per hour. (\textbf{A}) Network traffic on the last Thursday in phase \RNum{1}, March 4\textsuperscript{th}. The horizontal arrow marks the full-width-quarter-maximum length (FWQM). 
		(\textbf{B}) Downloaded gigabytes per hour on Thursday, March 18\textsuperscript{th}, the first Thursday in phase \RNum{3}.
		(\textbf{C}) Normalized cumulative activity for the day shown in panel A. The inset highlights the difference of the male and female curve. The grey shaded area marks the difference between the circadian rhythm of men and women, denoted $\Delta_{FM}$. (\textbf{D}) Same as in C, but for the curve in panel B. 
		(\textbf{E}) FWQM for men and women over time. 
		(\textbf{F}) As $r_{FWQM}$ does not deviate strongly from equality, we display
		$\Delta_{FM}$ over time. 
		For both genders the activity maximum in the late afternoon disappears and the length of the activity period is approximately 40 min shorter during the lock-down. A slight reduction in $\Delta_{FM}$ indicates that the circadian rhythms of men and women become more synchronized.
	}
	
	\label{fig:sleep_gb_down}
\end{figure*}

Figure \ref{fig:sleep_gb_down} shows our results on the downloaded gigabytes per hour.
Panels A and B depict how the activity peak in the evening disappears during the lock-down and is only weakly pronounced in the mornings for men. For women the download traffic in panel B does not show a pronounced maximum during the lock-down.
The downloaded gigabyte per hour in \ref{fig:sleep_gb_down} E show a reduction in FWQM by 40min from 14h 45min to 14h 5min. 
We only find a slight decrease in $\Delta_{FM}$ in the first week of phase \RNum{3}, otherwise the differences in the daily activity patterns of men and women stay constant.

\begin{figure*}[h]
	\centering
	\includegraphics[width=0.39\textwidth]{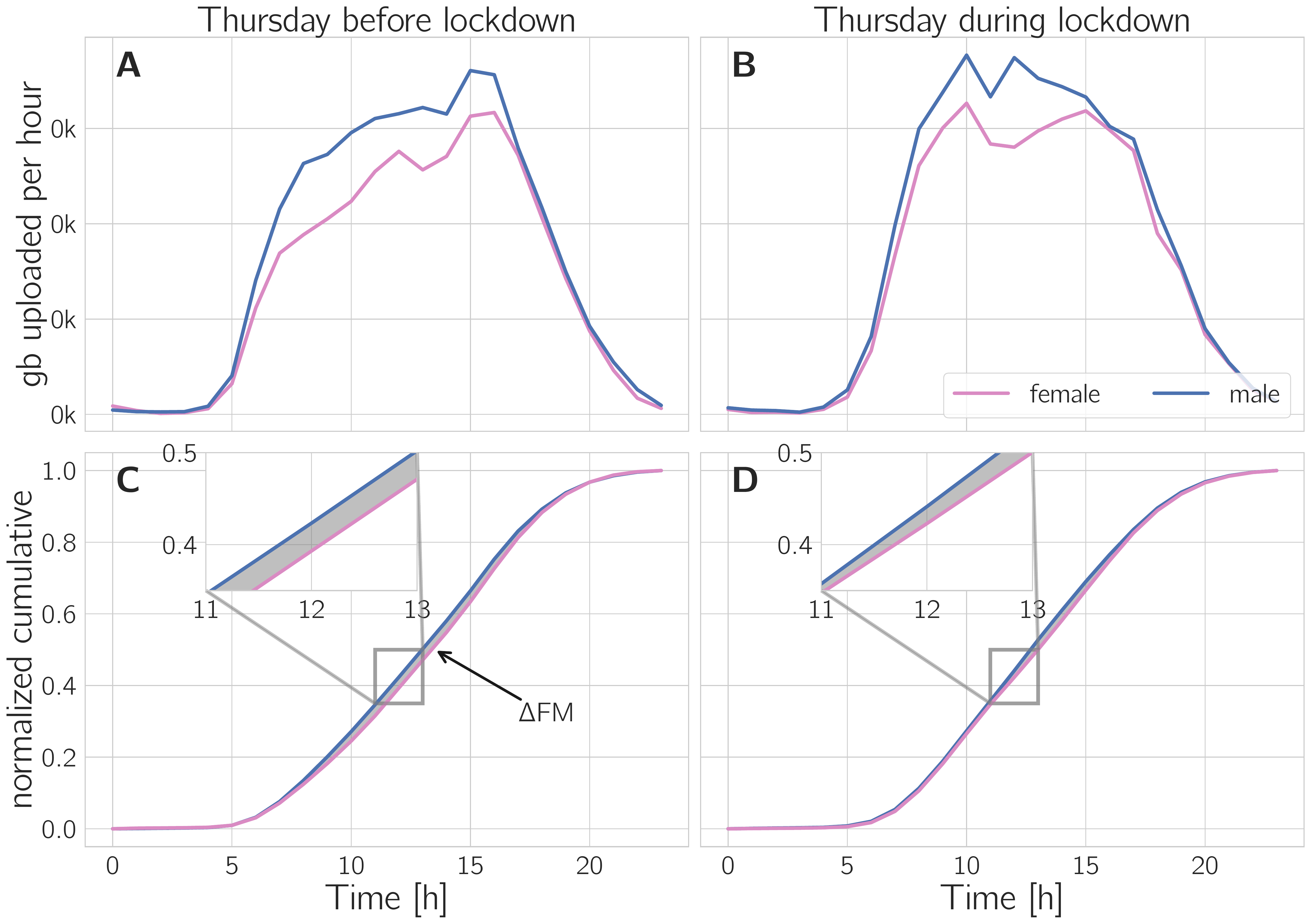} 
	\includegraphics[width=0.39\textwidth]{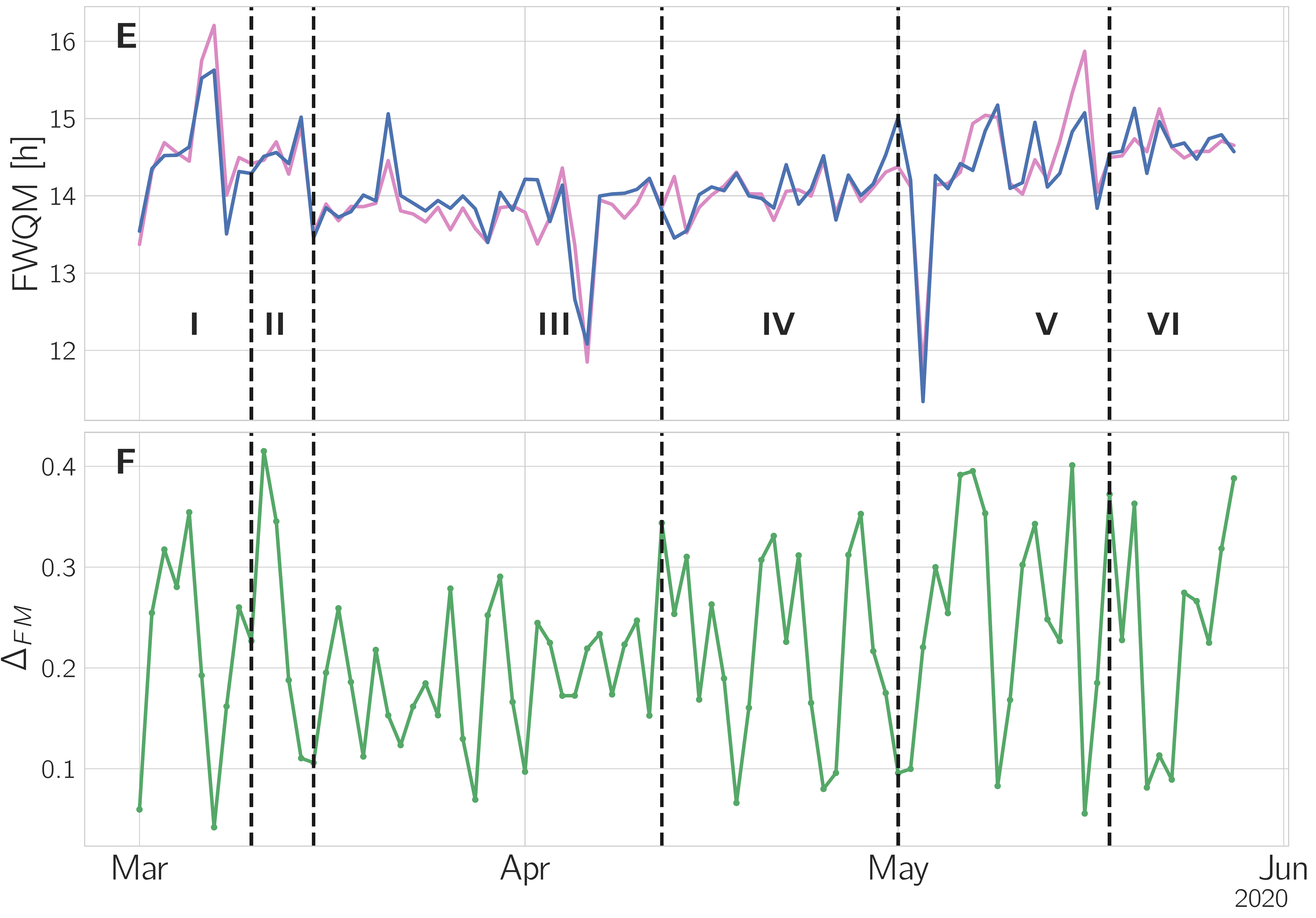}
	\caption{Changes in circadian rhythms during the lock-down measured by uploaded gigabyte per hour. (\textbf{A}) Network traffic  on the last Thursday in phase \RNum{1}, March 4\textsuperscript{th}. The horizontal arrow marks the full-width-quarter-maximum length (FWQM). 
		(\textbf{B}) Uploaded gigabytes per hour on Thursday, March 18\textsuperscript{th}, the first Thursday in phase \RNum{3}.
		(\textbf{C}) Normalized cumulative activity for the day shown in panel A. The inset highlights the difference of the male and female curve. The grey shaded area marks the difference between the circadian rhythm of men and women, denoted $\Delta_{FM}$. (\textbf{D}) Same as in C, but for the curve in panel B. 
		(\textbf{E}) FWQM for men and women over time. 
		(\textbf{F}) As $r_{FWQM}$ does not change much from one, we show $\Delta_{FM}$ over time. For both genders the activity maximum shifts from late afternoon to morning and the length of the activity period is approximately 45 min shorter during the lock-down. The reduction in $\Delta_{FM}$ indicates that the circadian rhythms of men and women become more synchronized.
	}
	\label{fig:sleep_gb_up}
\end{figure*}

Finally we consider the uploaded gigabytes per hour in Fig.  \ref{fig:sleep_gb_up}. The profile of the network traffic in panels A and B shift from a clear maximum in the evening to a stronger increase in the morning which results in a relatively flat profile during the day in panel B. As shown in panel E the FWQM drops by 45min from 14h 33min to 13h 48min. Panel F shows that $\Delta_{FM}$ drops from by approximately one third with the beginning of the lock-down and returns to pre-crisis levels around the beginning of May.

\end{document}